\newcommand{\aeff}{A_{\lambda}^{\rm e}}
\newcommand{\aeffs}{A_{\lambda,s}^{\rm e}}
\newcommand{\abeff}{A_{B}^{\rm e}}
\newcommand{\aieff}{A_{I}^{\rm e}}
\newcommand{\daeff}{\Delta A_{\lambda}^{\rm e}}

\documentclass[apj]{emulateapj}

\usepackage{amsmath}
\usepackage[FIGTOPCAP]{subfigure}

\begin{document}
\title{The 3-Dimensional Distribution of Dust in NGC 891}

\author{
Andrew Schechtman-Rook\altaffilmark{1}, Matthew A. Bershady\altaffilmark{1},
Kenneth Wood\altaffilmark{2,1}
}
\submitted{Submitted to ApJ 2011 August 8; accepted 2011 December 6}
\altaffiltext{1}{University of Wisconsin, Department of Astronomy, 475
N. Charter St., Madison, WI 53706; andrew@astro.wisc.edu}

\altaffiltext{2}{School of Physics and Astronomy, University of St. Andrews,
  North Haugh, St. Andrews, Fife, KY16 9AD, UK}


\begin{abstract}
  We produce three-dimensional Monte-Carlo radiative transfer models
  of the edge-on spiral galaxy NGC 891, a fast-rotating galaxy thought
  to be an analogue to the Milky Way. The models contain realistic
  spiral arms and a fractal distribution of clumpy dust. We fit our
  models to {\it Hubble Space Telescope} images corresponding to the B
  and I bands, using shapelet analysis and a genetic algorithm to
  generate 30
  statistically best-fitting models. These models have a
  strong preference for spirality and clumpiness, with average face-on
  attenuation decreasing from 0.24(0.16) to 0.03(0.03) mag
  in the B(I) band between 0.5 and 2 radial scale-lengths. Most of the
  attenuation comes from small high-density clumps with low
  ($\lesssim$10\%) filling factors. The fraction of dust in clumps is broadly
  consistent with results from fitting NGC 891's spectral energy distribution. Because of scattering effects and the intermixed nature of the dust and
  starlight, attenuation is smaller and less wavelength-dependent than
  the integrated dust column-density. Our clumpy models
  typically have higher attenuation at low inclinations than
  previous radiative transfer models using smooth distributions of
  stars and dust, but similar attenuation at inclinations
  above 70$^\circ$. At all inclinations most clumpy models have less
  attenuation than expected from previous estimates based on
  minimizing scatter in the Tully-Fisher relation. Mass-to-light
  ratios are higher and the intrinsic scatter in the Tully-Fisher
  relation is larger than previously expected for galaxies similar to NGC 891.  The attenuation curve changes as a function of
  inclination, with $R_{B,B-I}=\frac{A_{B}}{E(B-I)}$ increasing by $\sim$0.75
  from face-on to near-edge-on orientations. 
\end{abstract}

\section{Introduction}
Understanding the 3D structure of dust is crucial for studies of
spiral galaxies. In addition to detailing the complex structure of the
ISM an accurate representation of dust is necessary to correct
measurements of starlight. This is important for a variety of studies
including e.g., the Tully-Fisher (TF) relation, where the behavior of the
dust as a function of inclination is required to correct integrated
photometry \citep{Verheijen01b}; constructing rotation curves of highly
inclined galaxies, since attenuation can censor the rotation curve's
inner rise, leading to erroneous mass models \citep{Matthews01};
correcting measurements of disk kinematics for the asymmetric amount
of dust extinction above and below the disk of face-on spirals
\citep{Bershady10a,Bershady10b}; and producing accurate radial
surface-brightness profiles corrected for spurious broken-exponential
morphologies caused by dust absorption \citep{deJong96}.  Spirals with
inclinations at or near 90$^{\circ}$ (known as edge-ons) provide the
best information about the vertical structure of the dust, both
because the corrections for viewing angle are small but also because
the dust is projected into easy to see lanes that can attenuate the
midplane by more than 10 mag in the optical
\citep{Kylafis87}. As a result edge-ons are an important component to
our knowledge of dust structure. Separating the individual components
of a dust-light mixture of unknown composition is a difficult
endeavor, and is frequently done via radiative-transfer (RT)
modeling.

Monte Carlo (MC) RT models, used in astrophysics since the 1970s (for a
discussion of early models see \citealt{Witt77}), have been
extensively employed to probe the structure of edge-on spirals. Recently,
these models have come into prominent usage
(e.g. \citealt{Bianchi96,Kuchinski98,Matthews01,Baes03,Baes11}), spurred by
the need to precisely track multiple scatterings off dust grains.
While costlier than the direct analytical approach of \cite{Kylafis87}
in terms of CPU time, increases in processor speed and RAM limits
coupled with rapidly dropping costs now allow detailed Monte Carlo
models to be run on relatively short timescales. In addition, the
current trend in high-performance computing is towards increased
numbers of CPUs at slower clock speeds; this is also a benefit to
Monte Carlo methods, as they are exceedingly well suited to
parallelization and thrive in a distributed environment.

NGC 891 is the most studied nearby (d$\approx$9.5 Mpc) edge-on spiral galaxy in
the universe. It is almost exactly edge-on ($i\approx 89.7^{\circ}$
\citep{Xilouris98}), with a rotation speed (212 km/s) similar to that
of the Milky Way. Because the dust morphology of edge-ons appears to
change significantly between galaxies with rotation speeds above and
below 130 km sec$^{-1}$ \citep{Dalcanton04}, comparing the Milky Way
with other fast-rotating systems is especially relevant. NGC 891 is
also one of the first galaxies discovered to have high-latitude HI, a
discovery prompted by observations of high-latitude HI in our own
galaxy \citep{Gerard73,Strong78}. Indeed, NGC 891's potential as an
analogue of the Milky Way is what aroused much of the early interest
in this extragalactic system \citep{Bahcall83,vanderKruit84}. More
recently, evidence for a strong two-armed spiral pattern like that
seen in grand design spirals such as M51 has emerged based on
asymmetry in H$\alpha$ and B band emission
\citep{Kamphuis07,Xilouris98}. NGC 891 appears to have a bar
\citep{GarciaBurillo95} as well as a nearby companion (UGC 1807,
\citealt{Mapelli08}), both of which dramatically increase the
likelihood of the aforementioned grand design pattern being real
\citep{Elmegreen82}.

NGC 891 is also known for its abundance of high-latitude dust and
complex dust substructure. Given the significant high-latitude HI
found in NGC 891, it is not surprising to also find dust away from the
midplane. \cite{Howk97} used unsharp masks to study individual
high-latitude dust clumps in NGC 891, and later found evidence that
these extended substructures may be common to most spirals
\citep{Howk99}. If high-latitude dust caused by star-formation induced
outflows is indeed abundant in spiral galaxies, it would act as a
foreground screen preferentially over the star-forming spiral arms,
increasing the apparent optical depth in the arms without increasing total
dust content. While unsharp masks are very useful for enhancing clumpy
substructure the masking process destroys quantitative information
about the clumps. Therefore, a new procedure for highlighting the dust
that preserves this information is needed.

Because there are some indicators of spirality and significant
extraplanar dust component, NGC 891 is a clear example of the need for
advanced 3D modeling which takes into account these features. Indeed,
studies attempting to use smooth, axisymmetric models on NGC 891 have
had significant difficulty in fitting the data due to the asymmetry in
blue light. \cite{Xilouris98} split the galaxy in half and used an
infinitely long thin disk of stellar emission to reconcile the left
and right hand sides of NGC 891 in the B and V bands, while additional
dust and light components have been required by smooth models to fit
NGC 891's mid-infrared emission \citep{Popescu00,Bianchi08}. Some
groups have added clumpy/non-axisymmetric structure to RT models of
edge-on spirals
(e.g. \citealt{Kuchinski98,Mihos99,Matthews01,Misiriotis02,Bianchi08});
however they restrict their analysis to a small range of clumpy models,
with dust modeled by hand or based on results for the Milky Way.

In this work we quantitatively fit the {\it first} RT models to
include both dust clumping and realistic spirality to F450W and F814W
Hubble images of NGC 891. These RT models resemble real spiral
galaxies from any inclination angle, allowing us to make accurate
inferences about the dust properties of galaxies like NGC 891 as seen
at all inclinations. In Section \ref{sec:obsconsts} we present the
archival data used in this work and our methods of analysis. In
Section \ref{sec:model} we discuss the parameter space of our model,
our fitting algorithm, and our fitness metric. We present our results,
including a brief {\it post-facto} analysis of our shapelet-based
fitness metric, in Section \ref{sec:results}. In Section
\ref{sec:discussion} we discuss the photometric properties of our
models and compare them to the literature. Finally, in Section
\ref{sec:conclusion}, we present our concluding remarks.

\section{Observational Constraints}
\label{sec:obsconsts}
\subsection{Data}
We downloaded HST WFPC2 images of the central region of NGC 891 from
the Hubble Legacy Archive\footnote{Based on observations made with the
  NASA/ESA Hubble Space Telescope, and obtained from the Hubble Legacy
  Archive, which is a collaboration between the Space Telescope
  Science Institute (STScI/NASA), the Space Telescope European
  Coordinating Facility (ST-ECF/ESA) and the Canadian Astronomy Data
  Centre (CADC/NRC/CSA).} in the F450W and F814W bandpasses (hereafter
referred to a B and I band, respectively). For S/N per pixel $\approx$
10 the data go down to a limiting mag of $\sim$18.5 STMAG\footnote{Defined as
  ${\rm STMAG}=-2.5\log (F_{\lambda}) - 21.10$, where $F_{\lambda}$ has units of erg
  cm$^{-2}$ s$^{-1}$ ${\rm\AA}^{-1}$\citep{Sirianni05}.} 
mag arcsec$^{-2}$ for both B and I bands. The images were
rotated in order to make the galaxy midplane horizontal, and then
smoothed to a resolution of $\sim$0.9'' (roughly 42 pc at the adopted
distance of 9.5 Mpc) to match the resolution of the models. Due to the
large size of the galaxy on the WFPC2 images the sky had been
oversubtracted in the reduced, archival data. We corrected the sky
subtraction while scaling and centering the images by comparing the
data to smooth, axisymmetric RT models constructed exactly from the B
and I band parameters given in \citet{Popescu00}. This procedure was
iterative: starting from a guess of the offsets, we computed a linear
least-squares fit between individual pixels from the model images and
their analogues in the shifted data frames. The best-fitting shift is
the one required to center the image, while the slope and y-offset of
the best fit are the flux calibration and background offset,
respectively.  We then masked out bright foreground stars and
background galaxies from the images. These masks were homogenized
between the B and I bands, then preserved so we could identically mask
all of our models before fitting them to the data.

\subsection{Metrics of non-axisymmetric systems with clumpy dust distributions}

The aim of this study is to determine the clumpy and non-axisymmetric
distribution of star-light and dust in NGC 891. To do so, we introduce
two new metrics to characterize the data and our models.  These
metrics are designed to evaluate the performance of these models in
matching the data compared to axisymmetric models with smooth
star-light and dust distributions.  First, we motivate the concept of
attenuation, from which we construct a differential
index of the attenuation relative to a smooth model. This
index allows us to leverage the most important observable differences
between smooth and clumpy distributions of dust. Second, with the use
of models with clumpy distributions of dust, it is no longer tractable
to find a model that looks identical to the data (as done in all prior
smooth model analyses).  We apply a method from the literature for
making orthogonal decompositions of two-dimensional light
distributions. These characterizations are used to statistically
compare and contrast the observed and modeled differential attenuation maps.

\begin{figure*}
\begin{centering}
  \includegraphics[scale=0.9]{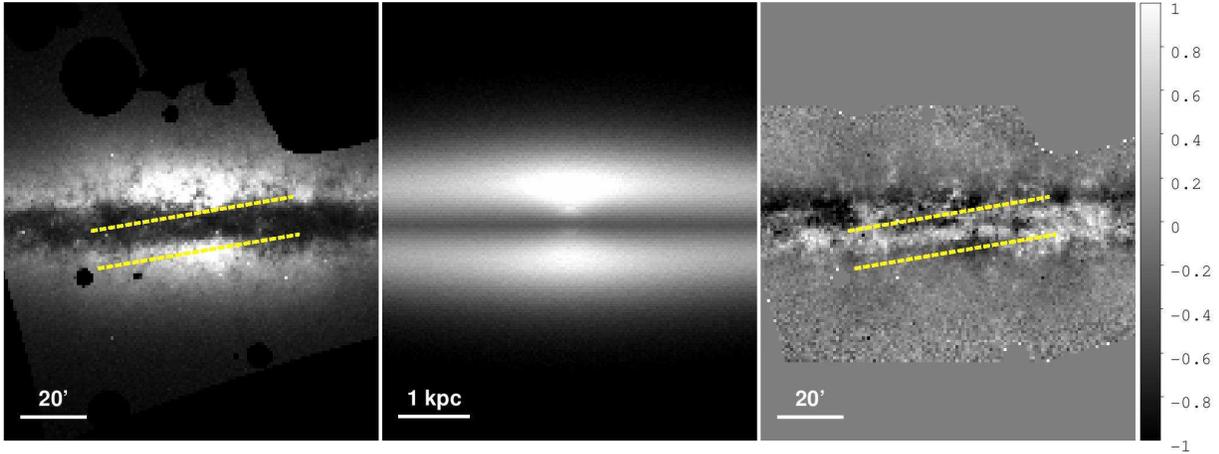}
  \caption{Demonstration of the $\Delta \aieff$ map. Left: smoothed I
    band HST image of NGC 891. Center: smooth model from
    \citet{Popescu00}. Right: $\Delta \aieff$ map. The color-map
    shows the differential attenuation in mag. The clumpy, non
    axisymmetric structures have been highlighed by the $\Delta \aieff$
    map while smooth structure (e.g. high-latitude bulge light) has been
    removed. Of special note is a line of increased attenuation running across
    the center of the galaxy but at a non-zero angle to the midplane
    (bracketed by dashed yellow lines). This could be indicative of a warped
    outer spiral arm.}
\label{fig:idataextmapdemo}
\end{centering}
\end{figure*}

\subsubsection{Differential Attenuation Maps}

Whenever dust grains are mixed with a distribution of stellar
emissivity the traditional notion of dust as a purely foreground
screen no longer applies; instead what we observe for systems such as
galaxies is the {\it attenuation}
$\aeff$ at wavelength $\lambda$.  The attenuation can be
defined in analogy to the foreground extinction, $A_\lambda$, by the
ratio of the observed or modeled line-of-sight flux in the presence of dust ($F_\lambda$) to the
line-of-sight flux (of a model) in the absence of dust
($F_\lambda^0$):
\begin{equation}
\aeff \ \equiv \ -2.5\log\left ( \frac{F_{\lambda}}{F_{\lambda}^{0}} \right ).
\label{Aeffeq}
\end{equation}
In general, $\aeff$ is dependent on the underlying distribution of
emissivity, absorption, and scattering.  Consequently, $\aeff$ is {\it
  not} related simply to the dust optical depth, and hence cannot be
used to derive directly the underlying dust column density.

Previous radiative transfer modeling of galaxies which fit models to images (e.g., \citealt{Xilouris98}) have constrained smooth distributions of stars and dust by
fitting directly to the observed, two-dimensional light distribution
at different wavelengths\footnote{Clumpy radiative transfer models of galaxies
  are generally compared to SEDs
  (e.g. \citealt{Gordon97}; \citealt{Popescu11}), light or color
  profiles (e.g. \citealt{Kuchinski98}), or smooth radiative transfer models (e.g. \citealt{Misiriotis02}; \citealt{Pierini04})
  instead of directly to images.}. 
To first order, our clumpy dust models must
reproduce this basic two-dimensional light distribution {\it on
  average}. However, if we try to fit clumpy models directly to the
observed light distribution, the solution will be degenerate since
there will be many combinations of clumpy dust and star-light that
either (a) will result in the same observed intensity along a given
line of sight; or (b) will result in similar net deviations (in a
$\chi^2$ sense) from the observed data. Because of these degeneracies
(whereby a wide range of astrophysically distinct models yield similar
goodness of fit to the data), fitting clumpy models to the observed
light distribution is not particularly meaningful. Instead, what is
meaningful is to determine the similarity of the {\it statistical
  deviations} of the light-distributions of the data and models with
respect to some fiducial distribution.  Since we are interested in
understanding the importance of clumpy dust, the best fiducial
distribution is a model consisting of a smooth light and dust
distribution that best approximates the data.

Hence, instead of directly comparing our clumpy models to the data, we use
differential attenuation ($\daeff$) maps. Similar to the
attenuation $\aeff$, these images are created by taking the
ratio of the observed flux or non-axisymmetric model flux to the flux
distribution of a fiducial, axisymmetric, smooth model:
\begin{equation}
\Delta \aeff \ \equiv \ -2.5 \log\left ( \frac{F_{\lambda}}{F_{\lambda,s}} \right ),
\end{equation}
where $F_{\lambda,s}$ is the flux from the fiducial model. The
fiducial model is comprised of starlight and dust, and can be
parameterized in terms of its own attenuation $\aeffs$ and the
same model without dust ($F_{\lambda,s}^{0}$):
\begin{equation}
F_{\lambda,s} = F_{\lambda,s}^{0}\, 10^{-0.4\aeffs}.
\end{equation}
The differential attenuation can then be expressed as a
function of the underlying stellar distributions and the attenuation of the data/clumpy model and the smooth model:
\begin{equation}
\daeff = -2.5 \log\left(
  \frac{F_{\lambda}^{0}}{F_{\lambda,s}^{0}} \right ) + \aeff - \aeffs.
\label{deltaeffextblowout}
\end{equation}
If the fiducial model's light distribution, integrated along the
line-of-sight in the absence of dust, is a reasonable approximation to
the astrophysical object or clumpy model (also in the absence of
dust), then the first term is negligible.  In general the choice of
fiducial model is relatively unimportant as long as it is held
constant for the entire analysis since the mismatch can be treated as
a pedestal value to $\daeff$.  We choose the axisymmetric, smooth
model described in \citet{Xilouris98} and Table 1 of \citet{Popescu00}
as our fiducial. The observed $\daeff$ and the ingredients are shown
for NGC 891 in Figure \ref{fig:idataextmapdemo}. The left panel shows
a smoothed I band HST image, the center panel shows an image of the
smooth model created using our MC RT software, and the right panel
shows the ratio of the other two panels --- the $\Delta \aeff$ map.

\begin{figure*}
\begin{centering}
\includegraphics[scale=0.8]{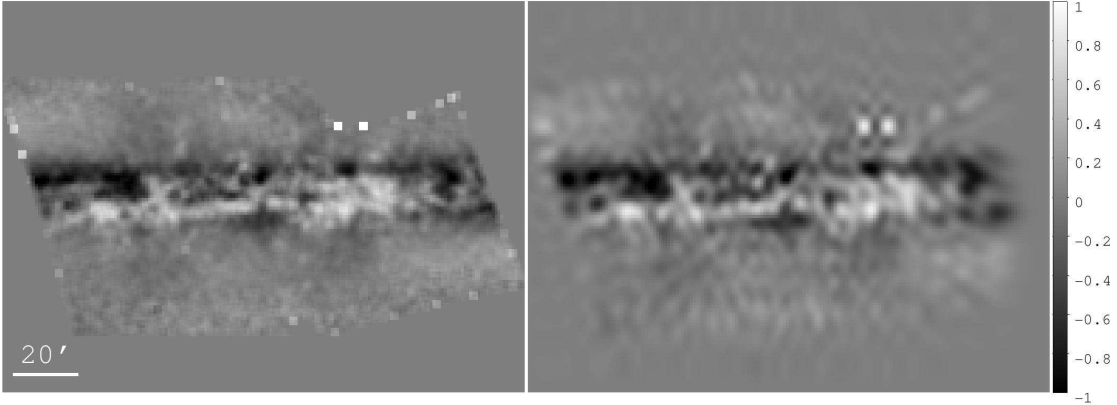}
\caption{Sample shapelet reconstruction of the observed I band $\Delta
  \aeff$ map. Left: $\Delta \aieff$ map. Right: reconstruction of
  $\Delta \aieff$ map using 50 orders in x and 25 orders in
  y. Color-bar indicates differential attenuation, in mag. The
  shapelet reconstruction effectively smooths the image with a boxcar
  of radius one pixel. Going to higher orders does not significantly
  improve this resolution, as it is impossible to perfectly
  reconstruct an image without using infinite orders.}
\label{fig:reconimg}
\end{centering}
\end{figure*}

It is important to note two things about the $\Delta \aeff$
maps. First, these maps do not represent the \textit{true} attenuation
because the axisymmetric models include a smooth dust disk: The
$\Delta \aeff$ map is really showing the excess or dearth of attenuation due to non-axisymmetric components (such as spiral arms and
clumps). For instance, the $\Delta \aieff$ map (and to a lesser
extent $\Delta \abeff$ map) shows a long, thin structure running
across the middle of the galaxy with greater than expected attenuation
(highlighted in Figure \ref{fig:idataextmapdemo}). This feature is
angled off of the midplane and resembles what would be expected from a
warped outer spiral arm. A warp in NGC 891 would help explain its
significant extraplanar gas emission from ionized and neutral gas
\citep{Rand90,Swaters97,Oosterloo07}, although \cite{Keppel91}
indicate that a warp is unlikely. By highlighting these structures the
$\Delta A_{\lambda\, eff}$ maps focus our model fitness parameters on
the higher order structure that we are searching to constrain.

Second, the $\Delta \aeff$ maps' ability to highlight dust
substructure is similar to the unsharp masks used by \citet{Howk99};
however, by using the attenuation instead of an unsharp mask
we are able to bring out the substructure \textit{without} destroying
the quantitative information in the image (a byproduct of any
algorithm involving smoothing). This is a significant advantage over
unsharp masking because it allows us to compute quantitative values
that can be directly compared to physical models.

\subsubsection{Shapelet Analysis}

The problem of identifying ways to characterize clumpy models and
compare them meaningfully to data has been taken up in the literature.
However, the solution to date has been to turn away from modeling the
observed spatial distribution of light (as done in all prior smooth
model analyses), and towards fitting the spectral energy distribution
(SED). The latter essentially compresses all the model images into a single dimension,
which can then be fit using a simple $\chi^{2}$ analysis. However this
method throws away all spatial information, which is essential for
testing physical models where stellar emission, absorption,
scattering, and thermal re-emission are not necessarily co-spatial
processes. By ignoring spatial information modelers also run the risk
of finding false minima, simulations that fit the SED but
do not look like real galaxies (for an example of this see
\citealt{Bianchi08}).

Another approach to constraining clumpy models which does not throw
away spatial information, is to construct a suitable statistical
descriptor of the galaxy image structure (here we use $\Delta \aeff$
for the structure image). One must find a way to measure the
statistical properties of the image that is able to equate two systems
with the same underlying {\it distribution} of structure. The human
eye is very good at this task, and ``by-eye'' image fitting has had
some success (e.g., \citealt{Matthews01}). However, in order to create
rigorous and well-constrained clumpy models a computerized,
quantitative tool to measure image structure is needed. As a first
step in this new direction we employ shapelet analysis to generate
model fitnesses.

Shapelet analysis, first described by \citet{Refregier03}, is a method for
decomposing an image into orthogonal basis functions using weighted Hermite
polynomials. A cousin of wavelets, shapelets are optimized for the more
circular features generally found in astronomy. The shapelet methodology
creates orthogonal basis functions based on Hermite polynomials:
\begin{equation}
\phi_{n}(x) = \frac{H_{n}(x)e^{-x^{2}/2}}{\sqrt{2^{n}n!\sqrt{\pi}}}
\end{equation}
where $H_{n}$ is the Hermite polynomial of order of positive integer
$n$. These 1D basis functions are combined to produce a dimensional 2D basis
function:
\begin{equation}
B_{n_{1},n_{2}}(x_{1},x_{2}; \beta) =\beta^{-1}\phi_{n_{1}}\left (\frac{x_{1}}{\beta}\right )\phi_{n_{2}}\left (\frac{x_{2}}{\beta}\right ).
\end{equation}
Here $\beta$ is a constant scaling factor used to scale the size of the
shapelets. The number of counts in an image pixel is then equal to the
infinite sum of the shapelet basis functions multiplied by the shapelet
coefficients, given by
\begin{equation}
f_{n_{1},n_{2}} = \int \int f(x_{1},x_{2})B_{n_{1},n_{2}}(x_{1},x_{2};\beta)\,\mathrm{d}x_{1}\,\mathrm{d}x_{2},
\end{equation}
where $f(x_{1},x_{2})$ are the counts for a given pixel. For a much
more thorough description see \citet{Refregier03}.  Traditionally
shapelets have been used to model spatially small galaxies at a low
number of fourier modes (e.g. \citealt{Kelly04,Massey04}; and
\citealt{Kuijken06}). We are trying to deconstruct objects on size
scales ranging from a few pixels (dust clumps) to the size of the
image (the bulge and disk) and therefore must go to much higher order
in shapelet space. We choose to go to 50th order along the major axis
and 25th order along the minor axis, based on a comparison of sample
reconstructions of the smoothed HST images. Going to these higher
orders allows us to capture much of the structural detail.  Exploring
such high order shapelets is possible because shapelet deconvolution
is a relatively efficient procedure, especially compared to the time
it takes to run our RT models. Additionally, parts of the process are
parallelizable (e.g. computing shapelet coefficients), a fact we take
full advantage of wherever possible. A sample reconstruction using the
selected orders is shown in Figures \ref{fig:reconimg} and
\ref{fig:shapeletrecon}. Reconstructing the image to less than
infinite orders imposes a penalty on image resolution; the
reconstructed image appears slightly smoothed, roughly equivalent to
boxcar-smoothing the image with a radius of one pixel. The
pixelization of the image, pixel-to-pixel noise, and the rapidly
increasing computational expense inhibit the use of higher
orders. Even though the shapelets we use are designed for circular
objects (cf. \citealt{Bosch10}) we find that they are able to
accurately reproduce the oblong features of edge-on spirals as well as
the clumpy dust.

\section{Modeling}
\label{sec:model}
Including both spirality and dust clumping adds many free parameters to
the RT models. Because many of these parameters are new to the modeling
literature, we cannot simply adopt values based on results from previous
studies. Consequently, our parameter space is very large
and so we employ a genetic algorithm to maximize our efficiency at finding
'good' solutions. Such an algorithm is well suited to comparing a statistical
goodness-of-fit constructed from the amplitude distribution of the shapelet
decomposition of $\Delta
\aeff$.

\subsection{The model}
We use the 3D scattered light Monte Carlo RT model developed by K. Wood, which has been
used in a variety of different astrophysical environments
(for example see \citealt{Wood96,Wood97,Wood99b,Wood00,Matthews01};
 and \citealt{Sankrit01}). We will summarize the relevant details
 here; for a full description of the code see \cite{Wood99a}. 

The Monte Carlo model tracks individual packets of photons through a model
consisting of a 3D Cartesian grid, where each cell has a fixed dust density. The packets are transmitted,
absorbed, or scattered in a given cell based on the results of random numbers
weighted by the physical properties of the dust and its density in that
cell. The temperature of the dust is not tracked; our scattered-light model does not re-emit
absorbed photons --- once they are absorbed, they are terminated. We ignored the effects of dust emission in this work, although
UV-excited extended red emission (the so-called ERE: \citealt{Perrin95,Pierini02}) might contribute modestly to the I-band
light of NGC 891 at high latitudes. The advantage of using a scattered-light RT code is
computational --- models can be run faster and with higher resolution than if we
used RT software which tracked absorption and re-emission of dust. 

Our model employs a forced first
scattering algorithm, where all photon packets are scattered at least once
\citep{Witt77}. We also use a 'peeling off' formula \citep{Yusef-Zadeh84}, which directs a fraction
of each photon packet's light toward the observer regardless of the packet's
nominal direction. Both of these modifications allow us to achieve a
higher S/N for a given number of emitted photon packets with minimal loss in
accuracy. 

We use a grid of 1000x1000x500 cells, which is mapped to a spatial volume of
40x40x20 kpc. This gives us a resolution of $\sim$0.9'' at a distance of 9.5
Mpc. Because our simulations are at such high resolution, we have parallelized
the Monte Carlo code to run on a 12-core processor. 

\subsubsection{Stellar Emissivity}

As described below, we include multiple components to characterize
both the stellar emissivity and dust density, using a very similar
parameterization to that of \citet{Popescu00} and
\citet{Misiriotis00}.

We model the stellar emission as a smooth spatial distribution with
non-axisymmetric components in the form of spiral arms. Further,
emission is treated completely separately for each band to take into
account the change in dominant stellar population as a function of
wavelength. This is in line with the models of \citet{Xilouris99} and
\citet{Popescu00}, among others. The emissivity is governed by the
following equation:
\begin{equation}
L(R,z)=L_{B,0}e^{-7.67B^{0.25}}B^{-0.875}+L_{D,0}e^{-\frac{R}{h_r}-\frac{z}{h_z}}\xi.
\label{eq:emissivity}
\end{equation}
The first part of this expression describes an elliptical de
Vaucouleurs bulge, where $L_{B,0}$ is the bulge emissivity of the
central cell and $B$ is an intermediate quantity containing
information about the bulge effective radius $R_{e}$, semimajor axis
$a$, and semiminor axis $b$:
\begin{equation}
B=\frac{\sqrt{R^{2}+z^{2}(a/b)^{2}}}{R_{e}}.
\end{equation}
The second part of equation (\ref{eq:emissivity}) controls a
double-exponential disk, where $L_{D,0}$ is the disk emissivity of the
central cell, and $h_r$ and $h_z$ are the scale-length and -height,
respectively. The third component, $\xi$, is the logarithmic spiral
disk perturbation, and expands into
%\begin{equation}\label{eq:spirality}
\begin{align}
\xi=&1-w+\prod_{n=2,n+2}^{N}\frac{n}{n-1}w\sin^{N}\times \nonumber \\
&\left(\frac{\ln(\sqrt{x^{2}+y^{2}})}{\tan(p)}
  -\tan^{-1}\left(\frac{y}{x}\right)+45^{\circ}\right)
\label{eq:spirality}
\end{align}
%\end{equation}
where $x$ and $y$ are the cartesian coordinates, $w$ is the fraction
of light (or dust, see \S\ref{sec:dust}) entrained in the arms, and
$p$ is the pitch angle.  Equation \ref{eq:spirality} is a modified
form of logarithmic spirality in \citet{Misiriotis00}; the main
changes are the inclusion of the even-numbered exponent $N$ and the
fixing of the spiral arms at two. The asymmetry of H$\alpha$, 60
$\mu$m, and CO emission point to NGC 891 being a classical grand
design spiral \citep{Kamphuis07,GarciaBurillo92,Xilouris98},
justifying the use of only two arms. In the Misiriotis spirality
formulation the arms and interarm regions have equal
size. \citet{Schweizer76} showed that this was not the case; the ratio
between arms to interarm regions is closer to 0.2 for grand design
spirals. Exponentiating the sine part of the function allows us to
(coarsely) adjust the arm-interarm ratio; $N=10$ roughly corresponds
to the correct ratio. We have chosen the form of the spirality so that
if $N=2$ it reduces to the equation in \citet{Misiriotis00}. The product
in front of the equation ensures that the spiral perturbation does not
change the total emissivity.

\begin{figure}
\begin{centering}
\includegraphics[scale=0.5]{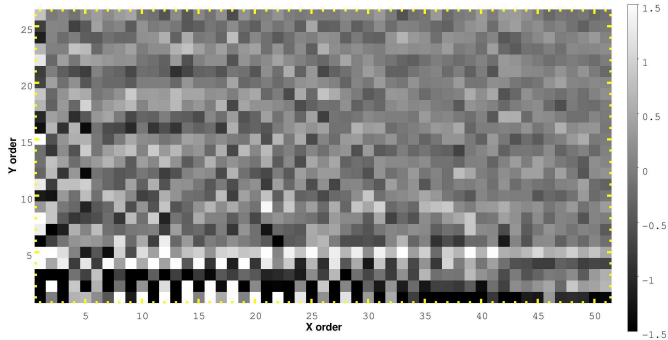}
\caption{Shapelet coefficients of the observed I band $\Delta \aeff$
  map. Horizontal orders are along the x-axis, while vertical orders
  are along the y-axis. Lower orders are on the left and bottom,
  respectively. The color-bar shows the amplitude of the shapelet
  coefficients.}
\label{fig:shapeletrecon}
\end{centering}
\end{figure}
\subsubsection{Dust}
\label{sec:dust}

We populate our models with the same dust geometry in all wavebands,
as the dust density distribution is independent of wavelength. The RT software
takes as an input the total dust+gas density, so we note that the assumption
that the dust follows the gas is implicit in our models. The
dust density distribution is assumed to be very similar to that of the
starlight in the disk on large scales, e.g. there is no dust bulge,
and hence
\begin{equation}
\rho_{d}(R,z)=\rho_{0,d}e^{-\frac{R}{h_r}-\frac{z}{h_z}}\xi\,\digamma(R,z),
\end{equation}
where $\rho_{0,d}$ represents the dust+gas disk density of the central
cell. The scale length and height can be different than those of the
stars, $\xi$ is the same basic formula as for emissivity (Equation
\ref{eq:spirality}) but with (potentially) different values for $w$,
and $\digamma$ represents a clumping modifier. Clumpiness is governed
by a fractal algorithm based on the distribution of molecular dust in
the Milky Way \citep{Elmegreen97} and is detailed in
\citet{Mathis02}. Because the fractal algorithm is designed to mimic
dust substructure on much smaller scales than we will probe, it is not
known {\it a priori} how well it can reproduce the large, chimney-like
structures usually taken as evidence for large scale outflows due to
e.g., supernovae. In the ideal case we would use a dynamical,
turbulent model of the ISM computed from hydrodynamical simulations
(like those created by \citet{Joung06}), but using these models on
galactic length scales is currently computationally infeasible. The
fractal algorithm is computationally efficient and can produce a wide
range of substructure.

For the physical properties of the dust grains we choose values consistent
with the diffuse interstellar medium of the Milky Way, based on the dust
model of \citet{Mathis77} as computed by \citet{White79}. The adopted parameters are fairly consistent with prior
efforts to fit RT models to images (e.g. \citealt{Xilouris99}). More recent studies (e.g. \citealt{Draine03}) have found somewhat different
dust properties, although we note that determining these parameters can be difficult
(\citealt{Lewis09} and references therein). Three parameters govern the dust physics relevant
to our models: $\chi$,
the total (dust+gas) opacity; $a$, the albedo; and $g$, the scattering asymmetry
parameter. For the B band we adopt $\chi=286$ cm$^{2}$ g$^{-1}$, $a=0.54$, and
$g=0.48$. In I we use $\chi=105$ cm$^{2}$ g$^{-1}$, $a=0.49$, and
$g=0.29$. These values for $\chi$ correspond to an $R_{V}=3.27$, marginally
larger than the value of $R_{V}$ for a screen of dust in the Milky Way \citep{Cardelli89}. The same dust
model is used for both the smooth and the clumpy components of the dust
distribution. 

\subsubsection{Free and Fixed Parameters}
\label{paramjustification}

We hold the bulge parameters constant using values from the smooth
model of \citet{Popescu00} (L$_{Bulge,B}=12\times 10^{30}$ erg s$^{-1}$
  pc$^{-3}$ str$^{-1}$, L$_{Bulge,I}=2.23\times 10^{30}$ erg s$^{-1}$ pc$^{-3}$ str$^{-1}$, b/a$_{B}$ = 0.6, b/a$_{I}$=0.54, R$_{e,B}$=1.12
kpc, and R$_{e,I}$=1.97 kpc), under the assumption that the bulge should
be relatively smooth and axisymmetric and is well characterized from
light outside the midplane where dust absorption is minimized. This
assumption is not perfect. For instance NGC 891 is known to have a bar
\citep{GarciaBurillo95}. We choose not to fit a bar to save CPU
expense in this large parameter space since our main area of interest
lies in the disk and not the bulge/bar.

We expect that the only change to the scale-lengths of the starlight
from those reported for smooth exponential models will be due to
azimuthal variations in the spiral pattern; therefore any deviation
from the \citet{Popescu00} values of the scale-length ($h_{r,B}=5.67$ kpc,
$h_{r,I}=4.93$ kpc) in one band will
be well correlated with deviations in the other, giving a single free
$h_{r,B}$ parameter. However, while any randomly generated models
follow this rule, we allow the scale-lengths to change just like the
other free parameters as the models 'breed' with each other. Therefore
the scale-lengths in both bands are generally highly correlated, but
not perfectly so. We leave the central emissivity $L_{D,B,0}$ and
$L_{D,I,0}$ and the scale-heights $h_{z,B}$ and $h_{z,I}$ as free
parameters because while they are largely insensitive to spiral
perturbations \citep{Misiriotis00} they are likely to be sensitive to
dust clumpiness.

\citet{Misiriotis00} show that with the addition of spirality
$L_{D,0}$ and $h_{r}$ can vary up to about 30\% (for large offsets
between stellar and dust arms).  We allow $L_{D,B,0}$, $L_{D,I,0}$, to
vary between 0.67 and 2 times the values given by \citet{Popescu00}
and $h_{r,B}$ to vary between 0.5 and 1.5 times the literature
value. As the effect of clumpiness on scale-heights is uncertain we
allow $h_{z,B}$ and $h_{z,I}$ to vary by up to a factor of two from
values in the literature. (For reference, \citet{Popescu00} use $L_{D,B,0}=2.66\times
10^{27}$ erg s$^{-1}$ pc$^{-3}$ str$^{-1}$, $L_{D,I,0}=3.44\times
10^{27}$ erg s$^{-1}$ pc$^{-3}$ str$^{-1}$, $h_{z,B}=0.43$ kpc, and $h_{z,I}=0.38$ kpc).

\begin{deluxetable*}{l c c c c c c c}
\tablewidth{0pt}
\tablecaption{Free Parameters}
\tablehead{\colhead{Parameter name} & \colhead{Symbol} & \colhead{Lower limit}
& \colhead{Upper limit} & \colhead{Median} & \colhead{10\%} & \colhead{90\%} &
\colhead{Unit}}
\startdata
Central disk emissivity (B band) & $L_{D,B,0}$ & 1.78x10$^{27}$& 5.35x10$^{27}$&2.69x10$^{27}$&2.04x10$^{27}$&4.03x10$^{27}$&erg s$^{-1}$ pc$^{-3}$ str$^{-1}$\\
Central disk emissivity (I band) & $L_{D,I,0}$ & 2.3x10$^{27}$& 6.9x10$^{27}$&3.39x10$^{27}$&2.74x10$^{27}$&4.65x10$^{27}$&erg s$^{-1}$ pc$^{-3}$ str$^{-1}$\\
Emission scale-length & $h_{r,B}$ & 2.84& 8.51&6.23&4.52&7.88&kpc\\
Emission scale-height (B band) & $h_{z,B}$ & 0.22&0.86 &0.46&0.37&0.59&kpc\\
Emission scale-height (I band) & $h_{z,I}$ & 0.19&0.76 &0.45&0.33&0.60&kpc\\
Central dust+gas density & $\rho_{0,d}$& 1.24$\times$10$^{-4}$&0.31 &1.6x10$^{-2}$&6.3x10$^{-3}$&8.8x10$^{-2}$&g cm$^{-2}$ kpc$^{-1}$\\
Dust scale-length & $h_{r,\rho}$ & 3.99& 11.96&6.41&4.44&9.34&kpc\\
Dust scale-height & $h_{z,\rho}$ & 0.135&0.54&0.24&0.16&0.34&kpc\\
B band spiral perturbation strength & $w_{B}$ & 0 & 1&0.56&8.7x10$^{-2}$&0.71&-\\
Dust spiral perturbation strength & $w_{\rho}$ & 0 & 1&0.49&0.20&0.74&-\\
Pitch angle & $p$ & 10&30 &20.50&15.17&26.6&degrees\\
%Position angle & $PA$ & 0 & 180&degrees\\
Number of initial fractal clumps & $N_{1}$ & 50&200 &129&75&163&-\\
Smooth fraction (dust) & $f_{\rho}$ & 0 & 0.8&0.42&0.15&0.59&-\\
\enddata
\label{tab:parameterspace}
\end{deluxetable*}

The dust parameters are generally treated very similarly to their
analogues in stellar emissivity. Changes to the scale-length of the
dust $h_{r,\rho}$ appear to be correlated with changes to the scale
length of the stellar emission \citep{Misiriotis00}; however this
correlation appears to break down for small pitch angles and so we
leave $h_{r,\rho}$ to vary to within 50\% of literature value (7.97 kpc). The
dust vertical scale height can vary by a factor of two from the
\citet{Popescu00} value of 0.27 kpc. Because
$\rho_{0,d}$ is a strong function of clumpiness \citep{Misiriotis02}
we allow it to vary up to 50 times literature value (we choose a fiducial
$\rho_{0,d}$ of 0.0062 g cm$^{-2}$ kpc$^{-1}$, a compromise between the B and
I band values of \citet{Popescu00}); while we do not
expect such large discrepancies from the literature in $\rho_{0,d}$
(or indeed, any of our free parameters) the ability of the genetic
algorithm to quickly find local minima allows us to be very ambitious
in selecting our parameter space.

We allow the dust perturbation strength ($w_{d}$) and the B band
perturbation strength ($w_{B}$) to vary freely; \citet{Schweizer76}
found that the ratio of arm strength between bands similar to B and I
was about 1.2, and we use this value to set $w_{I}$. The pitch angle
$p$ of the stellar spiral arms is also a free parameter, allowed to
vary between 5 and 30 degrees, while the pitch angle of the dust lag
is fixed to the emission so that the dust arms always are 5 degrees
apart from the stellar arms. Both the pitch angle range and dust
offset angle are based on studies of face-on spiral galaxies by
\citet{Kennicutt81}. We fix the position angle $PA$ (corresponding to the
$\phi$ direction of a cylindrical coordinate system where the r and z
directions are parallel to the radius and vertical extent, respectively, of the galaxy) because our data only
covers the innermost part of NGC 891 and is therefore unlikely to be sensitive
to changes in $PA$, although we do test each model for both spiral
orientations (left-handed or right-handed). Finally,
we allow the full range of spiral perturbation strengths (0 to 1) for
both stars and dust.

The major parameters of the fractal algorithm are the fractal dimension  $D$, governing the size scale of the clumps; $N_{1}$, the
number of clumps; and $1-f_{\rho}$, the fraction of the total dust
density trapped in the clumps. We follow \citet{Indebetouw06} and set $D=2.6$: while this is larger than the value used by
\citet{Elmegreen97} it is within the observational limits found by
\citet{Elmegreen96} and gives more ``fluffy'' clouds
(\citealt{Indebetouw06}; for a schematic illustration of the fractal
algorithm see Figure 4 of \citealt{Wood05}). Small values of $N_{1}$
lead to a few, isolated, clouds while large values correspond to
smoother, more uniform, structures. We therefore allow $N_{1}$ to vary
freely between 50 and 200, which (despite the aforementioned fact that
the fractal algorithm is designed for smaller structures) we
experimentally find produces realistic looking clumpy distributions. All
subsequent casts control the size and filling factor of individual
clumps and are set to 32, which we have found produces rich structure
down to our resolution limit without significant added CPU expense.

After determining the distribution of clumpy regions we set the
fraction of dust in the clumps. The smooth fraction $f_{\rho}$ is a
free parameter which is allowed to vary between 0 and 80\% of the
total dust mass, observing that smooth fractions larger than 80\% tend
to be indistinguishable from the entirely smooth models which are
uninteresting to probe in this study. The ranges of all the free
parameters used in the models is shown in Table
\ref{tab:parameterspace}.

\subsubsection{Additional Test Models}
With a total of 13 free parameters, we also wanted to understand how well
constrained our model was. To that end, we also ran a set of models with a
significantly reduced parameter space. We fixed both the scale-length and
scale-height for the stars and the dust at the values of our fiducial model
($h_{r,B}=5.67$ kpc, $h_{r,I}=4.93$ kpc, $h_{z,B}=0.43$ kpc, $h_{z,I}=0.38$ kpc, $h_{r,\rho}=7.97$ kpc, and $h_{z,\rho}=0.27$ kpc),
and set the stellar spiral perturbation strength $w_{B}$ (0.4) and pitch angle
$p$ (15$^{\circ}$)
equivalent to values consistent with Sb galaxies \citep{Schweizer76}. This leaves us with a total
of only six free parameters for our `constrained' models.

Additionally, we examined our decision to fix the $PA$ by running a subset of
8 models with the full set of 13 free parameters as well as varying the $PA$
between 0 and 180 degrees. These 8 models are heareafter referred to as the `free-PA' models. We investigate how changing our free parameters in this way affects our resulting models in \S\ref{consistency}.

\begin{figure*}
\begin{centering}
\subfigure{
  \includegraphics[scale=0.25]{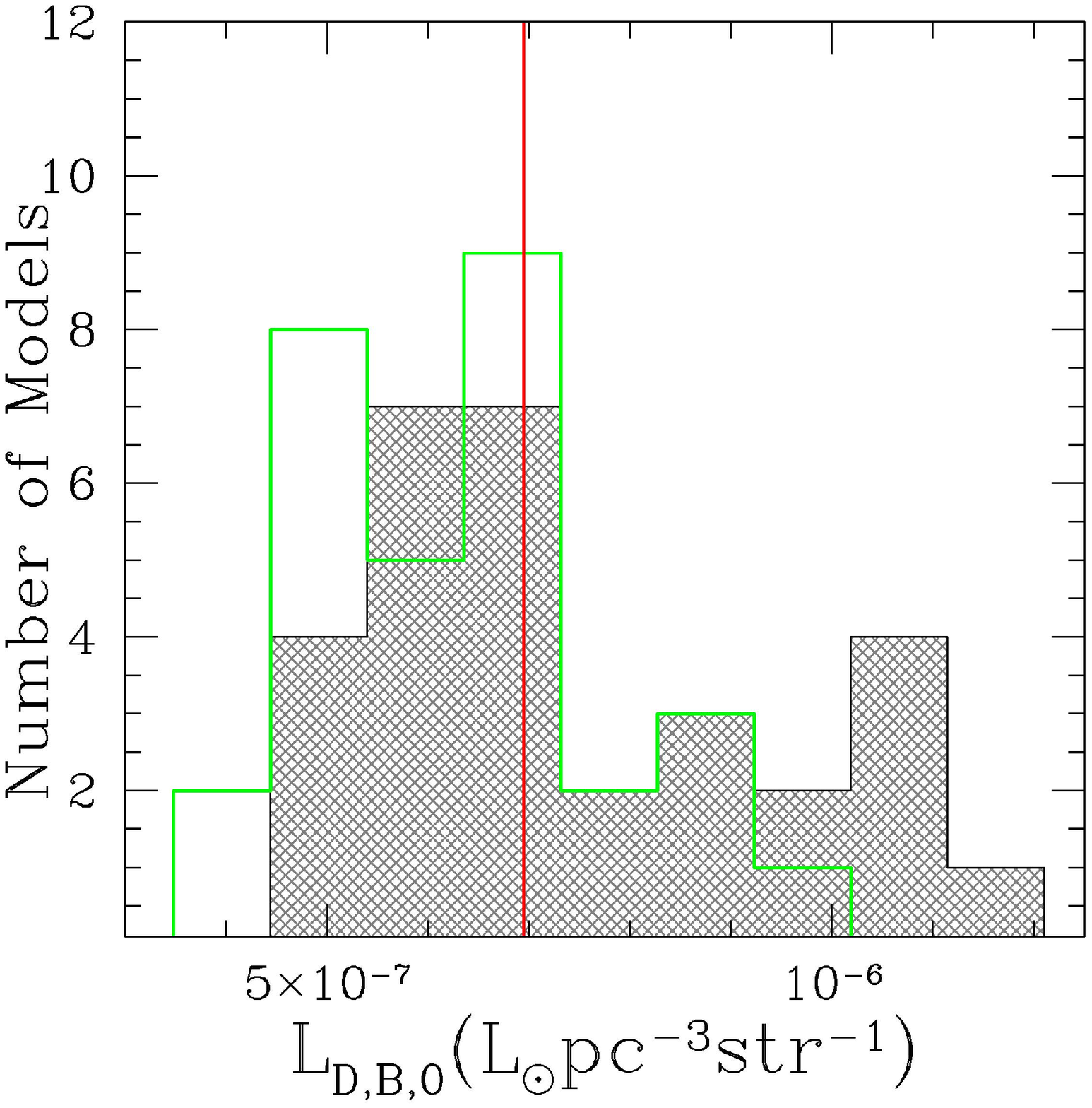}
}
\subfigure{
  \includegraphics[scale=0.25]{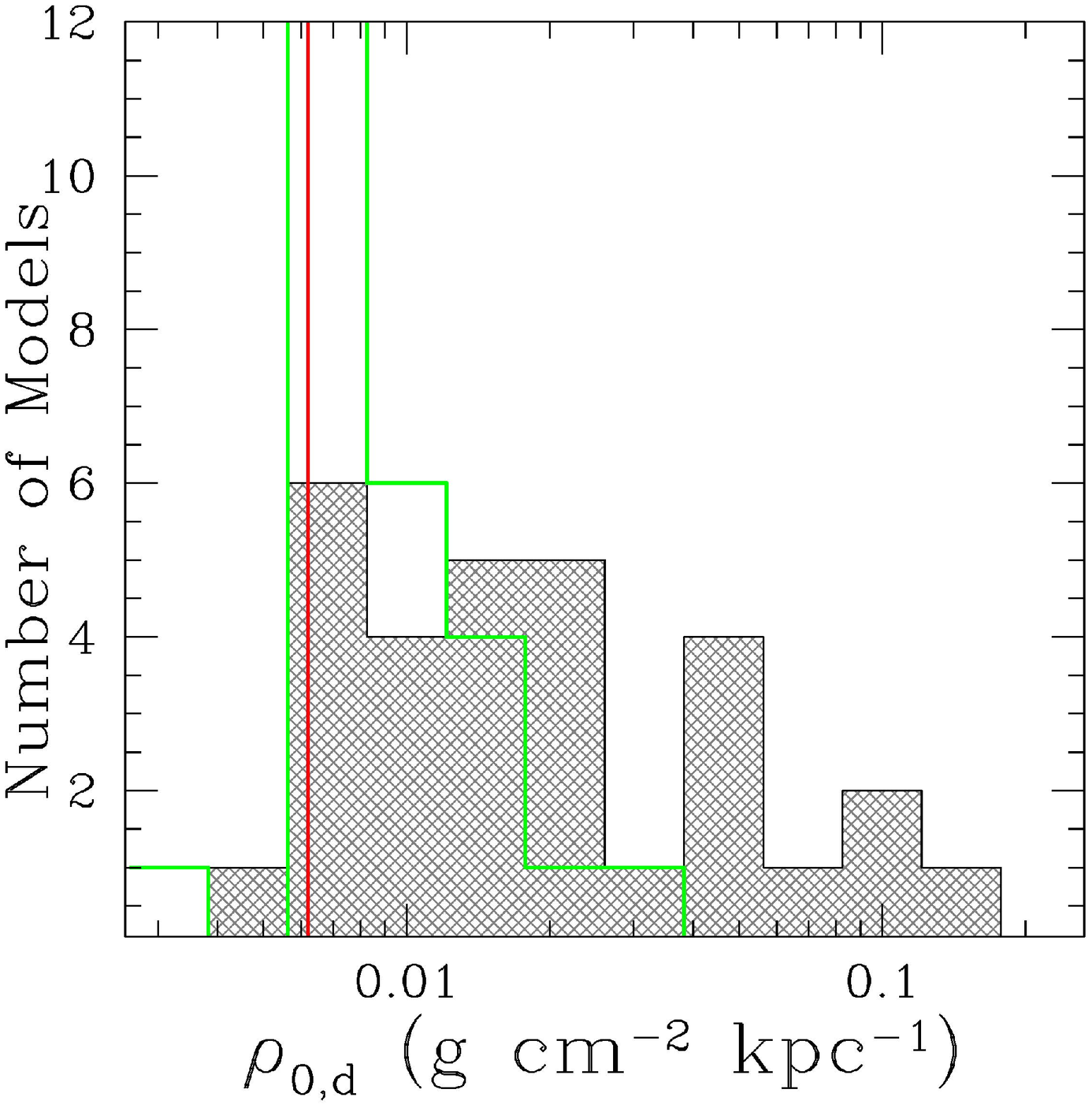}
}
\subfigure{
  \includegraphics[scale=0.25]{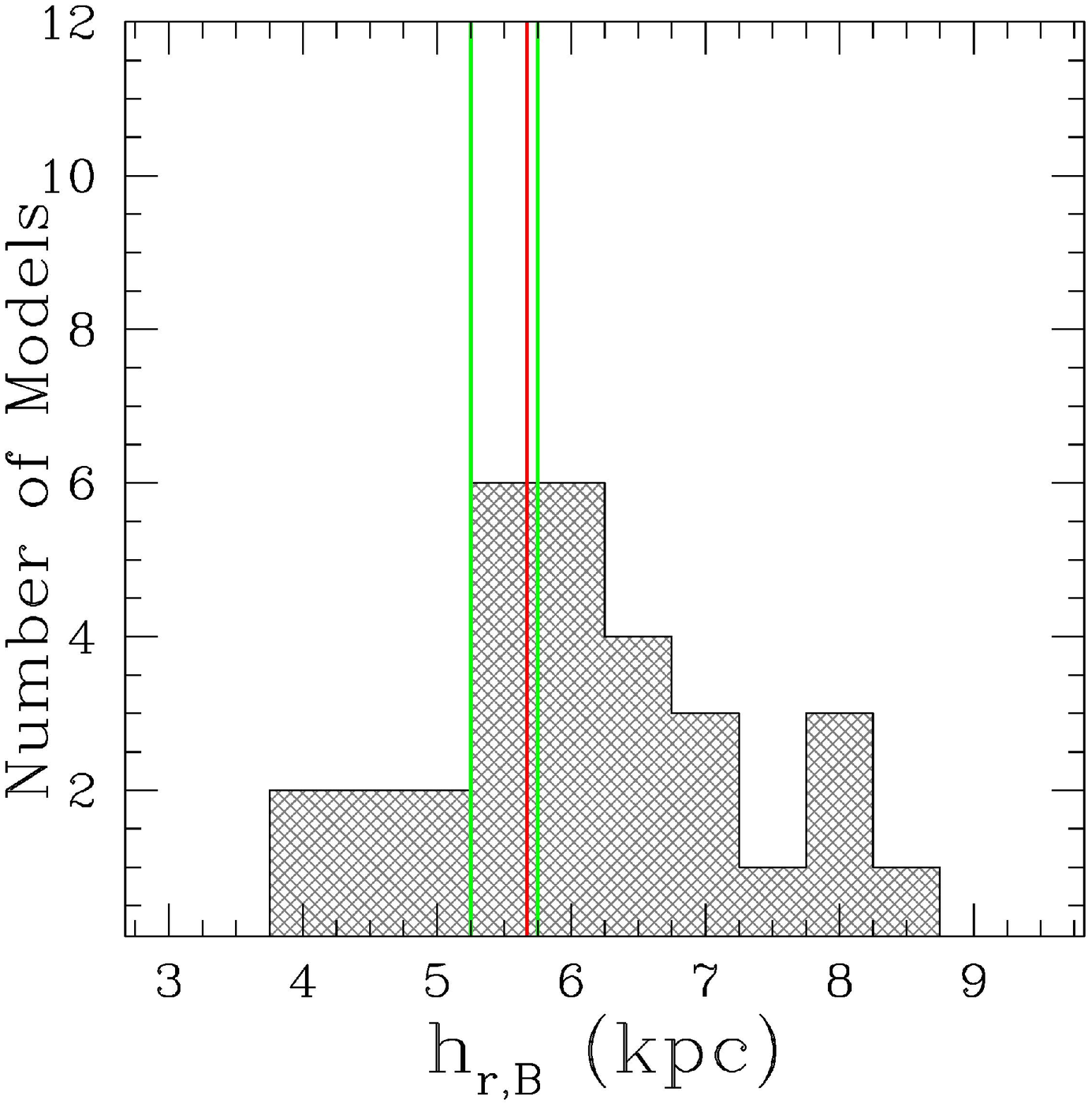}
}
\caption{Histograms of central B band emissivity (left), central dust+gas
  density (middle), and B band scale-length (right) for the 30
  best-fitting models. Gray shaded histograms are for the unrestricted sample,
  while green empty histograms show the restricted sample. Red vertical lines denote values from the
  simple smooth model of \cite{Popescu00}. }
\label{fig:fitnessplot1}
\end{centering}
\end{figure*}

\subsection{Genetic Algorithm}
Our total number of free parameters is 13. This presents an algorithmical
challenge, as it would take prohibitively long to do a brute-force search of
parameter space. What is required, then, is an algorithm that can efficiently
search through a parameter space without deeply probing regions
with only poor fits to the data. Multiple approaches exist for this problem, but we
choose a genetic algorithm both for its simplicity and efficiency. 

Genetic algorithms take their inspiration from the evolutionary
principles of natural selection. A population of models is created,
then their fitnesses are computed. The model with the best fitness is
cloned, and passed on unchanged to the next generation. Individual
models then pseudo-randomly pair off to 'breed' to populate the next
generation with the same number of models, where models with better
fitnesses are statistically more likely to breed more often. The gene
pool is refreshed either through the occasional influx of new models
or through random mutations of existing models, thereby avoiding
population stagnation.

If the fitness criterion is adequate at discriminating between `good'
and `bad' models a genetic algorithm will rapidly converge to a local
minimum. Due to the continuous mixing of parameters and introduction
of new ones the genetic algorithm is capable of skipping lightly over
poor local minima and quickly finding a relatively deep local
minima. After this point increasing the number of generations has
little effect, as the minimum is too deep for mutations or breeding
with new models to easily climb out and find a better one. At this
point the genetic algorithm becomes little better than a Monte Carlo
analysis. Therefore, the optimal way to run a genetic algorithm is to
determine roughly how long it takes for the algorithm to converge on a
good minimum (usually through running empirical test models), then run
multiple iterations of the algorithm only up to that number of
generations to map out the deep local minima. 

Because the genetic algorithm spends little time searching through bad regions
of parameter space, adding unconstrainable free parameters to the model does
not significantly increase computing time. Therefore, while we may not expect
all 13 of our free parameters to be well-constrained (e.g. the spirality
parameters) including them does not strongly affect run-time or `pollute'
well-constrained parameters. 

Our genetic algorithm is based on that of \citet{Howley08}, who used
it to study the orbit of NGC 205, a satellite of M31. Our algorithm is
very similar to theirs; the reader is referred there for more
detail. Our modifications are small: we use eight models in each
generation instead of five, we replace all but the best model with
random ones every eight generations to prevent population stagnation,
and we use a single fitness indicator instead of trying to combine
several as they do.

\begin{figure*}
\begin{centering}
\includegraphics[scale=0.75]{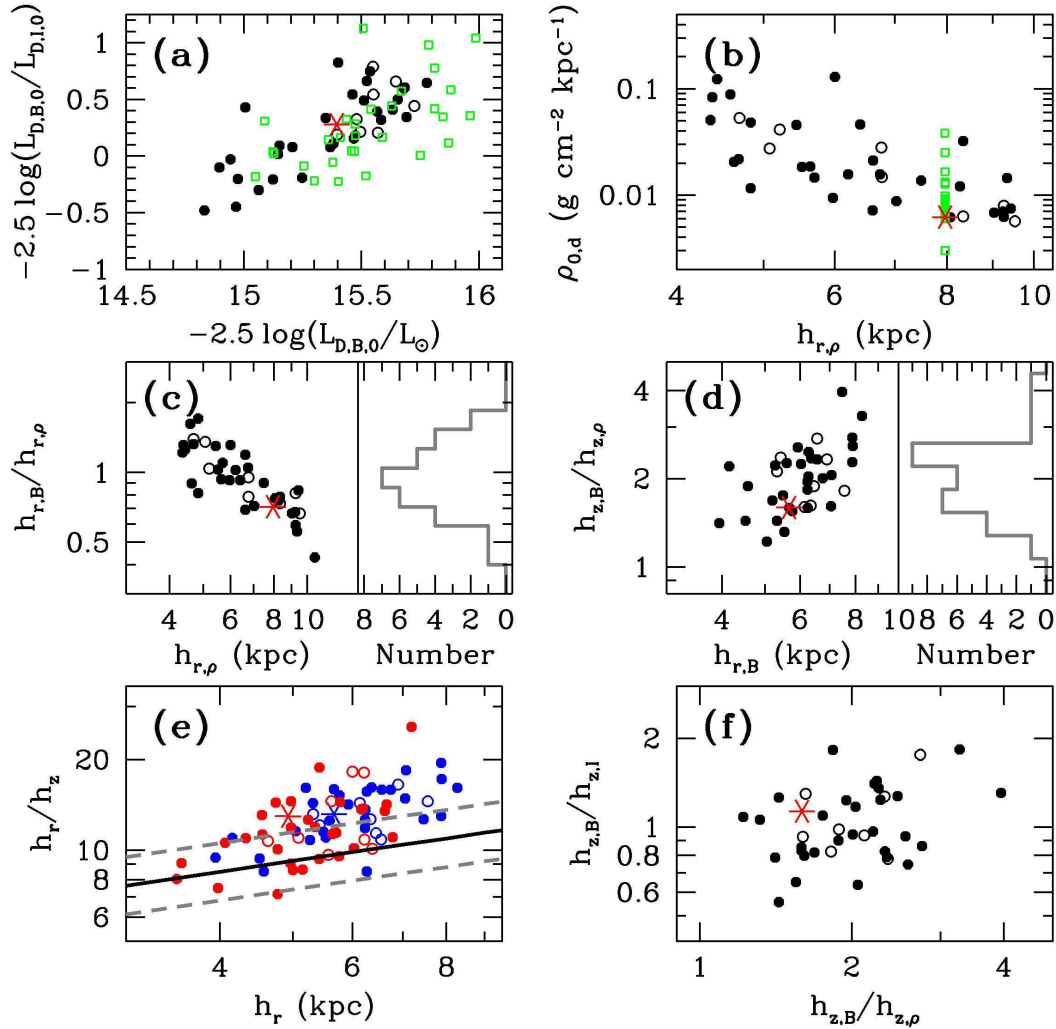}
\caption{Correlations between free parameters shown in Figure
  \ref{fig:fitnessplot1} and other axisymmetric parameters for our
  best fitting clumpy models and the simple smooth model of
  \cite{Popescu00}. (a) Central emissivity color as a function of
  B band central emissivity. (b) Central dust+gas density versus dut scale-length.  (c) Ratio of the
  B band to dust scale-lengths versus the dust scale-length.  (d)
  Ratio of the B band and dust scale-heights against the B band
  scale-length.  (e) Ratio of scale-length to scale-height for the B
  (blue) and I (red) bands against the B and I band
  scale-length. (f) Ratio of B to I band scale-heights as a function
  of the ratio of B band to dust scale-heights. Black dots denote clumpy
  models, red stars denote the simple smooth model, open black circles denote the
  subset of clumpy models with variable position angles, and open green
  squares (in a and b) denote the models with restricted parameter space. When necessary, blue
  denotes the B band and red denotes the I band. The thick black line
  shows the empirical correlation and 1$\sigma$ scatter between
  scale-length and scale-height from \citet{Bershady10b}.}
\label{fig:fitnessplot2}
\end{centering}
\end{figure*}

\subsection{Fitness}
In our approach, shapelet analysis produces a map of the contribution
to the reconstructed $\Delta \aeff$ image at all possible combinations
of orders. However, we do not use this map to model $\Delta \aeff$
images for comparison on a pixel-by-pixel basis with the observed
$\Delta \aeff$ image. This is because the detailed spatial variations
of $\Delta \aeff$ are dependent on random variables (in reality due to
the stochastic nature of star formation; in the models by
construction). Consequently a direct comparison in image-space is
swamped by the mis-match in random structure. In contrast, however,
the relative amplitudes of shapelet coefficients do capture the power
in structures at different spatial frequencies, and hence provide a
good statistical descriptor of the physical distribution of emission,
absorption, and scattering. Therefore we use the shapelet coefficients
directly to compute a fitness statistic:
\begin{equation}
\mathcal{F} = \sum_{n=0,m=0}^{N,M}\left |C(n,m)_{model} - C(n,m)_{data}\right|
\end{equation}
where $n$ and $m$ are the shapelet orders, and $C(n,m)$ is the
shapelet coefficient magnitude at order n,m.  Equal weights are given
to each coefficient, and by not adding in quadrature we prevent the
largest coefficients (usually in the lowest orders, governing the
large-scale structure) from dominating the fitness. The fitnesses of
each filter are added in quadrature, and the geometric orientation
with the best combined fitness is used in the genetic algorithm. Using
this merit function the genetic algorithm essentially matches each
coefficient to the same absolute tolerance, automatically adjusting
the tolerance from large to small as it improves the fit over the
course of several generations.

\begin{figure*}
\begin{centering}
\subfigure{
  \includegraphics[scale=0.25]{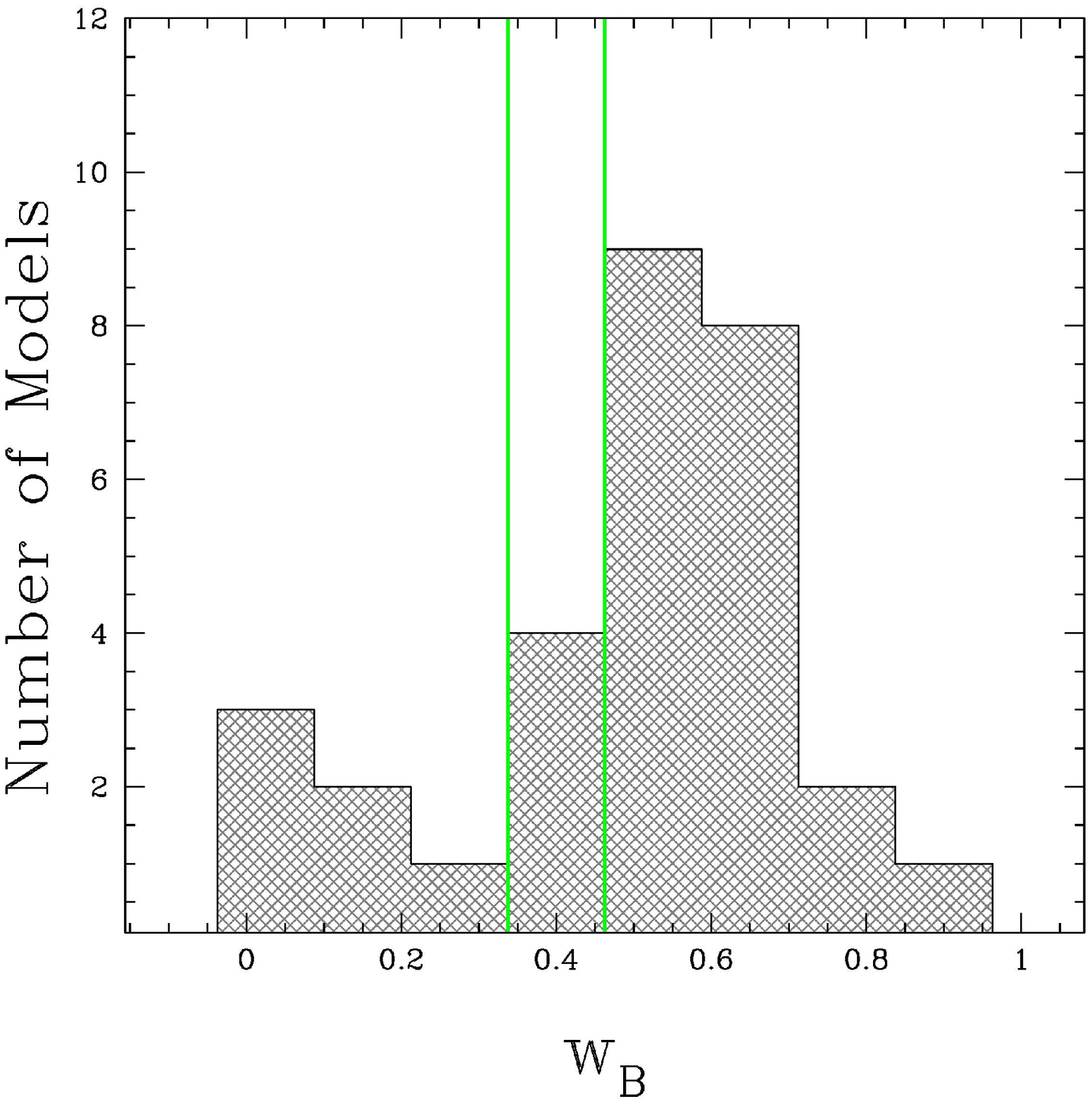}
}
\subfigure{
  \includegraphics[scale=0.25]{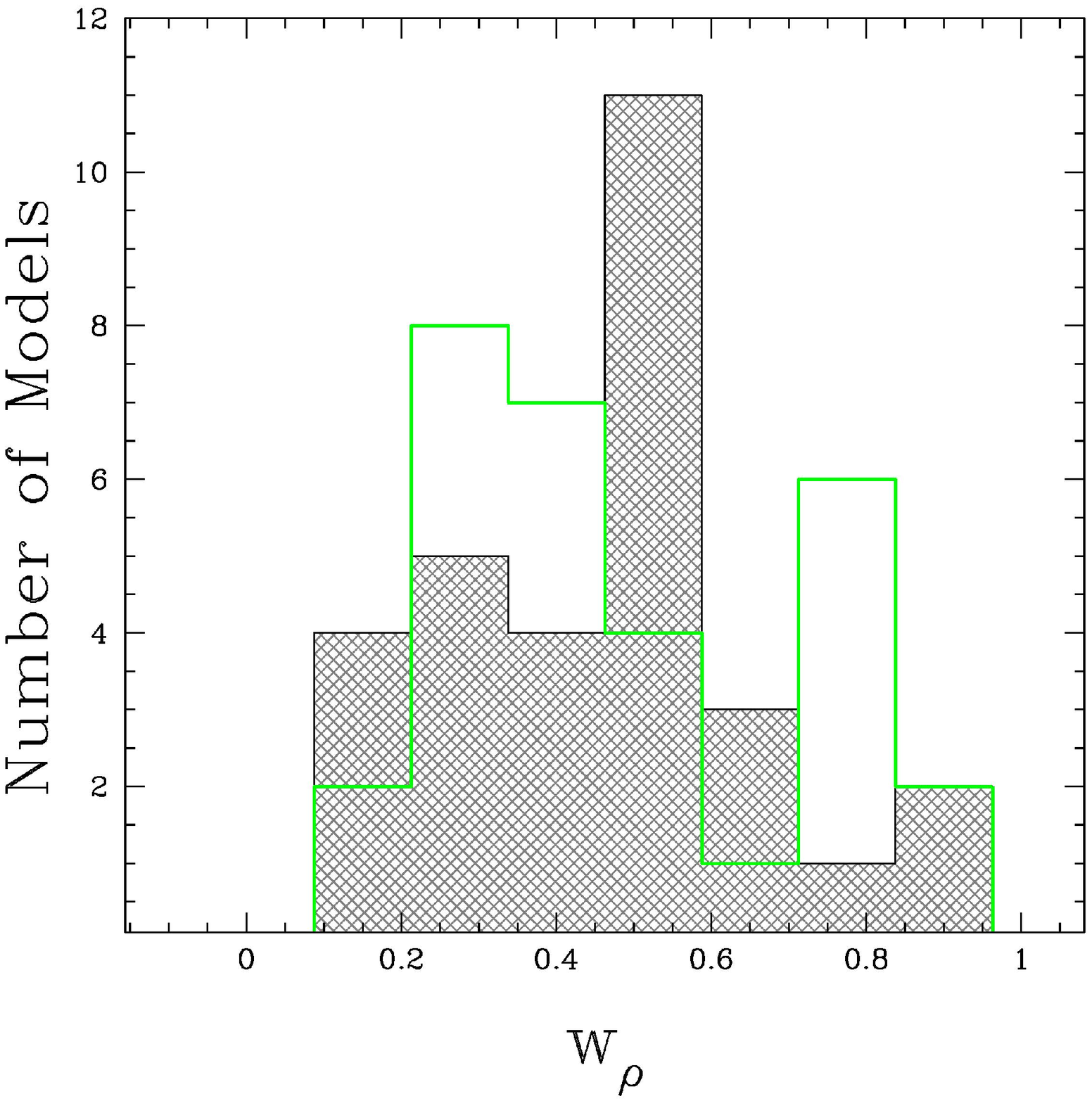}
}
\subfigure{
  \includegraphics[scale=0.25]{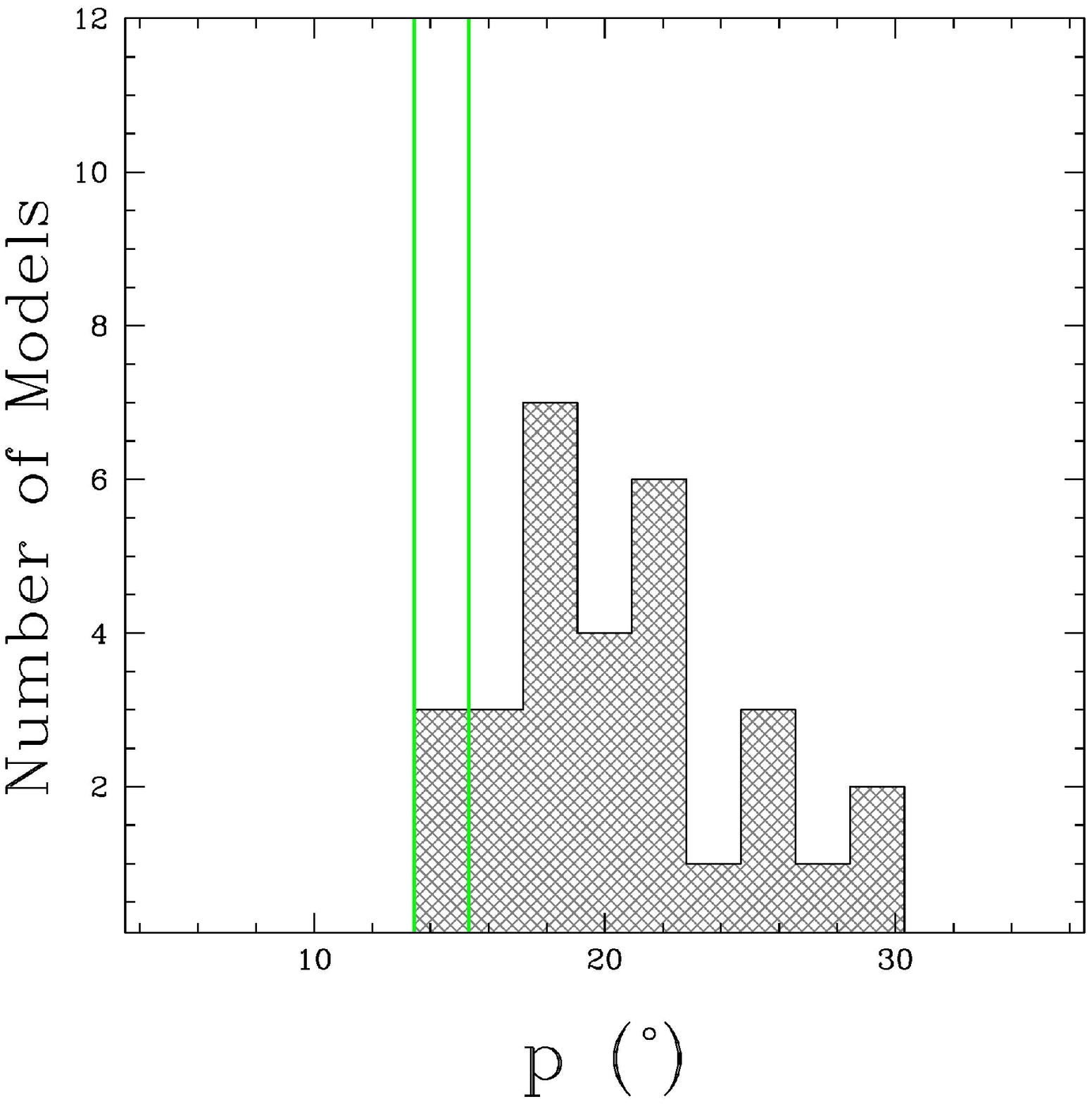}
}
\subfigure{
  \includegraphics[scale=0.25]{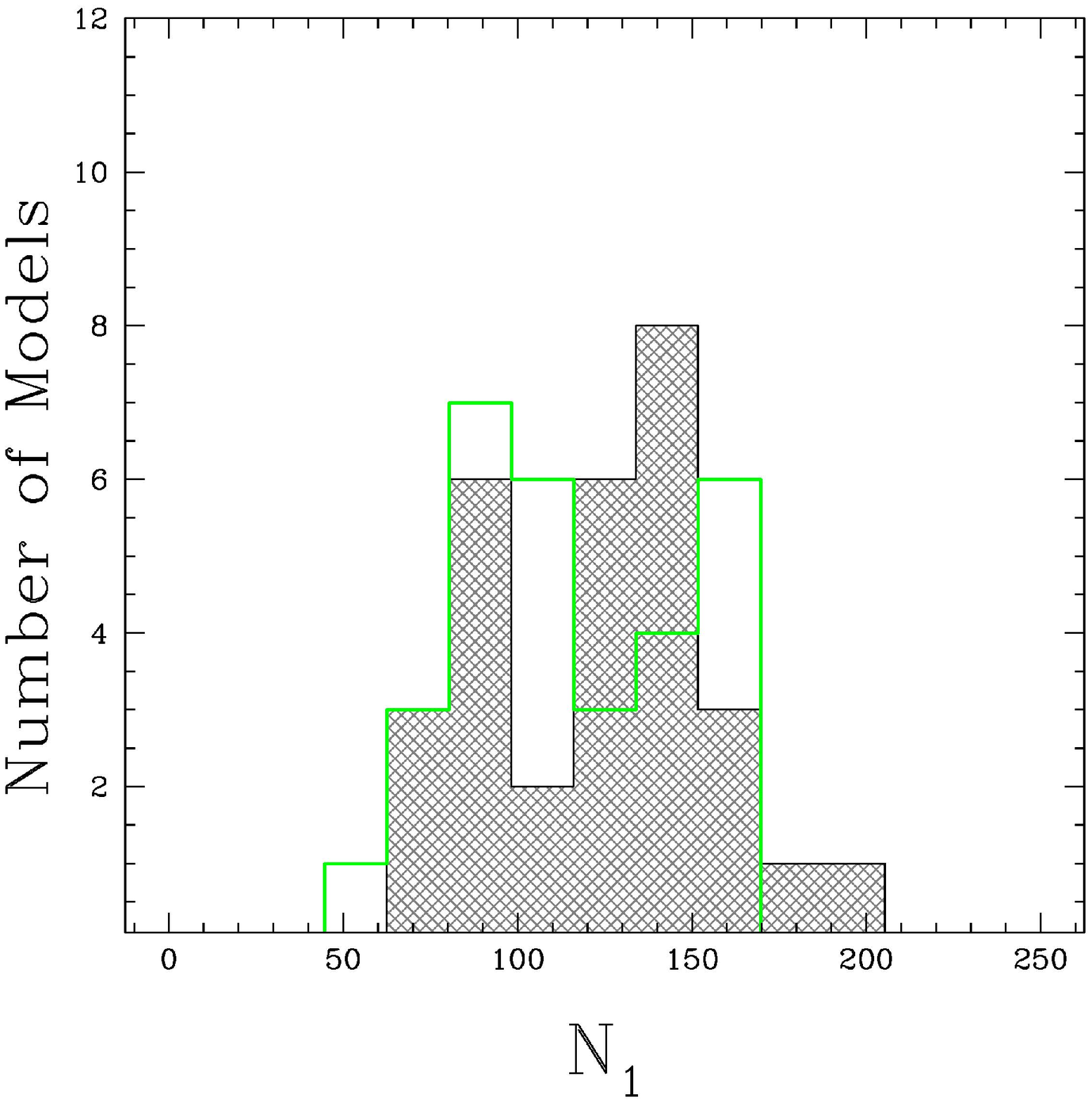}
}
\subfigure{
  \includegraphics[scale=0.25]{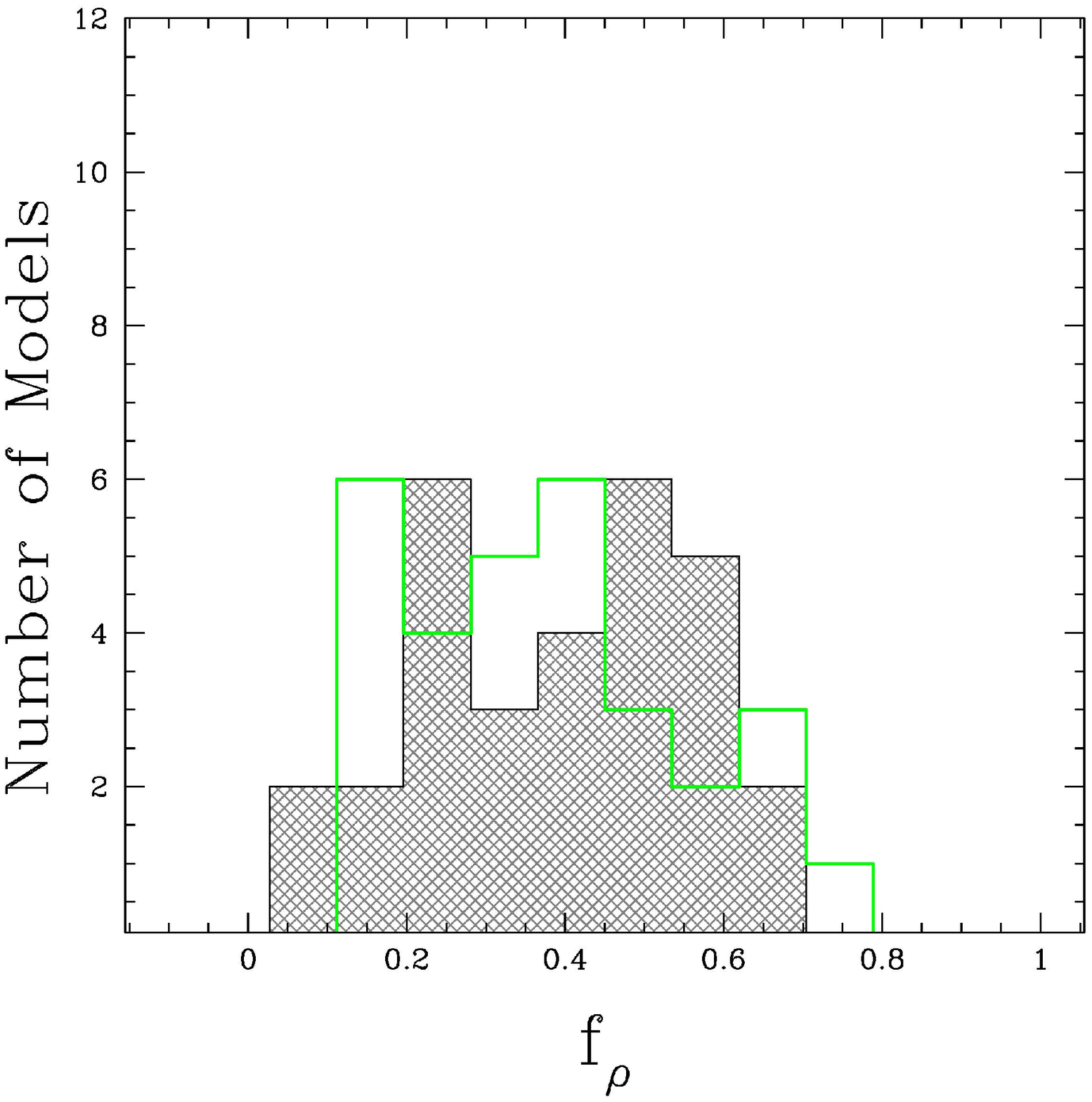}
}
\caption{Histograms of non-axisymmetric parameters for the
  best-fitting clumpy models for unrestricted sample (gray shading) and
  restricted sample (green lines). Top left: B band spiral perturbation
  strength. Top middle: dust+gas spiral perturbation strength. Top right: pitch angle. Bottom left: number of fractal
  clumps. Bottom right: smooth dust fraction.}
\label{fig:fitnessplot3}
\end{centering}
\end{figure*}

\section{Results}
\label{sec:results}
The HST images only cover the inner $\sim$9 kpc of NGC 891 so we only
output model images that cover a corresponding area. This results in
images that are $230\times230$ pixels, where 1 pixel = 1 projected
cell. We choose to simulate a much larger volume than we need so that
we can produce images showing the full model galaxy with the same
random clumping behavior (because of the semi-random nature of the clumps
changing the grid volume would change the clump distribution, even when
holding the random seed constant). However, in our genetic algorithm we only
allow photons to be emitted along the central 10 kpc of the disk (5
kpc in radius) in both the y and z directions (where x is along the
line of sight) which allows us to obtain higher S/N with fewer
photons. Empirically, we determine that a model with 5 million photons
emitted with the above restriction produces acceptable images.

On a 2.93 GHz 12-core machine each band for NGC 891 takes
approximately 2-3 minutes to run for a single set of model parameters,
which is by far the time-limiting step in the genetic algorithm. We
find that the genetic algorithm has largely settled into a local
minimum after 50 8-model generations, so we truncate the algorithm
there. Even with only 50 generations each run of the genetic algorithm
takes about a day, so while it would be ideal to use more generations
per run (\citealt{Howley08} use 1000) we opt to limit the number of
generations and therefore have time for more runs. Even so, time
constraints limit our final sample to only 30 runs. Each run contains
up to 351 distinct models ($>10^{4}$ models total), but we only select
the best model from each run for our analysis. While 30 runs does not
allow us to make the detailed analysis of parameters found in
\citet{Howley08}, it does allow us to make a broader survey of the
parameter space. For our parameter space tests, we used 8 runs for the models
with variable
$PA$ added to the normal 13 free parameters and a full 30 runs for the constrained set of models containing only 6 free parameters. 

\begin{figure*}
\begin{centering}
\includegraphics[scale=0.8]{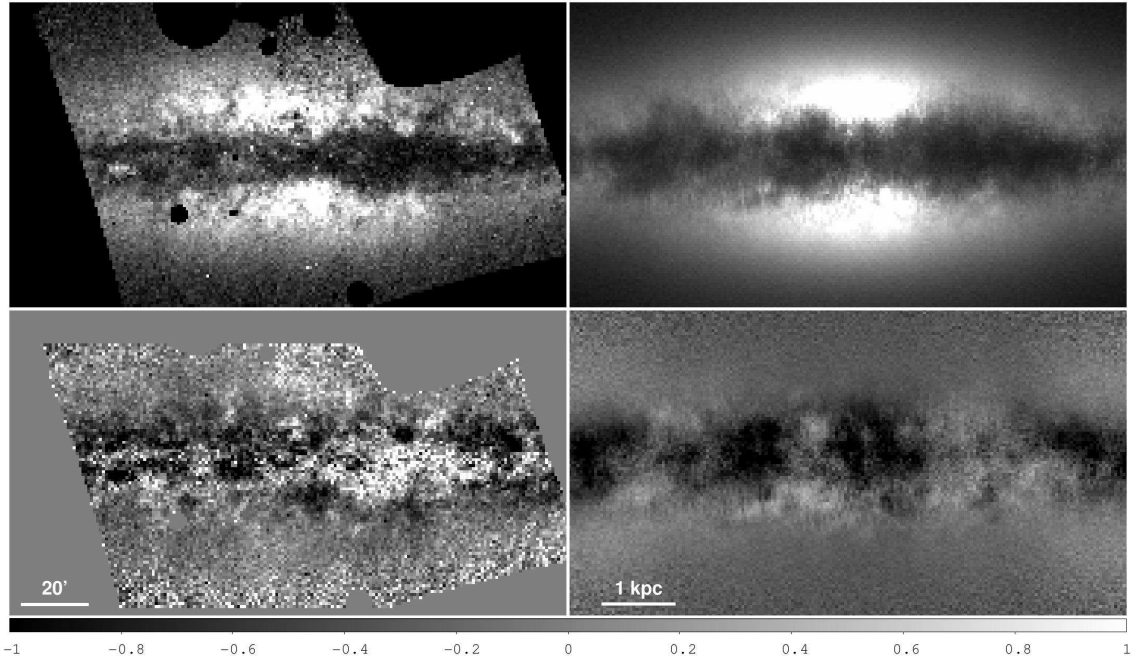}
\caption{Comparison of one of our best fitting non-axisymmetric RT models with data for
  the B band. Top left: smooth HST image. Top right: model
  image. Bottom left: $\Delta \abeff$ map for data. Bottom right:
  $\Delta \abeff$ map for model. The color-bar shows the
  differential magnitudes of the $\Delta \abeff$ maps.}
\label{fig:bestbband}
\end{centering}
\end{figure*}

\begin{figure*}
  \begin{centering}
    \includegraphics[scale=0.8]{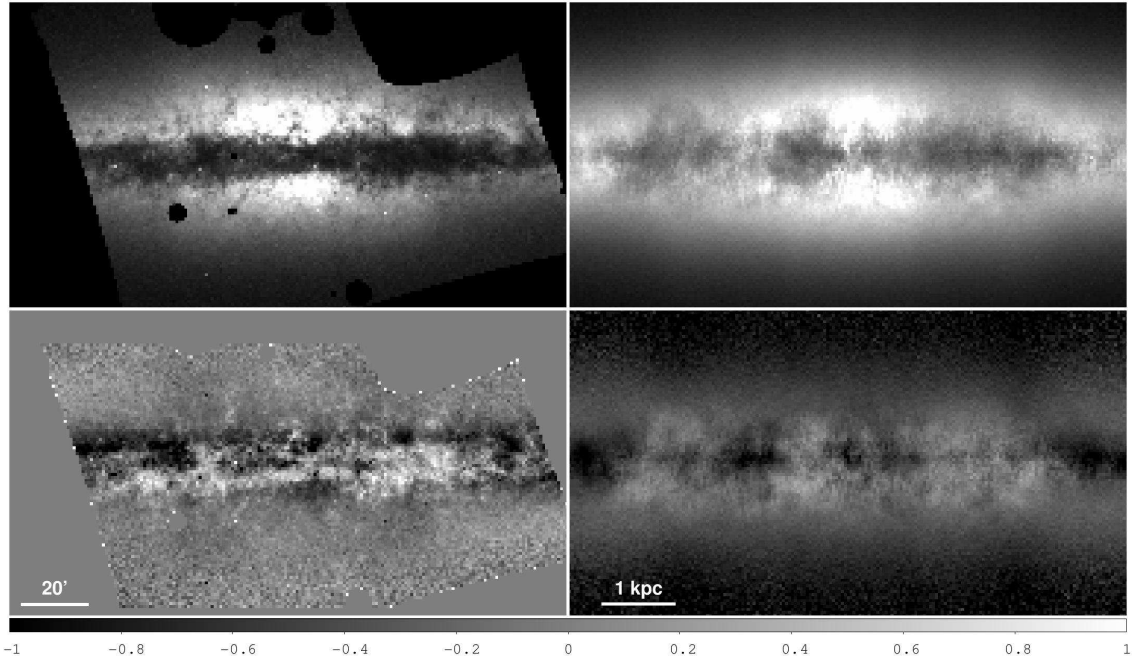}
  \caption{Same as Figure \ref{fig:bestbband}, but for the I band.}
  \label{fig:bestiband}
\end{centering}
\end{figure*}

\subsection{Best Fitting Models}

In Figures \ref{fig:fitnessplot1}---\ref{fig:fitnessplot3} we show
histograms and plots of the free parameters in our clumpy
models, as well as (where appropriate) our `free-$PA$' and `constrained' models. Figure \ref{fig:fitnessplot1} contains histograms of the
central B band emissivity $L_{D,B,0}$, central dust+gas density
$\rho_{0,d}$, and B band scale-length $h_{r,B}$ overplotted with the
appropriate values of the smooth, single-disk model of
\citet{Popescu00} which we used as the basis for our clumpy
models. While there is significant scatter in all three parameters, on
average our models prefer slightly lower central B band emissivities,
slightly higher central dust densities, and B band scale-lengths
roughly equivalent to those of the smooth model. 

There is significant scatter for most of the individual parameters. While
initially this may seem to indicate that our models are unconstrained, a
closer inspection reveals that the scatter for individual parameters hides
deeper correlations between multiple parameters, indicating that there are
several strong joint constraints.

In Figure \ref{fig:fitnessplot2} we plot several noticeable
relationships between the parameters shown in Figure
\ref{fig:fitnessplot1} and the other emissivity and scale
parameters. Values for the \cite{Popescu00} model are shown with red
stars, or blue (B band) and red (I band) stars when data from both
bands are plotted simultaneously. In all cases the values for the
smooth model fall within the bivariate distributions of our clumpy
models. For reference, the data from the `free-PA' and `constrained' subsets of models are
shown (where appropriate) as open circles and open squares, respectively.

First, there is a correlation between
redder central emissivities and fainter central B band emissivity
(Figure \ref{fig:fitnessplot2}a), but the correlation is in excess of
what would arise for a constant I band emissivity. In other words,
model disks with fainter central B band emissivity tend to have
relatively higher central I band emissivity.

Figure \ref{fig:fitnessplot2}b shows that as the radial size of the dust disk
is inversely proportional to the central dust density: larger dust disks have
more diffuse dust. In terms of the relative size scales of the dust and stellar disks,
the B band radial scale-length is comparable to the dust scale-length
(Figure \ref{fig:fitnessplot2}c), with a mean ratio of
$\sim$1. However, as the size of the dust disk increases the relative
size of the B band stellar disk decreases, with the largest dust disks
having $\frac{h_{r,B}}{h_{r,\rho}}\approx 0.5$. The stellar vertical
scale-heights of our clumpy models are typically twice that of the
dust, with a positive correlation between B band radial scale-{\it
  length} and the ratio of B band scale-height to the dust
scale-height. (Figure \ref{fig:fitnessplot2}d). Hence the dust
oblateness is roughly half that of the stars. The oblateness of the
stellar disk in the B (blue) and I (red) bands increases modestly
with the stellar radial scale-length (Figure \ref{fig:fitnessplot2}e),
departing somewhat from the empirical relationship between oblateness
and scale-length derived by \cite{Bershady10b}.  The stellar disks are
slightly larger in the B band than the I band, and both bands are
somewhat flatter than typical galaxies of comparable size and Hubble
type. We find little degeneracy in the models between intrinsic
vertical color gradients in the stellar emissivity and varying
scale-heights of dust to stars: In Figure \ref{fig:fitnessplot2}f
there is little correlation between the ratio of the B and I band
scale-heights and the ratio of the B band scale-height to the dust
scale-height (bottom right).

Finally, we plot histograms of parameters related to the spirality and
clumpiness in Figure \ref{fig:fitnessplot3}. While there is
significant scatter, fully two-thirds of
our models have at least 50\% of their B band emission and dust density tied up in
spiral arms with a preferred pitch angle of about 20$^{\circ}$. However, we
also find that our models don't show a strong preference for the {\it
  direction} of spirality (left- or right-handedness). It therefore seems
likely that the spirality is not well constrained in or simulations, and as a
result we do not present a thorough analysis of the spirality parameters.

Our
models do show a strong preference towards having a significant amount of dust
entrained in clumps; 27 of the 30 best fitting models have smooth dust
fractions $f_{\rho}$ smaller than 0.6, which means that 90\% of the
best fitting models have $\ge 40\%$ of their total dust mass arranged
into clumps. The median clumpiness fraction $1-f_{\rho}$ is 58\%, which is
a factor of only 1.2 smaller than current clumpy estimates fitting to the
energy balance solely through SEDs \cite{Bianchi11}. The number of clumps (governed by $N_{1}$) is generally
fairly large, with a slight preference for $N_{1} \ge 125$. These
parameters combine to give our models significant amounts of
non-axisymmetric structure.

\begin{figure*}
  \begin{centering}
    \includegraphics[angle=90,scale=1.2]{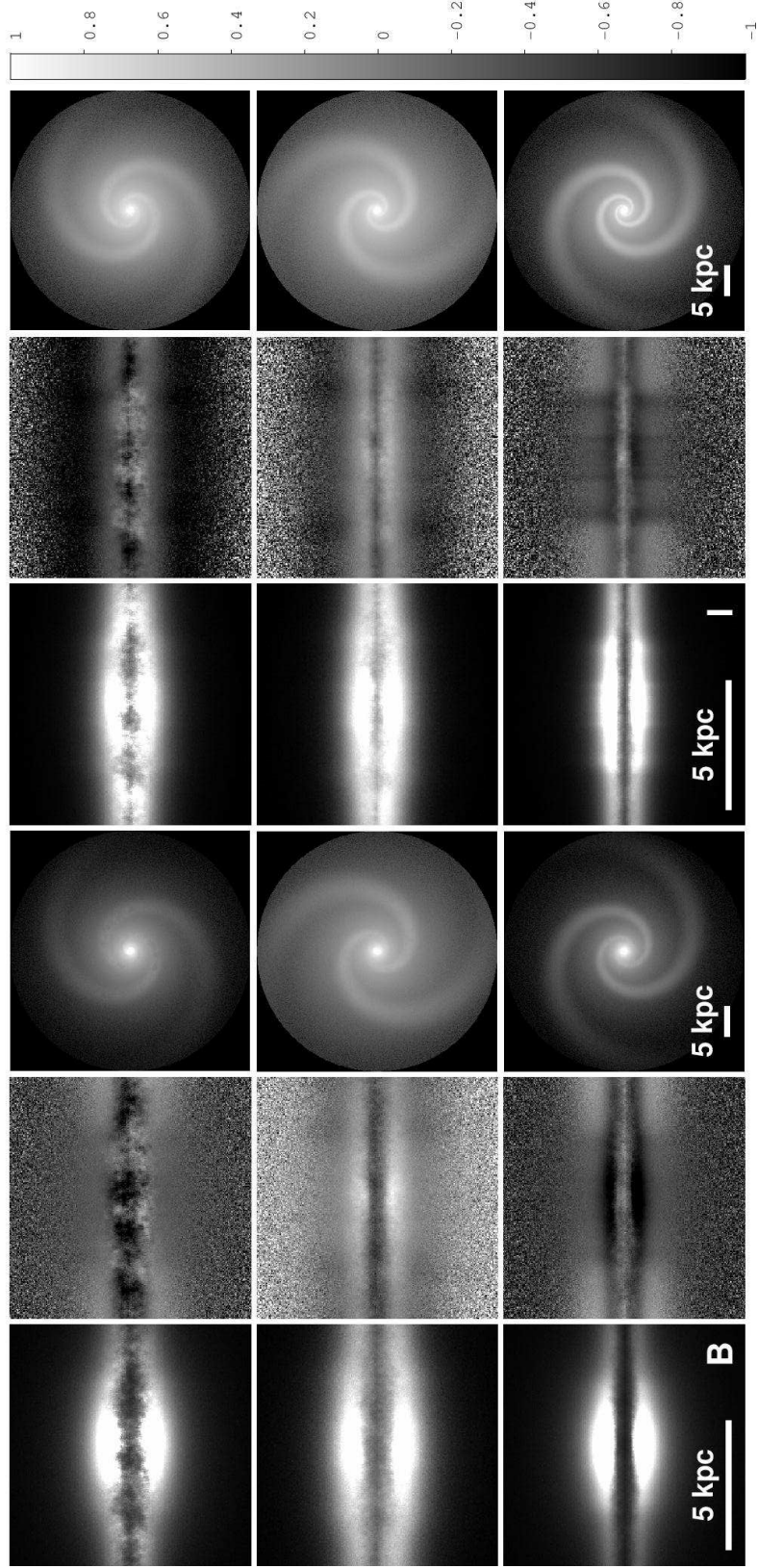}
    \caption{Representative sample of the best fitting clumpy
      models. From left: edge-on B band images, edge-on $\Delta \abeff$
      maps, face-on B band images, edge-on I band images, edge-on $\Delta \aieff$
      maps, face-on I band images. Color-bar denotes $\Delta \aeff$
    for B and I bands, in mag. Each model was fit both as a clockwise and counter-clockwise rotating spiral. While the orientations with the best fits are shown here, generally the difference in fitness between orientations was small. Additionally, some of the models (even with the same orientation) appear to have different $PA$s, however this is an illusion caused by the varying pitch angles as well as the self-similar nature of the logarithmic spirality.}
    \label{fig:modelimages}
  \end{centering}
\end{figure*}

\addtocounter{figure}{-1}
\begin{figure*}
  \begin{centering}
    \includegraphics[angle=90,scale=1.2]{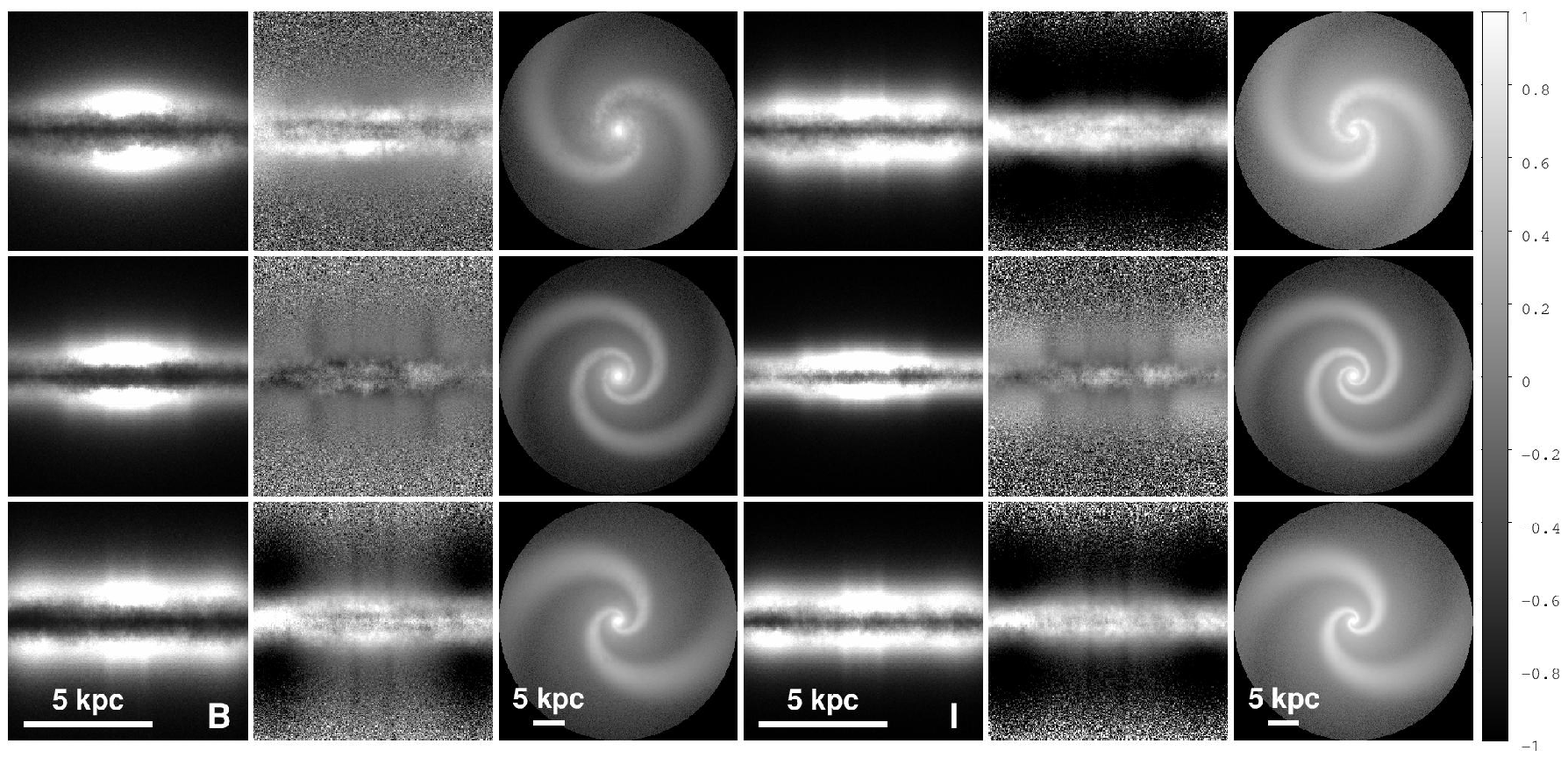}
    \caption{Continued.}
    \label{fig:modelimagescont}
  \end{centering}
\end{figure*}

\subsubsection{Consistency Checks}
\label{consistency}
The subset of models with variable position angle converge at $PA$s varying
between 61 and 170$^{\circ}$, with a median value of 109$^{\circ}$. The lack
of models with $PA$s between 0 and 60$^{\circ}$ is intriguing, as it indicates
that even with the restricted image coverage of HST our models may have some
sensitivity to spiral parameters. However, as seen in Figure \ref{fig:fitnessplot2}, the `free-PA' models do not have different physical parameters from the full
sample, indicating that our assumption of a fixed $PA$ does not significantly
affect our results. 

The `constrained' sample, however, does appear to have a distribution
of central luminosities that is, on average, fainter than for the full
sample. While initially surprising, on closer inspection this discrepancy
is a function of fixing the scale-lengths of the models: The literature value
for $h_{r,\rho}$ is larger than most of the fitted scale-lengths of the full
sample (Figure \ref{fig:fitnessplot2}c). This drives the central dust density
down (Figure \ref{fig:fitnessplot2}b), which also lowers the central
luminosities. It is worth noting that our full, unrestricted sample appears to
produce a tighter correlation between color and luminosity (Figure
\ref{fig:fitnessplot2}a) than the `constrained' sample. This, along with the lack of significant systematic differences
between the `constrained' and full samples for any of the non-axisymmetric
parameters (Figure \ref{fig:fitnessplot3}) leads us to conclude that our full
models are as well-constrained as the `constrained' sample and a better match
to the observational data. 

\subsection{Analysis of the shapelet coefficients}
Because of the large variations in the parameters of our best fitting
models, we check to make sure the shapelet coefficients are properly
optimizing our models. In Figures \ref{fig:bestbband} and
\ref{fig:bestiband} we show the best fitting model out of all the
runs, compared to the HST images. Our model is optimizing towards a
plausible representation for the global light and dust distribution of
NGC 891.

To investigate how well our shapelet-amplitude statistic performs as a
fitness indicator (and to get a better overall view of the range of
models selected by the genetic algorithm) we show a sample (20\%) of
the best-fitting clumpy models in Figure \ref{fig:modelimages}. The
images are representative of variation seen in the 30 models from all
the genetic algorithm runs.  Overall it is clear that the
shapelet-derived fitness in the $\Delta \aeff$ images is doing an
adequate job of picking models that globally resemble NGC 891. The
luminous bumps in $\Delta \aeff$ seen to the left and right of the
bulge in some of the models (see e.g., the bottom panels of Figure
\ref{fig:modelimages}) are due to the spiral arms. These images also
show that our formalism can create models with significant clumpy,
non-asymmetric structure. However, even the clumpiest-looking models
in our sample lack the high-z tendrils often described as vertically
oriented ``chimney''-like structures. This discrepancy may indicate
that the fractal algorithm is unable to produce the necessary
substructure to mimic those seen in massive, edge-on galaxies, and
highlights the need for physically motivated dynamical models for the
ISM (such as \citealt{Wood10}).

While the dust chimneys seen in images of NGC 891 and other edge-on
galaxies are visually striking, their effect on the global photometric
properties of galaxies is likely to be small. The chimneys' apparent
stochasticity and likely origin in large outflows implies that
properly modeling them would not significantly change the free
parameters such as dust mass, scale-height, or central
emissivities. Similarly, they are likely to be very localized,
foreground structures and as such should play a relatively minimal
role in determining total attenuation corrections at all
inclinations.

\begin{figure*}
\begin{centering}
   \epsscale{1}\plottwo{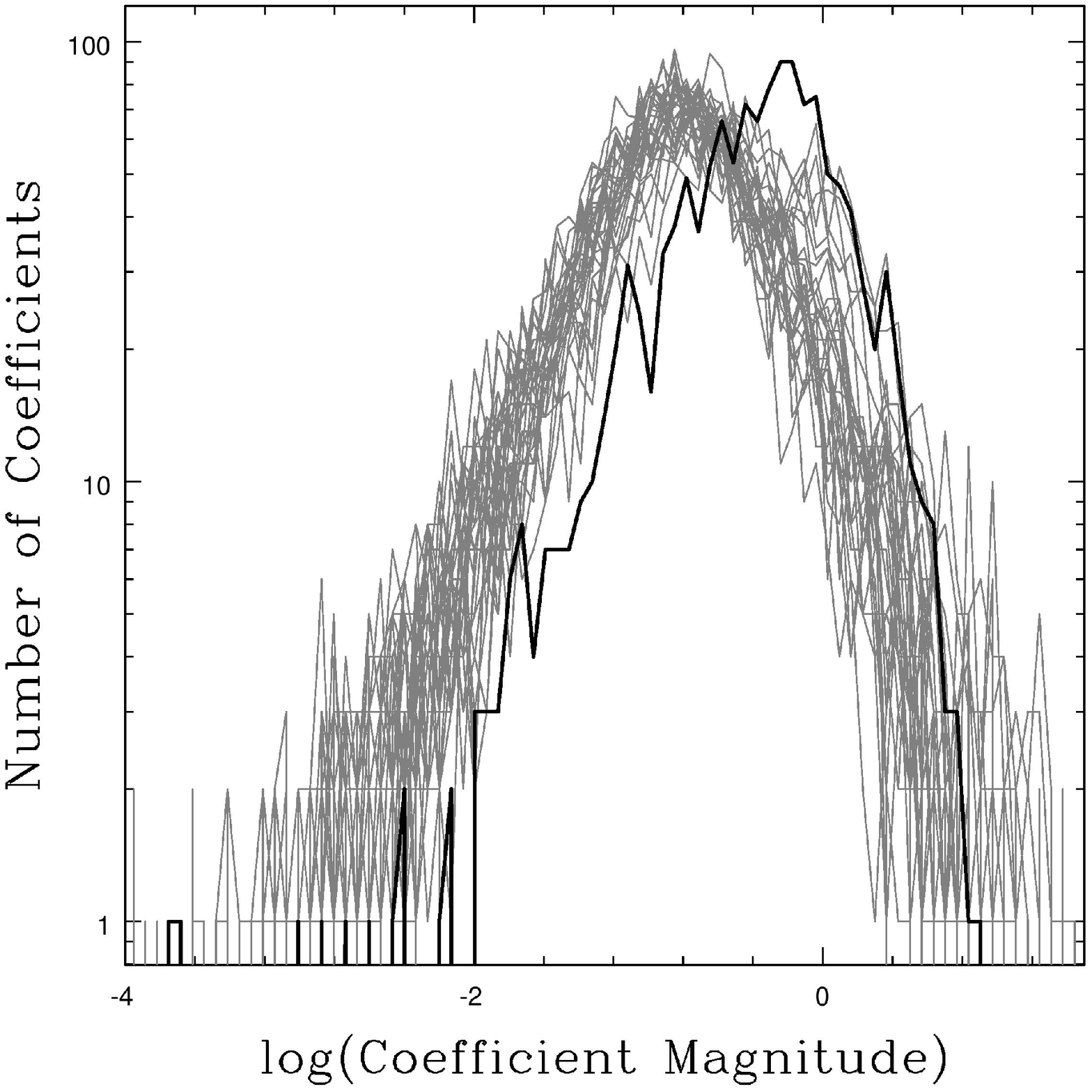}{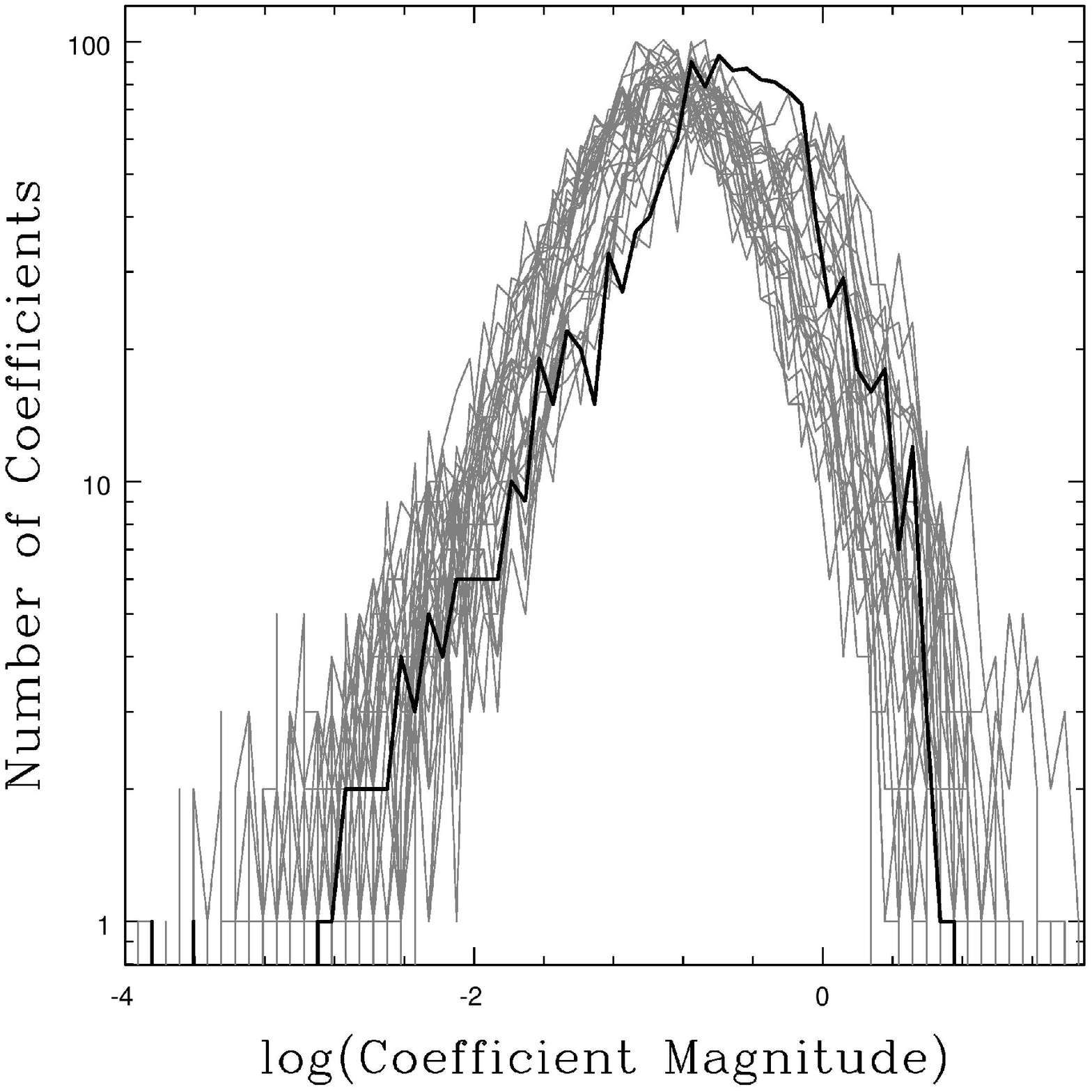}
   \caption{Histograms of shapelet coefficient magnitudes that
     reproduce the observed (black) and best fitting clumpy models'
     (gray) $\Delta \aeff$ maps. Left: B band. Right: I band.}
   \label{fig:bandhists}
\end{centering}
\end{figure*}

In an attempt to gain more insight into how our algorithm matches and
mismatches the observed complex dust structures, we have examined the
shapelet amplitude distributions of the 30 best-fitting models. We
plot histograms of the magnitudes of the shapelet coefficients at all
x,y orders for the HST images and our models in Figure
\ref{fig:bandhists}. The peak of the coefficient histograms of the
clumpy models occurs at a lower coefficient magnitude than the peak
for the reconstruction of the HST data in both B and I, although the
peaks are less separated in the I band --- possibly due to the B
band's increased sensitivity to dust attenuation. The clumpy model
reconstructions in both bands also have more coefficients at the
largest magnitudes. The lack of medium-magnitude coefficients in the
clumpy models means that most of the power lies in just a few
coefficients with large magnitudes, producing less rich substructure
than seen in reconstructions of the data.

\begin{figure*}
\begin{centering}
   \epsscale{1}\plottwo{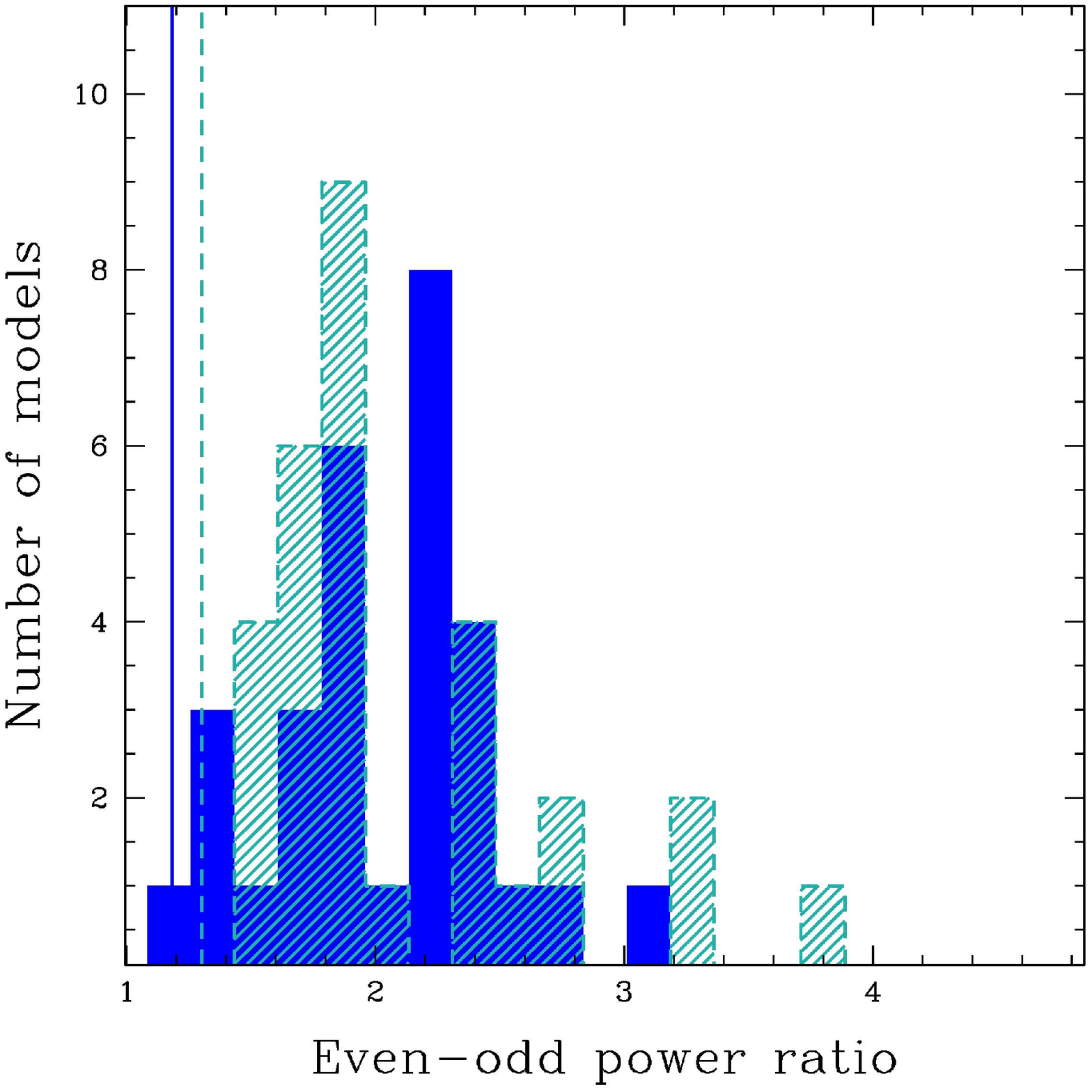}{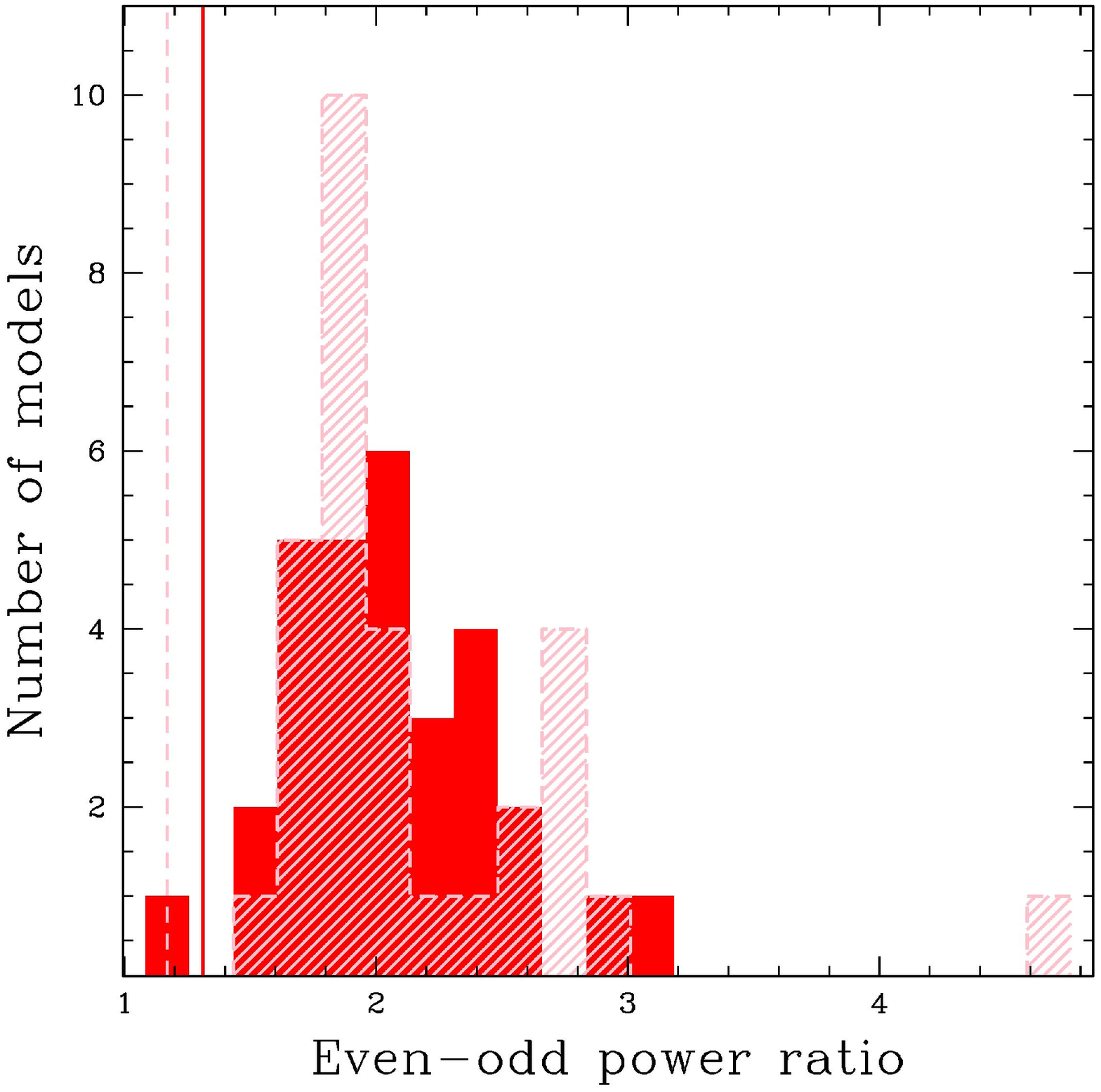}
   \caption{Distribution of the total power ratio in even and odd
     shaplet coefficients that reproduce $\daeff$ of the 30
     best-fitting models. The power ratio is computed by summing the
     magnitude of every even coefficient and dividing by the summed
     magnitude of odd coefficients. Left panel: B band. Right panel: I
     band.  Solid histograms denote ratios computed by selecting
     coefficients at even x (horizontal) values, while the dashed
     histograms show ratios for even y (vertical) coefficients. Solid
     and dashed vertical lines show the ratios selecting for x and y
     even coefficients, respectively, for the $\daeff$ constructed from
     the observed images of NGC 891.}
   \label{fig:evenvodd}
\end{centering}
\end{figure*}

We also investigated the even vs. odd distribution of the shapelet
coefficients, by comparing the sums of the coefficient magnitudes in
the even orders with those in the odd orders. This was prompted by
inspection of shapelet coefficient maps like the one shown in Figure
\ref{fig:shapeletrecon}, which has a sawtooth-like pattern of
coefficients. We find an even-to-odd ratio of $\sim$1 for both
observed bands. Our models, however, tend towards much higher ratios,
which can be seen in Figure \ref{fig:evenvodd}. This trend is apparent
when coefficients that are even in the x (horizontal) or y (vertical)
dimension are used, and is another indicator that it may be possible
to further exploit the shapelet analysis to find more realistic
models.

\section{Discussion}
\label{sec:discussion}

\subsection{Face-on appearance of the models}

The goal of this work was to create physically realistic RT models
that would resemble real galaxies at any inclination, something that
the current literature on axisymmetric structure and/or hand-added
clumpiness has difficulty doing. It is clear from the results of our
genetic algorithm that our spiral structure formalism is able to
reproduce the general edge-on appearance of NGC 891, which is not a
surprise based on the results of \citet{Misiriotis00}. But how
realistic do our best-fitting models appear at other inclinations?

To answer this question we reran our RT code, using the same inputs
for all 30 of our best-fitting models (even down to the same random
seed) but changing the inclination angle to zero and allowing photons
to be emitted from the outer parts of the disk. A representative
sample of the results is shown in columns three and six of Figure
\ref{fig:modelimages}. Virtually all of the models have prominent
spiral features, although there is disagreement in the handedness
of the spirality. Given the restricted radial range of the data, it
is unsurprising that direction of the spiral arms is poorly
constrained; while we display the orientation with the best fitness
most models had comparable fitnesses for spiral arms in both
directions. There is also some variation in pitch angle,
but for a given handedness it is relatively consistent, giving a
fairly uniform look to many of these models. It is likely that at
least some of the variation in spiral parameters is due to the
restricted areal coverage of the HST images; using a lower-resolution
image of the full galaxy might yield stronger constraints.

Whether or not the spiral parameters are well constrained, our prescription for spirality works fairly well, producing
realistic-looking spiral arms. Our models produce face-on images with
much brighter spiral arms compared to the bulge than the 'typical
spiral galaxy' model of \citet{Misiriotis00}. Compressing our spiral
arms by a factor of five greatly increases their relative brightness
for the same axisymmetric model parameters, preventing our models with
larger pitch angles from looking like the oblong blobs in the top
panel of their Figure 1.

\subsection{Face-on extinction and attenuation}

\subsubsection{Attenuation}
The general effect of compressing the dust into lagging spiral arms is
to create lanes of absorption along the inner edges of the luminous
arms. For the models with smoother dust this can be difficult to see
in Figure \ref{fig:modelimages}. Images of the face-on attenuation, $\aeff$, of the models from Figure \ref{fig:modelimages}
are shown in the first and third columns of Figure \ref{fig:allfaceon}
for the B and I bands, respectively. These attenuation maps
are computed by comparing models with dust to models with no dust,
computed using identical parameters. Red regions
are areas of high attenuation, while blue indicates regions
with no attenuation. Most ($\sim$70\%) of the models show a maximum
face-on $\aeff < 0.6$ mag in both bands. Of the few models with
$\abeff > 1$ mag, none have a peak attenuation $\gtrsim$2 mag, while
the peak attenuation in the I band is $\sim$1.5 mag. All
models show very low attenuation at radii larger than $\sim$1
$h_{r}$, even for models with strong spiral features at larger radii
between the center of the galaxy and the observation direction in
edge-on orientation (off the bottom of the face-on images).  Areas of
peak attenuation are highly confined to the spiral arms: the
average filling factor of the clumpy models for $\aeff \ge 0.5$ mag at
0.5, 1, and 2 $h_{r}$ is, respectively, 12.1, 5.2, and 0.4\% in the B
band and 5.5, 2.5, and 0.1\% in the I band. Filling factors at other
thresholds are given in Table \ref{tab:efffilfactors}.

We measured the average face-on attenuation to be
$\abeff = 0.24$, 0.15, and 0.03 mag and $\aieff = 0.16$, 0.10, and
0.03 at 0.5, 1, and 2 scale-lengths respectively.  In analogy to $R_V$
with a typical value of 3.1 measured for foreground dust screens in
the Milky Way, we define an attenuation curve parameter $R_{B,B-I}$ as
\begin{equation}
R_{B,B-I} = \frac{A_{B}}{E(B-I)}.
\end{equation}
With this definition, a foreground screen of Milky Way-type dust has
$R_{B,B-I} = 1.56$ \citep{Cardelli89}, and empirical estimates of
the attenuation of star forming galaxies yields $R_{B,B-I} =
1.76$ \citep{Calzetti00}.  In contrast the $R_{B,B-I}$ values for
our clumpy models are much larger ($R_{B,B-I} = 3$ at 0.5 and 1
$h_{r}$) and become undefined (positively infinite) at larger radii.
In other words, the attenuation is much grayer in our clumpy
models than for a foreground screen, and appears to become grayer with
increasing radius (due to scattering). The increase in the ratio of total to
selective extinction for
clumpy models relative to smooth models was noted for spherical systems by
(for example) \cite{Witt96} and \cite{Witt00}, as well as in more realistic
models of disk galaxies by \cite{Pierini04}. 

\begin{deluxetable}{cccrrrr}
\tablewidth{0pt}
\tablecaption{Filling Factors for Attenuation}
\tablehead{\colhead{} & \colhead{} & \multicolumn{5}{c}{Filling
    Factor Threshold (mag)}\\
\cline{3-7} \\
\colhead{Band} & \colhead{Radius ($h_{r,B}$)} & \colhead{$<0.1$} & \colhead{$<0.3$} & \colhead{$<0.5$} & \colhead{$<1.0$} & \colhead{$<3.0$}}
\startdata
B&0.5&47.4\%&23.8\%&12.1\%&2.5\%&0.0\%\\
I&0.5&34.1\%&11.1\%& 5.5\%&1.0\%&0.0\%\\
B&1.0&34.9\%&11.3\%& 5.2\%&1.2\%&0.0\%\\
I&1.0&22.1\%& 5.0\%& 2.5\%&0.3\%&0.0\%\\
B&2.0&13.3\%& 1.4\%& 0.4\%&0.0\%&0.0\%\\
I&2.0&15.7\%& 0.9\%& 0.1\%&0.0\%&0.0\%\\
\enddata
\label{tab:efffilfactors}
\end{deluxetable}

\subsubsection{Extinction as a foreground screen}
\begin{figure*}
  \begin{centering}
    \includegraphics[scale=0.8]{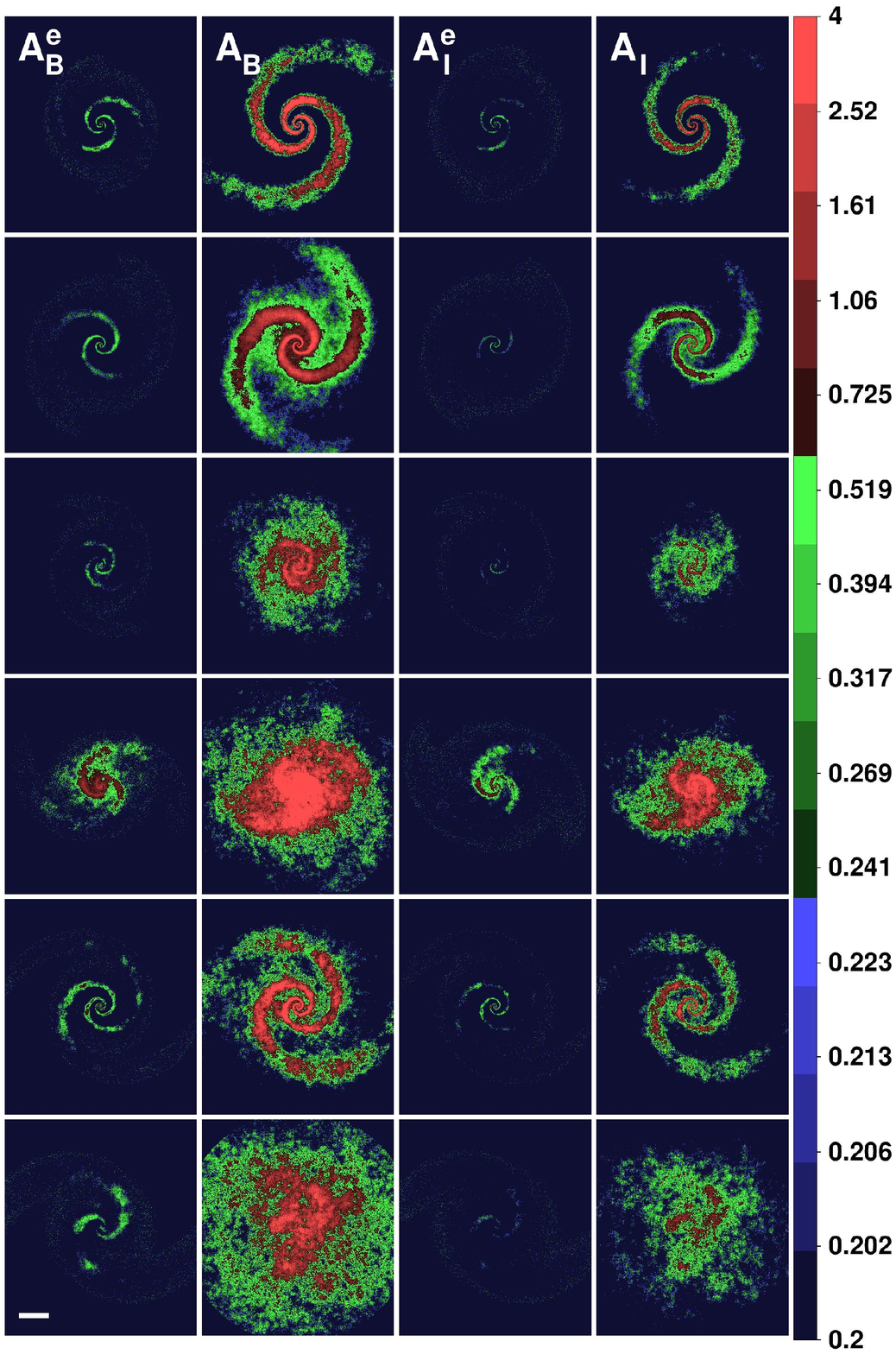}
    \caption{Face-on attenuation maps for the models shown in Figure
      \ref{fig:modelimages}. From left: B band attenuation,
      B band foreground screen extinction, I band attenuation, I band foreground screen extinction. The white bar
      in the lower left represents a distance of 5 kpc, while the
      color scale is in mag.}
    \label{fig:allfaceon}
  \end{centering}
\end{figure*}

We are also able to compute the face-on dust optical depth of our clumpy
models. The extinction from this optical depth (here called $A_{\lambda}$)
is very different than the attenuation: While $\aeff$ is the
attenuation of the admixture of dust and stars in a galaxy the
optical depth governs the extinction on objects behind the galaxy, where the
dust functions exclusively as a foreground screen. Foreground
extinction images are shown in the second and fourth columns of Figure
\ref{fig:allfaceon}. We find that this foreground screen extinction is
significantly higher than the attenuation, with an average
$A_{\lambda}$ at 0.5, 1, and 2 $h_{r}$ of 2.35, 1.35, and 0.45 mag in
the B band and 0.91, 0.55, and 0.22 mag in the I band. In magnitudes,
the scaling between $A_B$ and $\abeff$ is roughly a factor of 11,
while between $A_I$ and $\aieff$ there is a magnitude scaling of
roughly a factor of 6. The difference in these two scale factors imply
that the foreground screen is less gray, as expected.  We find foreground
screen values of $R_{B,B-I} \sim 1.8 \pm 0.2$, comparable to the
\citealt{Calzetti94} attenuation curve.  The foreground screen extinction
is much higher than $\aeff$ due to a combination of scattering and the
overlapping nature of the stellar emission and dust, a discrepancy which has
been noted previously (e.g. \citealt{Pierini04}). To confirm this we ran a
version of one of our clumpy models with the dust offset to 5 kpc in front of
the galaxy and scattering turned off, which produced a 'pseudo foreground
extinction' image very similar to the actual foreground screen
extinction. Removing the scattering even without offsetting the dust was
enough to increase $\aeff$ by about 0.5 mag in both B and I bands, a large
enough amount to make some regions change from optically thin to thick.

\begin{deluxetable}{cccrrrr}
\tablewidth{0pt}
\tablecaption{Filling Factors for Screen Extinction}
\tablehead{\colhead{} & \colhead{} & \multicolumn{5}{c}{Filling
    Factor Threshold (mag)}\\
\cline{3-7} \\
\colhead{Band} & \colhead{Radius ($h_{r,B}$)} & \colhead{$<0.1$} & \colhead{$<0.3$} & \colhead{$<0.5$} & \colhead{$<1.0$} & \colhead{$<3.0$}}
\startdata
B&0.5&99.1\%&92.7\%&82.6\%&58.8\%&18.7\%\\
I&0.5&95.2\%&68.4\%&50.2\&&23.4\%& 4.9\%\\
B&1.0&98.3\%&84.2\%&65.1\%&37.4\%& 8.4\%\\
I&1.0&89.5\%&51.3\%&30.6\%&11.4\%& 2.1\%\\
B&2.0&89.7\%&45.1\%&27.2\%& 9.4\%& 0.9\%\\
I&2.0&60.6\%&19.2\%& 8.6\%& 2.3\%& 0.1\%\\
\enddata
\label{tab:screenfilfactors}
\end{deluxetable}

We can compare these results to measurements in the literature for
other spirals.  Using the HyperLEDA database \citep{Paturel03} and
assuming D=9.5 Mpc we find $R_{25}\approx 18$ kpc and therefore
$h_{r}\approx 0.3 R_{25}$. At 0.3 $R_{25}$ \citet{Holwerda05} find an
average $A_{I}$ at this radius of roughly 0.5-1 mag (see the
top left panel of their Figure 7), in agreement with our results.
However, while we also find that the largest source of dust
column-density comes from the spiral arms, \citet{Holwerda05} report
$A_{I}$ of $\sim$1 mag at one scale-length. Our models contain less
dust extinction between the arms. For example, at one scale-length the
I band filling factor for extinction $\ge$0.5 mag in our models is
only 31\%. On the other hand, \cite{White00} and \cite{Domingue00} find similar values for
the foreground screen extinction in spiral arms in the B and I bands
(albeit with a spread of $\gtrsim$1 mag for both bands) but interarm
extinction values of $0<A<1.4$ mag, with most galaxies in their sample
having interarm extinction closer to zero. \cite{Holwerda09} report optical
depths almost entirely lower than 0.5 from B to I band in the outer regions of
a backlit spiral. These results are consistent with our models, where we find the average filling factor
for foreground screen extinction with a threshold of 0.3 mag is 84.2\%
and 51.3\% at 1 scale-length in the B and I bands,
respectively (additional optical depth values and filling factors for our models of NGC 891 are given in
Table \ref{tab:screenfilfactors}). Our models also consistently predict large central face-on optical depths,
with values ranging from $\sim$2.5 to $>$4 in the B band. However,
\cite{Kuchinski98} find central optical depths of 0.5-2.0 in the V band, which
for Milky Way-type dust is equivalent to $\tau_{B}\approx$0.7-2.7. This
discrepancy is likely due to the inability of optical models to see to the
center of the galaxy. 

An inspection of Figure \ref{fig:allfaceon} shows a wide range of dust
column-density morphology (and therefore foreground screen extinction)
but a much smaller range of appearances for the attenuation.
For example, in rows two and six, the model attenuation looks
fairly similar, with low levels of patchy attenuation tracing out
narrow spiral arms. In contrast, the foreground screen maps are vastly
different: The dust in row two is confined largely to prominent spiral
arms while the dust in row six shows little coherent structure. These
degeneracies imply that estimates of the attenuation based on measurements of a galaxy's
extinction as a foreground screen are not well constrained --- unsurprising
given the complex interplay between dust and starlight in these galaxies.  As
important, measurements of galaxies' dust content observed as a
foreground screen (e.g., \cite{White00}, \citet{Holwerda05}) yield
estimates of attenuation that are many times too large and not
gray enough.

\begin{figure*}
  \begin{centering}
    \includegraphics[scale=0.8]{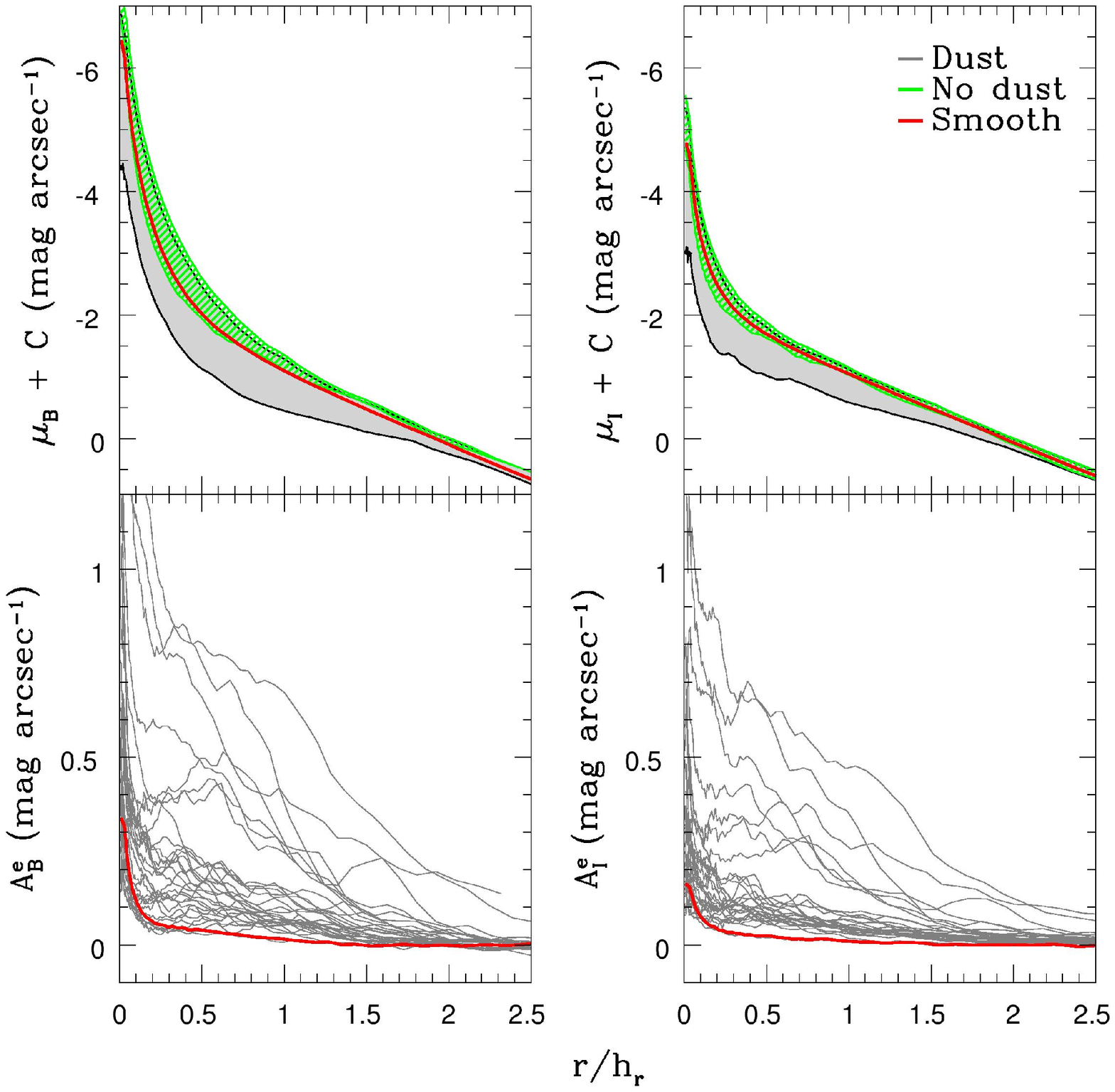}
    \caption{Face-on surface brightness and attenuation profiles of the 30 best fitting clumpy-dust models with spiral
      structure, in the B (left) and I (right) bands. Upper panels
      show the range of model surface brightness profiles with dust (gray full shaded region)
      and without dust (green dashed shaded region).  The red line
      indicates the profile for our fiducial model with a smooth dust
      and light distribution (see text). Lower panels show the
      residuals between the profiles of models with and without dust,
      i.e., an azimuthally averaged attenuation profile.}
\label{fig:allfaceonsbprofs}
\end{centering}
\end{figure*}

\subsection{Face-on surface brightness and color profiles}
While the attenuation maps presented show where the attenuation occurs, and we can quantify the filling factor of this
attenuation, neither tell us directly the light-weighted
attenuation as a function of radius, i.e. what fraction of
the total light in a given annulus is attenuated.  Therefore we
construct surface brightness profiles both with and without dust for
our best fitting models, as well as the smooth model from
\citet{Popescu00}. We plot these surface brightness profiles and the
difference between the models with and without dust in Figure
\ref{fig:allfaceonsbprofs}. The residual between models with and
without dust ($\mu-\mu_{\rho_{d}=0}$, where $\mu$ and
$\mu_{\rho_{d}=0}$ denote the surface brightness in the B or I bands
for models with and without dust, respectively), is a measure of the
azimuthally-averaged attenuation as a function of radius. We find that most models have very low attenuation at all radii, with 90\% of the models having $A_{B,I}^{e}
= \mu_{B,I}-\mu_{B,I,\rho_{d}=0} \le 0.1$ mag
at $r/h_{r} = 1.5$.  Average $\aeff$ values at 0.5, 1, and 2 $h_{r}$
are 0.20, 0.12, and 0.03 mag in the B band and 0.12, 0.08, and 0.04
mag in the I band. In general, the clumpy dust models have both higher
attenuation and more variability at all radii than the
fiducial model with a smooth dust distribution and no spiral
structure.  Additionally, the models with more attenuation also
appear to have type-II \citep{Freeman70} surface brightness
profiles. The low amounts of face-on attenuation imply that
measured kinematics of face-on galaxies are minimally affected by dust
attenuation \citep{Bershady10b}.

\begin{figure*}
  \begin{centering}
    \epsscale{1}\plottwo{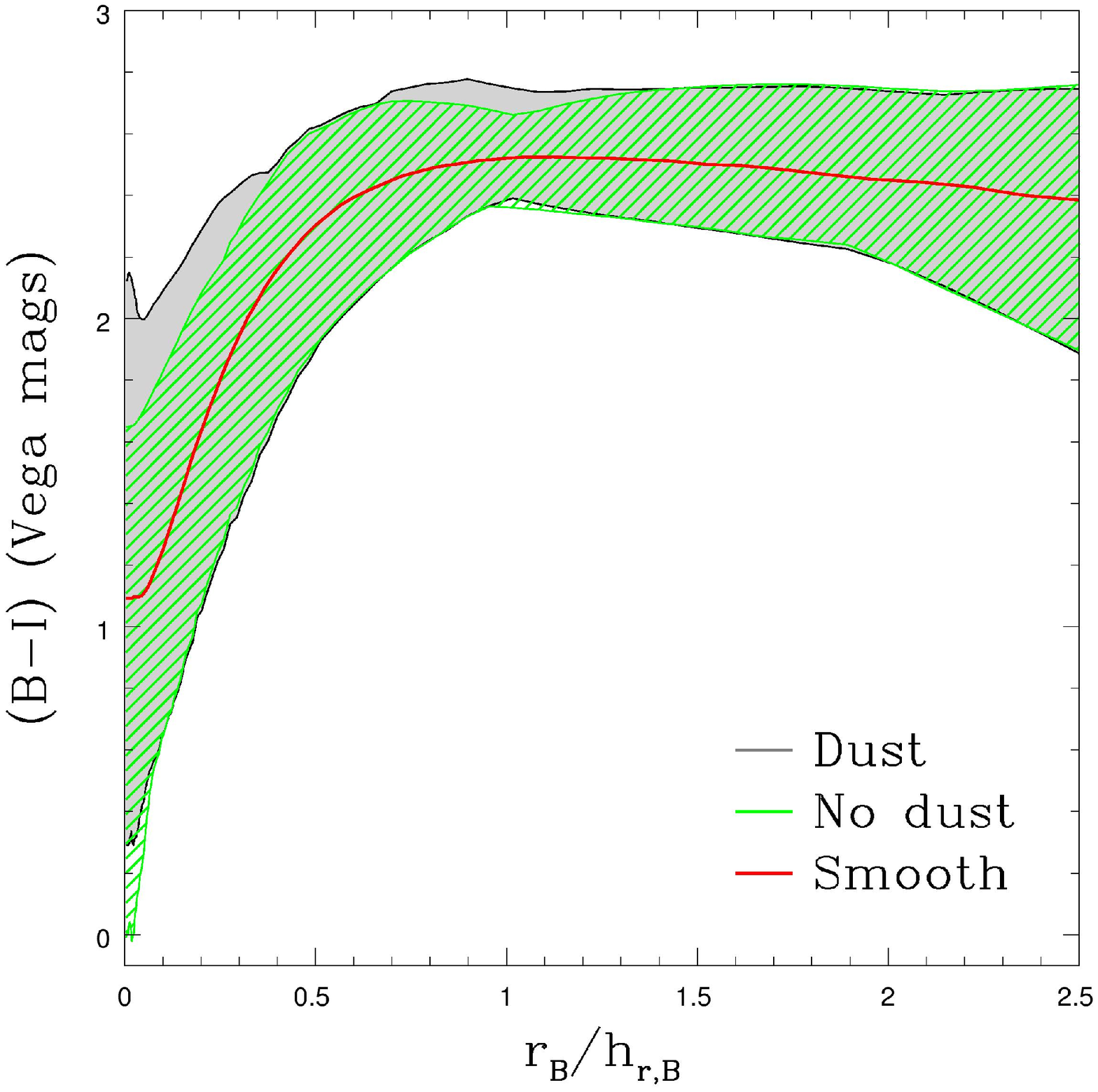}{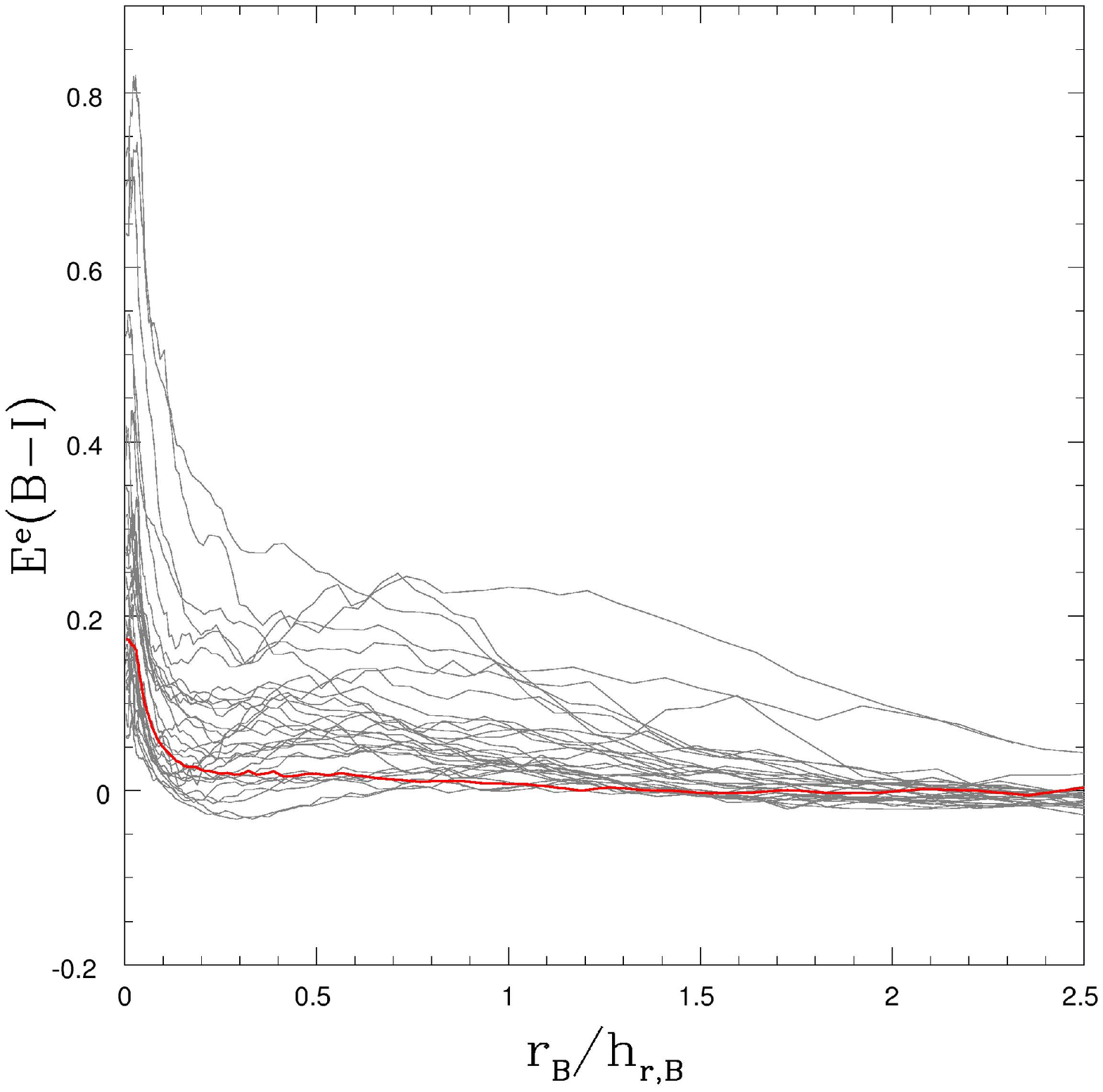}
    \caption{Color profiles and reddening: (a) Radial B-I color
      profiles of the clumpy-dust models, with and without dust, and
      the fiducial model illustrated in Figure \ref{fig:allfaceonsbprofs}
      with the same legend. (b) Effective color-excess profile in
      B-I.}
\label{fig:allfaceoncolorprofs}
\end{centering}
\end{figure*}

\begin{figure}
  \begin{centering}
    \includegraphics[scale=0.4]{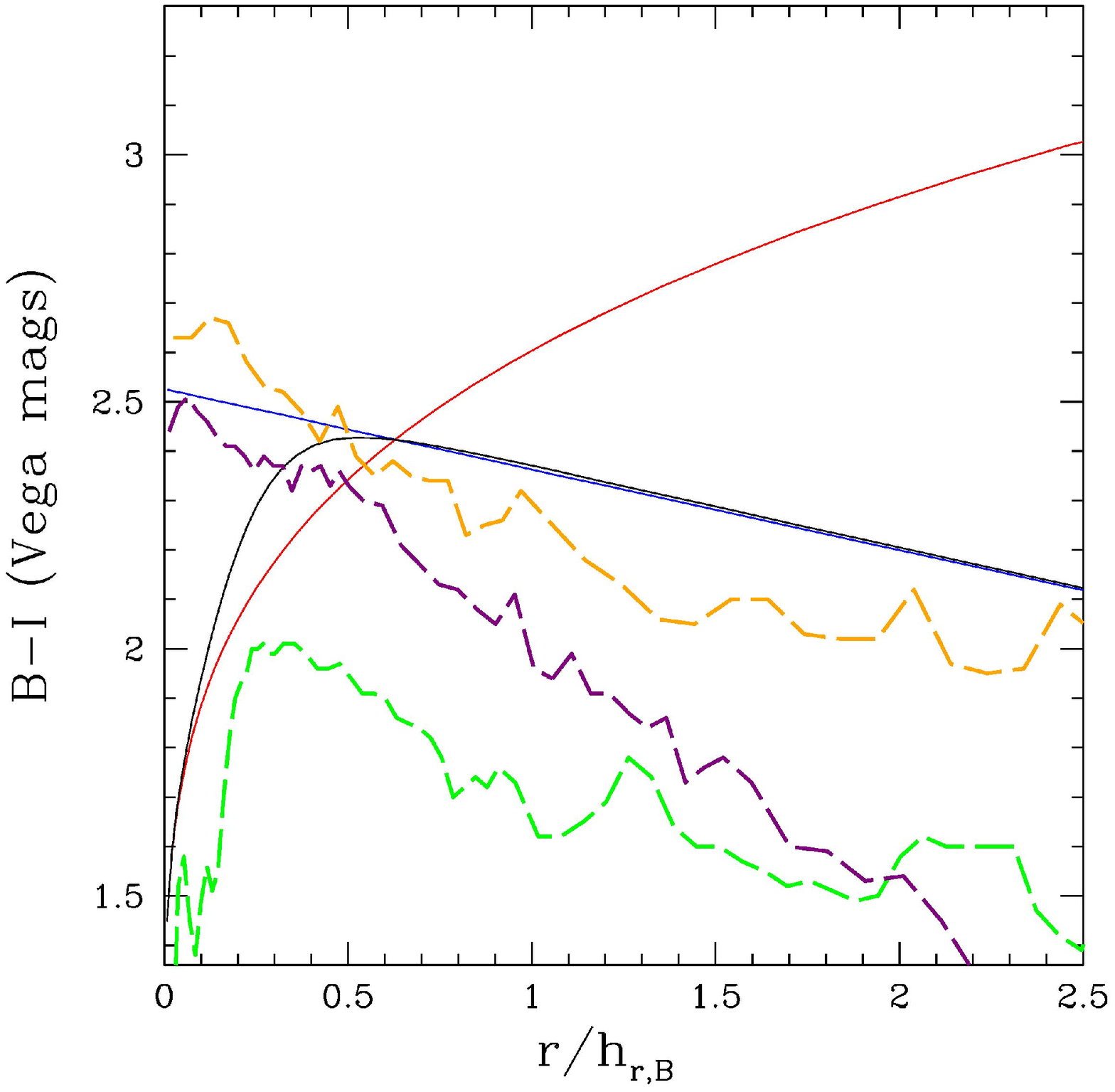}
    \caption{Color profile contributions without dust for the fiducial
      model with smooth dust and no spiral structure, from Figure
      \ref{fig:allfaceonsbprofs}, compared to observed color profiles
      for low-inclination spiral galaxies.  Model bulge and disk color
      profiles are shown by the red and blue solid lines,
      respectively. The total model color profile is in black. Three
      observed galaxy color profiles from \cite{deJong96} are overlaid
      with dashed lines: UGC 7450 (green), UGC 2064 (orange), and UGC
      11628 (purple). These examples are discussed in the text.}
    \label{fig:smoothcolorplot}
  \end{centering}
\end{figure}

\begin{figure*}
  \begin{centering}
    \subfigure{\includegraphics[scale=0.256]{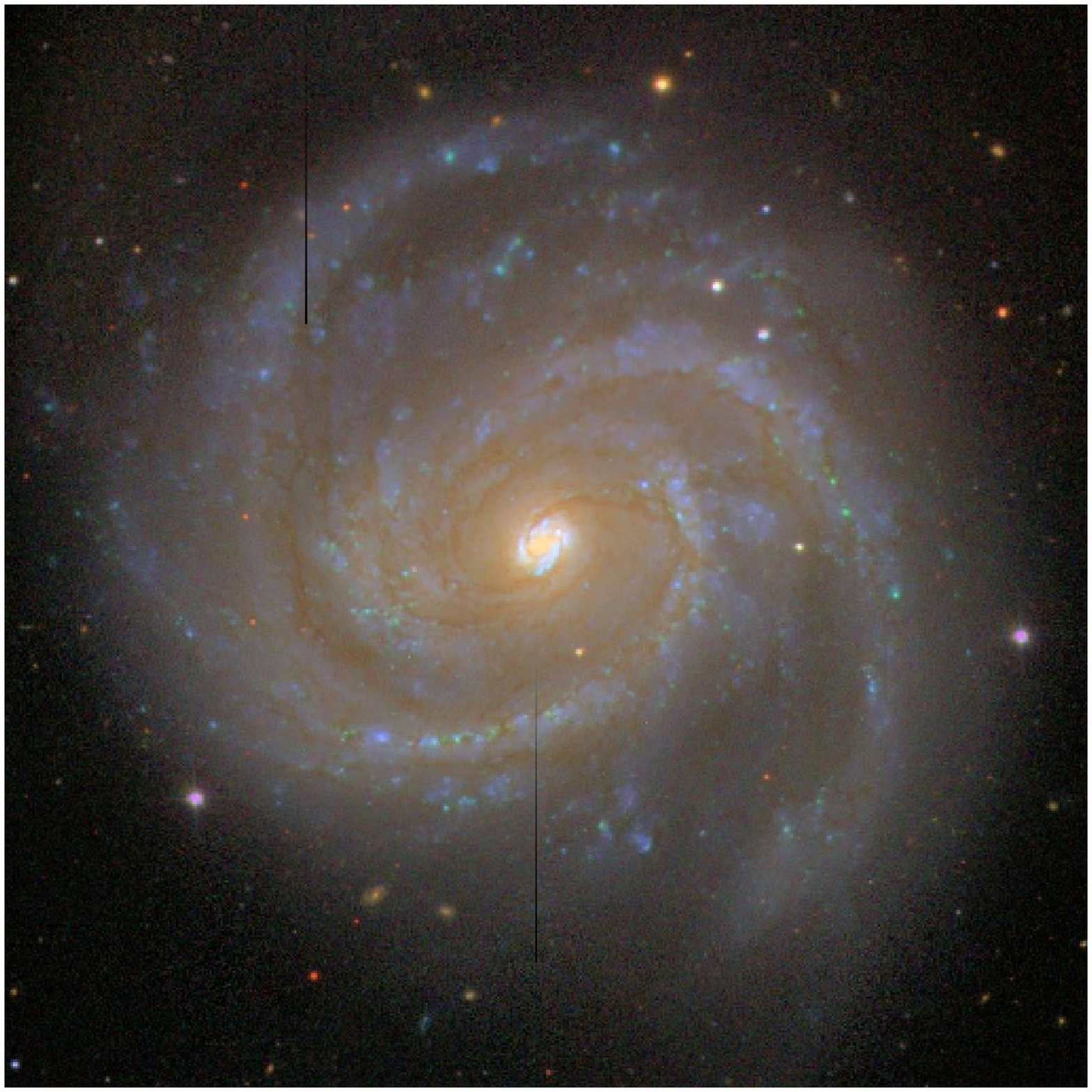}}
    \subfigure{\includegraphics[scale=0.265]{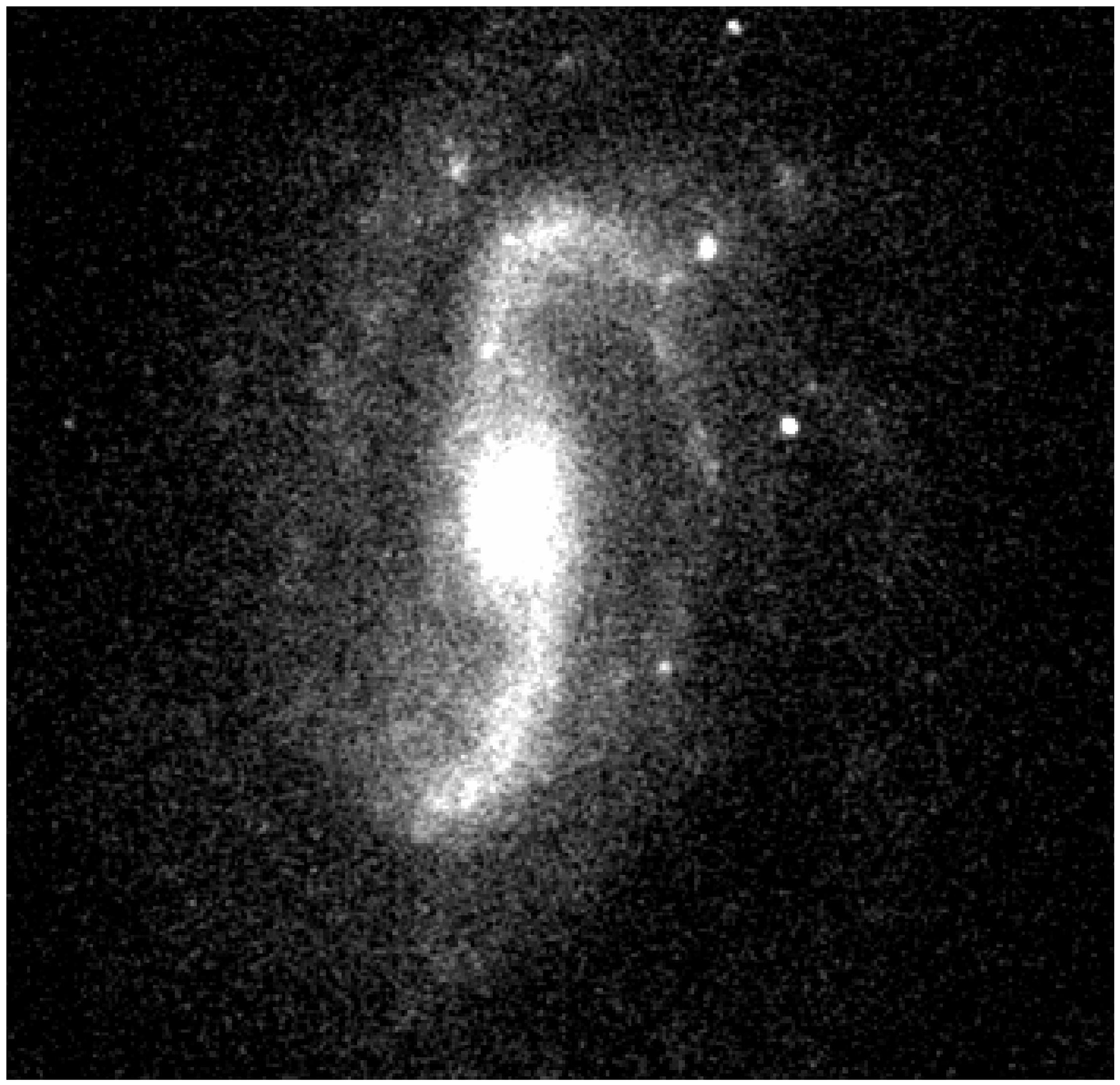}}
    \subfigure{\includegraphics[scale=0.256]{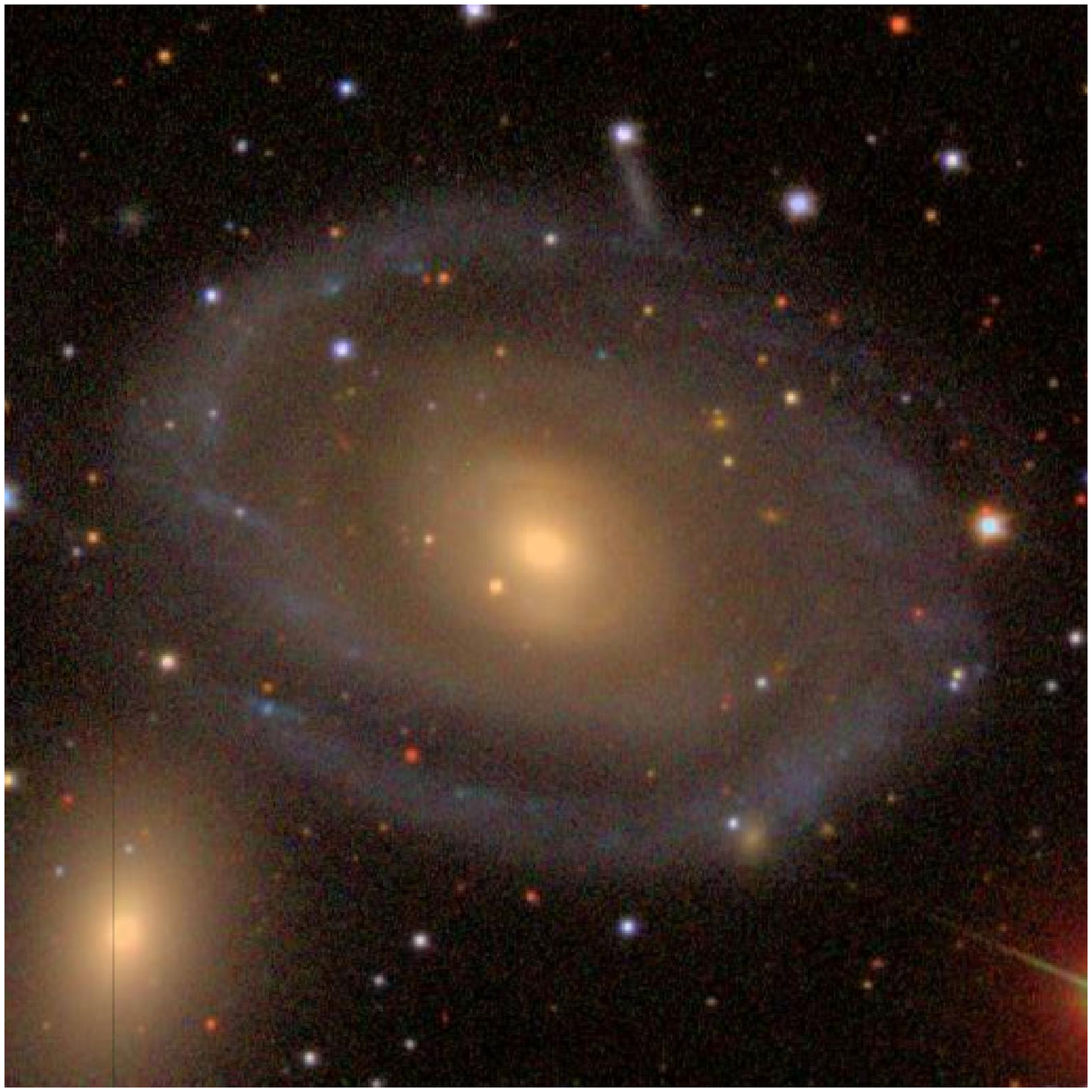}}
    \caption{Images of the three low-inclination galaxies with color
      profiles shown in Figure \ref{fig:smoothcolorplot}: UGC 7450
      (left; SDSS color composite, 400 arcsec field of view), UGC 2064
      (middle; V band image from \citealt{deJong96fits}, 100 arcsec
      field of view) and UGC 11628 (right; SDSS color composite, 200
      arcsec field of view).}
\label{fig:counterparts}
  \end{centering}
\end{figure*}

Radial profiles of $B-I$ color and the attenuation color excess $E^{\rm
  e}(B-I) = (B-I) - (B-I)_{\rho_d=0}$ are plotted in Figure
\ref{fig:allfaceoncolorprofs}. The inner region is unusually blue,
with a strong gradient toward bluer central colors. Recall that we
have simply adopted the bulge parameters from the literature for
smooth-dust models of NGC 891. The global (disk + bulge) color profile
rapidly reddens with increasing radius until $\sim$1 disk
scale-length, at a rate of roughly $\sim3 (r/h_r)$ mag in $B-I$ for
$r/h_r<0.5$. At larger radii, all but two models then have color
profiles which become monotonically bluer with increasing radius at a
modest rate of $\sim 0.1 (r/h_r)$ mag, where we expect the light from
the disk dominates. However, the $B-I$ color remains quite red ($>2$
mag) even at large radii, while $E^{\rm e}(B-I)$ remains very small
given the small values for the attenuation seen in Figure
\ref{fig:allfaceonsbprofs}. A few ($\sim$5) profiles with dust become
bluer than their dust-free counterparts at radii $<$ 1 B band
scale-length, due to scattering.

\subsection{Low-inclination Counterparts}

Since a blue bulge and red disk are not commonly found in face-on
spiral galaxy samples, it is worth exploring whether these results
found both from smooth and clumpy dust models are an artifact of
inaccuracies of the modeling and radiative transfer or an insight into
unusual properties of this extragalactic system viewed edge-on. First,
as the left panel of Figure \ref{fig:allfaceonsbprofs} shows, both the
smooth-dust model from the literature for NGC 891's disk and our
clumpy-dust models produce comparable colors (bulge colors are the
same by construction). Hence at least the outer portion of these color
profiles are not due solely to adding clumpiness and spirality to the
models.

To further investigate this issue we plot the color contributions of
the bulge and the disk independently for the fiducial model with
zero dust in Figure \ref{fig:smoothcolorplot}. As we anticipated, the
blue central color is entirely due to the bulge, while the disk, with
a central $B-I$ color a little redder than 2.5 mag, starts dominating
the color profile at $r > 0.5 h_r$. For a bulge as red as an
early-type stellar system (e.g., gm spectral-type;
\citealt{Bershady95}) we might expect $B-I\sim2.3$ mag. In contrast,
for a typical disk we might expect $B-I$ between 1.5 and 2 mag
\citep{deJong96}. However, galaxies of a given morphological type have
a large scatter around any mean color values. As examples, we overlay
color profiles from \cite{deJong96} in Figure
\ref{fig:smoothcolorplot} for UGC 7450, UGC 2064, and UGC 11628.
These are three low-inclination systems of type SABab/SABbc, not
unlike the putative morphological type of NGC 891 classified from an
edge-on view. Images of these systems are shown in Figure
\ref{fig:counterparts}.

UGC 2064 and UGC 11628 both have red disks consistent with that of NGC
891, at least in the inner regions. The disk of UGC 2064, however has
comparable color to that of the model for NGC 891 out to well beyond
$2h_r$, while UGC 11628 has enhanced star-formation at large radii.
Both appear to have sufficiently strong spiral structure at larger
radii to explain the asymmetry in star-formation indicators seen in
NGC 891, if these systems were viewed edge-on. In contrast to the
central colors of these galaxies, UGC 7450's central color profile is
much bluer than the disk, and is consistent with the model for NGC
891. The reason UGC 7450's central light profile is so blue is because
there exists a modest, nuclear ring of star formation, visible in
Figure \ref{fig:counterparts}).  While this feature is not a bulge per
se, it could be mimicked by a bulge component with a strong inner
gradient, as seen in the smooth model of NGC 891. While the unusual
color profile in the models are therefore {\it possible}, it is impossible to
discount the (perhaps likely) possibility that it is 
due simply to the inability of optical images to probe the dust-obscured inner part
of the galaxy. 

\subsection{Inclination corrections to integrated magnitudes}
\subsubsection{The Tully-Fisher Relation}

The attenuation of a disk galaxy is dependent on that galaxy's inclination in
addition to the detailed geometry of the dust (see \citealt{Witt92}
for a forceful discussion of this issue). When computing a disk
galaxy's integrated luminosity for, e.g. TF studies, this
attenuation must be corrected for. Multiple approaches to computing the
attenuation as a function of inclination exist in the literature.
\citet{Tully85} constructed an attenuation correction based on
theoretical formalism assuming a smooth dust slab mixed homogeneously
with the stars in the slab, with a fraction of starlight above and
below the slab.  This model can be parameterized as
\begin{align}
A_{\lambda}^{i}&=-2.5\times \log \nonumber \\
&\left [ f(1+e^{-\tau_{\lambda}\sec i})+(1-2f)\left (\frac{1-e^{-\tau_{\lambda}\sec i}}{\tau_{\lambda}\sec
        i}\right )\right ],
\end{align}
where $f$ is the fraction of light outside the slab of dust and
$\tau_{\lambda}$ is the optical depth. \citet{Giovanelli94} created an
empirical formalism for the attenuation correction, parametrized by the
inclination-corrected HI line-width W$_{R,I}^{i}$ \citep{Tully98},
and given by 
\begin{equation}
A_{\lambda}^{i} = A_{\lambda}^{i=0} - (\alpha_{\lambda} + \beta_{\lambda}(\log
W_{R,I}^{i}-2.5))\log \left (\frac{a}{b}\right )
\end{equation}
and
\begin{equation}
\frac{b}{a} = \sqrt{\cos^{2}(i)\,(1-q^2) + q^2}.
\end{equation}
Here $A_{\lambda}^{i=0}$ is the face-on attenuation, $(a/b)$ is the
axis ratio, $\alpha_{\lambda}$ and $\beta_{\lambda}$ are constants
that depend on the bandpass, and q is the intrinsic axial ratio. Both functions are optimized to minimize
scatter in the observed TF relation.

To investigate the inclination-dependence of the attenuation
on integrated magnitudes for NGC 891-like galaxies, we photometered all
of our 30 best-fitting clumpy-dust models rendered at a range of
inclinations, with and without dust. As before, the flux ratio between
a model with and without dust allows us to compute the attenuation. We perform the same analysis for our fiducial
smooth model.

\begin{figure*}
  \begin{centering}
    \epsscale{1}\plottwo{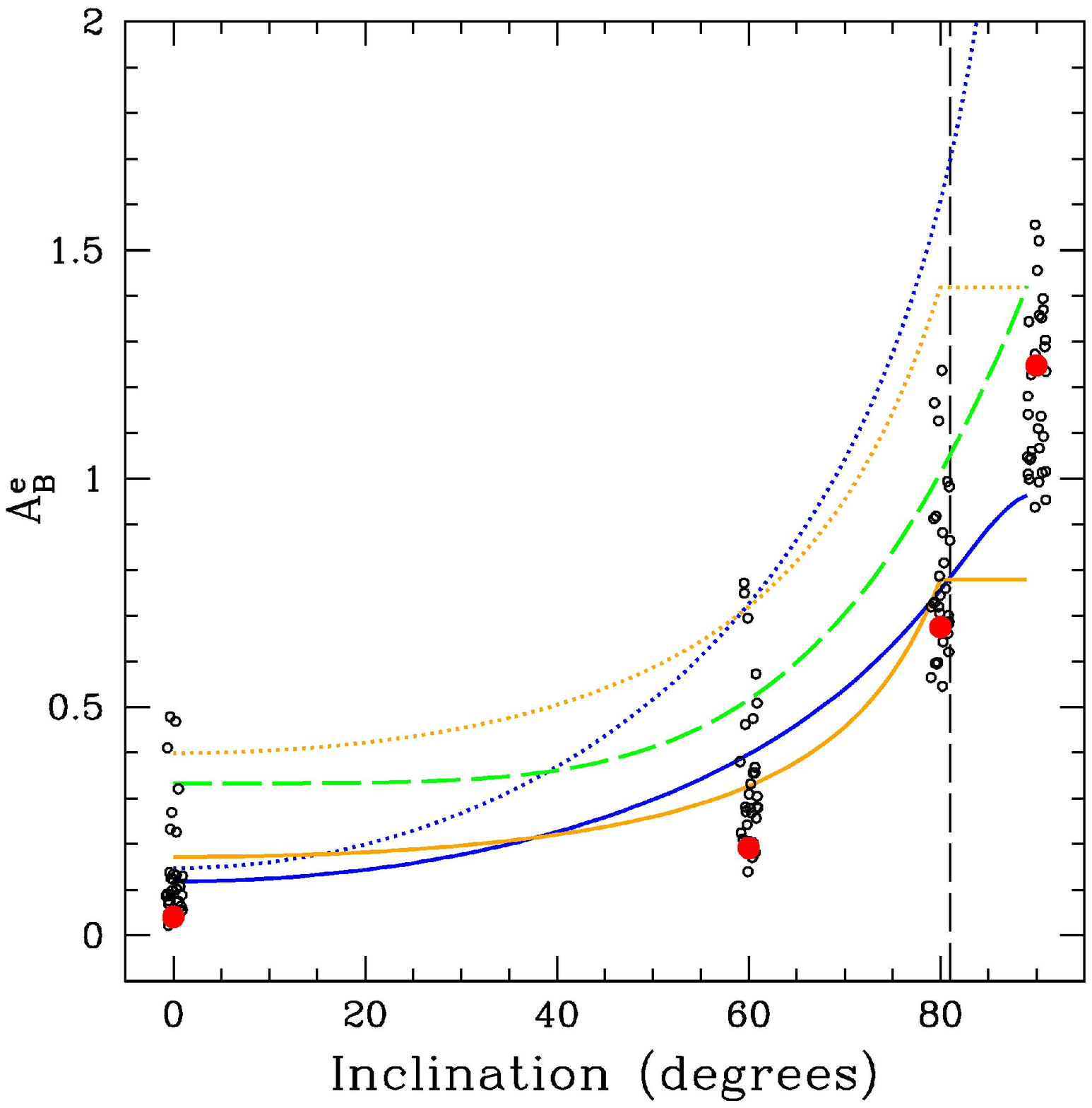}{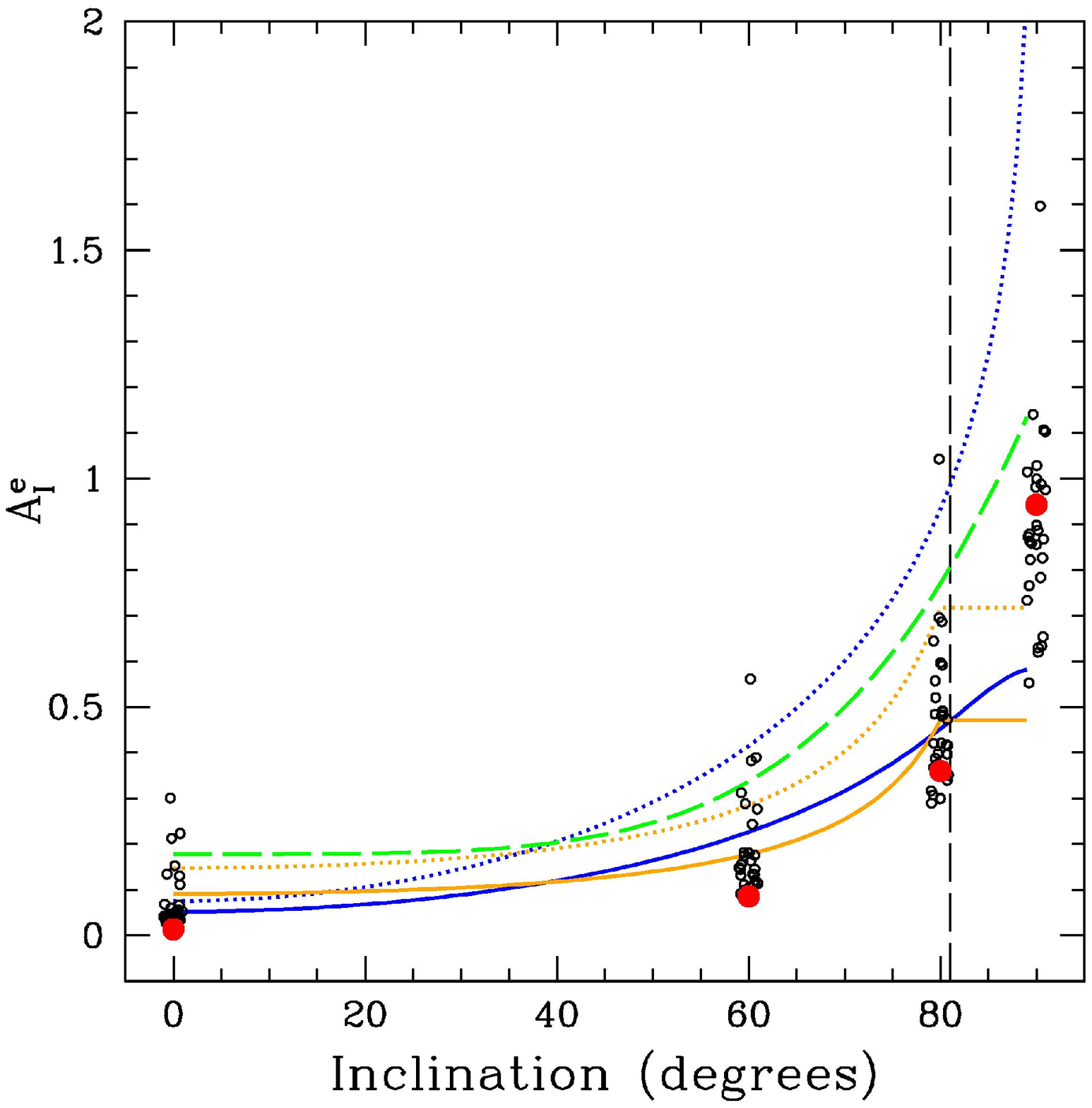}
    \caption{Inclination-dependent attenuation in B (left) and
      I (right) bands. Black open circles are clumpy models and the
      red dots are the smooth model. Horizontal scatter in the black
      open circles is random noise added to increase the visibility of
      overlapping data points. Orange lines denote model predictions
      using the \citet{Tully85} formalism, while blue lines use the
      \citet{Giovanelli94} formalism as given in
      \citet{Verheijen01b}. Solid lines represent fits of these
      analytic formalisms to the clumpy models between 0$\le$i$\le$80
      degrees; dotted lines represent these analytic formalisms for
      parameters in the literature (e.g., \citet{Verheijen01b}),
      relevant for NGC 891 (see text). The green dashed line shows the
      results for NGC 891 using a model that does not include
      spirality but has a star-forming thin disk
      \citep{Tuffs04,Driver08}.}
  \label{fig:inclexteffects}
  \end{centering}
\end{figure*}

We plot the inclination-dependent attenuation in both HST
bands in Figure \ref{fig:inclexteffects}. The clumpy models have
statistically higher attenuation at lower inclinations than
the smooth model, while at high inclinations they are comparable. This
indicates that fitting smooth radiative transfer models to edge-on
galaxies statistically underestimates the true amount of face-on
extinction by an average of 0.1 and 0.06 mag and a maximum of 0.45 and
0.32 mag in B and I bands, respectively.

We have also fit both the empirical \citet{Giovanelli94} and
theoretical (slab-model) \citet{Tully85} attenuation functions to the
clumpy models. Essentially this is equivalent to a calibration (in a
least-squares fitting sense) of the functional parameters of these
formalisms to our RT models. The fitting is done only for inclinations
$<$80$^{\circ}$, since the theoretical function is discontinuous and
the empirical function is strongly dependent on the axial ratio at
purely edge-on inclinations. While both functions do a credible job of
fitting the models at inclinations between 0 and 80$^{\circ}$, neither
adequately predict the increase in attenuation at inclinations
$>$80$^{\circ}$. This is not particularly worrisome for TF
studies; such high inclinations are dis-favored on the prejudicial
suspicion that indeed attenuation-corrections based on these formalisms
in this inclination regime are problematic.

However, what is relevant for TF studies are the differences between
the attenuation values used in the literature (dotted lines
in Figure \ref{fig:inclexteffects}), and the values we find using the
same formalisms calibrated to our radiative transfer models of NGC 891
(solid lines). It is also astrophysically interesting to note
how the functional parameters change.  For example, we find that in
both bands the theoretical formalism based on the simple slab model
used in the literature  differs from that
calibrated against our clumpy models almost entirely in the choice of
the face-on attenuation: The fraction of light above and below
the dust slab, $f$, is almost unchanged between the fit to the clumpy
models and that used in \cite{Verheijen01b}. In contrast, the face-on attenuation is roughly doubled in the literature for both the B and I
bands, from our fit value of 0.33 to 0.81 in \cite{Verheijen01b} for $\tau_B$
and from 0.17 to 0.28 for $\tau_I$.

For the empirical formalism we also plot in Figure
\ref{fig:inclexteffects} the attenuation function for literature
parameters of $\alpha_{\lambda}$ and $\beta_{\lambda}$ from
\citealt{Verheijen01a} ($\alpha_{B}=1.57$, $\alpha_{I}=0.92$,
$\beta_{B}=2.75$, and $\beta_{I}=1.63$) and a value of
W$_{R,I}^{i}$ appropriate for the observed rotation curve of NGC 891;
and also the attenuation function calibrated to our clumpy models, fixing
$\alpha_{\lambda}$ and $\beta_{\lambda}$ at the literature values 
but allowing W$_{R,I}^{i}$ to
vary. For both inclination corrections we follow \citealt{Giovanelli94} and convert between $i$ and
$(a/b)$ using $q=0.13$. For
the empirical fit using literature parameters we adopt $W_{R,I}^{i} =
424$ km s$^{-1}$, twice the value for the maximum gas velocity
in NGC 891 given by hyperLEDA \citep{Paturel03}.  In contrast, our
best calibration of this formalism requires $W_{R,I}^{i} = 189\pm 25$
km s$^{-1}$ and $202\pm34$ km s$^{-1}$ in the B and I bands,
respectively. This value for $W_{R,I}^{i}$ is less than half of what
is expected for NGC 891 based on the observed H~I gas velocity but
consistent between the two bands within the uncertainties. In this formalism
allowing $W_{R,I}^{i}$ to vary is equivalent to varying $\alpha_{\lambda}$
while holding $W_{R,I}^{i}$ and $\beta_{\lambda}$ constant, and we find that for $W_{R,I}^{i} =
424$ km s$^{-1}$ our clumpy models predict $\alpha_{B}=0.61$ and
$\alpha_{I}=0.40$, significantly lower than values from the literature. 

In both cases our results imply significant revision downward for
dust-corrections in TF studies. In the $B$ band for NGC 891-like
galaxies observed at $i = 60^\circ$ this is equivalent to a
$\sim$60\% decrease in the mass-to-light ratio, and intrinsic $B-I$
colors which are $\sim$30\% bluer. More fundamentally, our results
imply that the implicit assumption in the literature, namely that the
TF-scatter can be minimized as a metric for constraining other
astrophysical quantities, may be incorrect in general. Certainly this
assumption appears to fail in the case of determining the attenuation for galaxies like NGC 891.

\subsubsection{Co-moving Star-formation Rates and Stellar Mass Density}

Many estimates in the literature of star-formation rates and stellar
mass density in cosmological volumes are based on counting the number
of UV through near-infrared photons emitted from galaxies (see
\citealt{Driver08} for a recent summary). It is well known that these
estimates are subject to uncertainties in the attenuation,
which, as we have seen, are inclination dependent. Because most
star-formation appears to take place in a thin, dusty disk layer, the
details of radiative transfer and the impact of a clumpy dust
distribution are likely to be particularly important in predicting the
emergent flux and its SED.  In the context of
matching optical and far-infrared and sub-millimeter observations to
obtains bolometric estimates of star-formation rates, \citet{Driver08}
have explored the impact of embedding a thin, UV-bright disk into
disk+bulge systems like the ones we have modeled here for NGC
891. Given this recent development, it is relevant to compare them to
the models we have created here.

In Figure \ref{fig:inclexteffects} we overplot a combined disk+bulge
attenuation function derived for a smooth, axisymmetric RT model of NGC 891 from
\citet{Tuffs04} (green dashed lines). This model features a thin,
UV-bright disk in addition to the bulge, thick disk, and dust disk components.
 While all the disk components are represented by smooth
double-exponentials, the effects of clumpy star formation in the
thin disk are approximated by artificially reserving a fraction of the input
UV luminosity to be re-emitted by the dust grains. Spiral arms are not
included in either the dust or light components. The general form of the inclination dependence
comes from \citet{Driver08}, while the prescription for combining the
bulge and disk attenuation into a total attenuation for NGC 891 follows the method and data given in
\citet{Tuffs04}. 

This model does a much better job than either the
\citet{Verheijen01b} or \citet{Giovanelli94} parameterizations, and
shows that the addition of a star-forming, thin disk is roughly equivalent to adding clumpy dust in
spiral arms for our models with higher attenuation. This likely implies that modeling either just a thin disk
{\it or} spiral arms and clumpy dust is not enough to constrain the total
attenuation. 

We can robustly conclude that, for galaxies with dust distributions like
NGC 891, previous estimates of attenuation corrections to smooth
stellar light distribution were too large. While it is tempting to
infer that therefore so too were the resulting estimates for
star-formation rates previously over-estimated, this inference
requires a better understanding of the contribution of star-light that
is not smoothly distributed. It would not be surprising that in order
to truly determine the attenuation of spiral disks both the
detailed structure of star-forming regions and spirality must be
considered simultaneously.

\subsection{Inclination dependence to the attenuation curve}

\begin{figure*}
  \begin{centering}
    \includegraphics[scale=0.8]{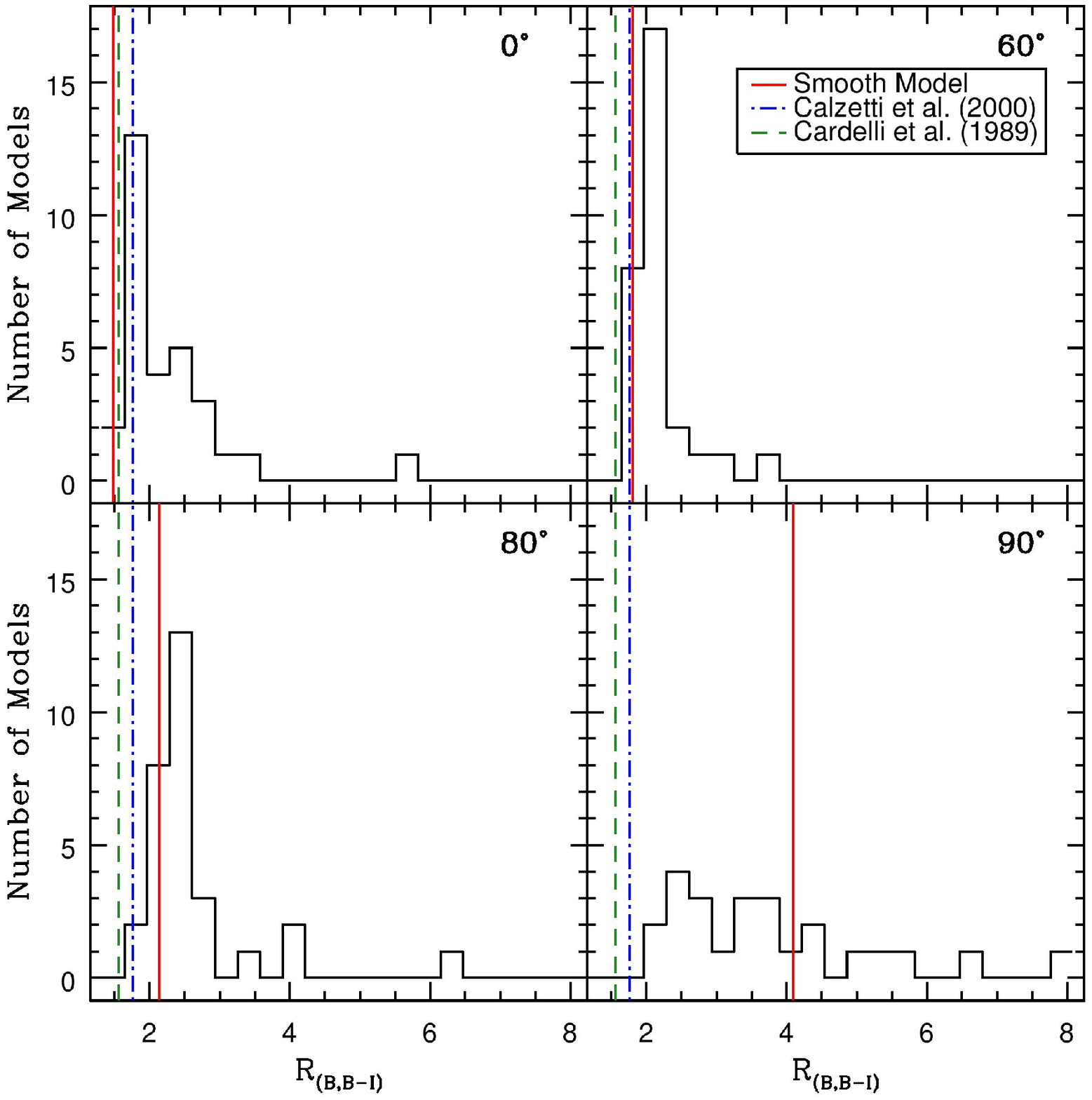}
    \caption{R$_{B,B-I}$ distribution for our 30 best-fitting clumpy
      models at i=0, 60, 80, and 90$^{\circ}$. The smooth model is
      shown as a red line. The galactic foreground screen model of
      \citet{Cardelli89} is plotted as a green dashed line. The star
      forming galaxy model of \citet{Calzetti00} is plotted as the
      blue dot-dash line. }
\label{fig:rbiplot}
\end{centering}
\end{figure*}

In Section 5.2 we considered the values of the attenuation curve
parameter R$_{B,B-I}$ at several different radii.  Here, we consider
the value that applies for the integrated light of a galaxy as a
function of inclination.  Using the same definition of the attenuation
curve parameter R$_{B,B-I}$ as before, in Figure \ref{fig:rbiplot} we
plot R$_{B,B-I}$ histograms for our 30 best fitting models at the four
inclinations of Figure \ref{fig:inclexteffects}.  We also include the
values for R$_{B,B-I}$ of the smooth model as well as literature values for a foreground
screen \citep{Cardelli89} and star-forming galaxies \citep{Calzetti00}. We first note the dependence of R$_{B,B-I}$ with
inclination, with the peak of the R$_{B,B-I}$ histogram increasing by
$\sim$0.75 between i=0$^\circ$ and 80$^{\circ}$. This dependence is
due to a combination of scattering and projection effects, as well as
the saturation of the color excess $E(B-I)$ at higher values of
$A_{B}$ \citep{Matthews01}. The smooth model shows this trend as well,
although it has a slightly lower R$_{B,B-I}$ value at all non-edge-on
inclinations.

At edge-on inclination R$_{B,B-I}$ is much more variable between our
models. This is due mainly to geometrical projection effects (from
having large amounts of relatively unattenuated flux above and below
the midplane) which cause the color excess to become very small for
some ($\sim$5) models. For these models R$_{B,B-I}$ becomes very
large. Rather than dilute the meaning of the figure we choose not to
plot these models. Even excluding the models with the largest values
of R$_{B,B-I}$, there is a much larger dispersion in the attenuation curve parameter at edge-on inclination, with a general preference toward larger
values of R$_{B,B-I}$. This is true even for the smooth
model, which at lower inclinations has a lower-than-average
R$_{B,B-I}$ but at 90$^{\circ}$ has a grayer attenuation curve than most of the
non-axisymmetric models.

The peak of the R$_{B,B-I}$ histogram occurs at larger values than the
foreground screen at all inclinations, from a difference of $\sim$0.25
at face-on inclinations to $\sim$1 at 80$^{\circ}$. This means that
for a given attenuation the internal reddening is grayer (i.e. less
wavelength-dependent) than expected for a simple screen. We find good
agreement between the \citet{Calzetti00} attenuation model and our
simulations at face-on inclinations, but at larger inclinations our RT
models have systematically larger values of R$_{B,B-I}$. This
difference is likely due to selection effects in the empirical
estimates, both because there are more galaxies with inclinations $<
60^\circ$ than there are with i $> 80^\circ$, and the
\citet{Calzetti00} sample is selected based partly on high UV emission
which is most likely to be from galaxies at or near face-on
orientation. At edge-on inclinations the attenuation curve has
a low dependence on color, but the grayness of the curve is highly
model-dependent.

\section{Conclusion}
\label{sec:conclusion}

We have optimized a 3D Monte Carlo scattered light RT model to HST
F450W and F814W images (essentially B and I band) of NGC 891. Our
model is the first RT model involving realistic dust clumping and
spirality to be quantitatively fit to imaging data. Dust is treated using a
fractal algorithm, while spirality is included using a logarithmic
spiral parameterization with a coarsely adjustable arm-interarm width
ratio.

In order to enhance the effects of non-axisymmetric dust we
constructed $\Delta \aeff$ maps using smooth models from the
literature, then used the coefficients of a high-order shapelet image
reconstruction to compare our clumpy models with the data. Computing
Monte Carlo models at a high resolution is computationally expensive,
so we used a genetic algorithm to find 'good' solutions in our
many-dimensional parameter space in a reasonable amount of time. The
efficiency of the genetic algorithm also allowed us to include a wide
range of parameter values in order to fully probe our parameter space.

We computed $30 \times 50$-generation runs of the genetic algorithm
and find a fairly wide variety of parameter values in the best-fitting
models. Some of this variety is likely due to complex
interdependencies of the free parameters, (e.g. central disk surface
brightness, scale length, and scale height for dust and stars), while
the rest is the result of degeneracies from choosing such a large
parameter space.

Despite these degeneracies, the best-fitting models from our 30
genetic algorithm runs do a credible job at mimicking the general
features of the edge on profile of NGC 891. Perhaps more importantly,
our models appear to be realistic simulations of \textit{face-on}
galaxies as well. While some of this is due to our more advanced
treatment of spirality, which automatically creates more
realistic-looking spiral features than previous work, it is important
to note that the amount of spirality and clumping were free
parameters; in 29 out of 30 runs our algorithm chose minima with clear
spirality (despite not constraining the direction of the spiral structure) and the median fraction of dust placed in clumps is 58\%. This is
the first quantitative measurement of the clumpiness of dust in edge-on spiral
galaxies that is not dependent on empirically matching IR SEDs but rather on observable dust morphology, and is in general
agreement with clumpy fractions obtained through the SED fitting method.

The preference for visible spiral features and clumpy
dust has significant implications for the face-on attenuation in spiral
galaxies. We find that while overall our models prefer low (A$_B$ and
$A_I < 1$) face-on attenuation, clumping and spirality
produce localized regions of enhanced dust absorption, creating Type
II surface brightness profiles as well as very small ($\lesssim 10\%$)
high attenuation filling factors. These clumps result in
larger total attenuation than a smooth model at low
inclinations while having comparable attenuation when viewed edge-on,
where the clumps overlap to more closely mimic a smooth dust
distribution.

Additionally, we probed the effect of the dust when viewed as a
foreground screen on background light sources. While still highly
concentrated in spiral arms, we find significantly higher optical depths
than would be suspected when only considering the attenuation
--- on average 1.2 mag larger in the B band and 0.5 mag
larger in the I band at one scale-length. The discrepancy between the
optical depths and the attenuation comes from both the admixture of
dust and starlight as well as the effect of scattering. The optical depths of our models are broadly in agreement with some
measurements from the literature but not others, which may be a result
of a large range in the optical depths in real galaxies.

We find that our models predict a blue bulge and red disk for NGC 891
when viewed face-on, regardless of the attenuation. The
smooth, axisymmetric fiducial model from the literature which we use to construct our $\Delta
\aeff$ maps also has this unusual morphology, and while we use bulge
parameters from the the literature the red disk is present in our
models even though we allowed the central disk emissivities and
scale-lengths and heights to vary. While perhaps not common, an
inspection of surface brightness profiles of a sample of face-on
galaxies shows that blue, star-forming cores and red disks do exist in intermediate-type spirals. However, we observe
neither of these features in the Milky Way, despite the frequently
claimed similarity between NGC 891 and our own galaxy.

Even with the increased face-on attenuation in the clumpy models
relative to smooth dust models, virtually all of our simulations have
smaller attenuations at all inclinations than would be expected
given corrections based on simple radiative transfer models and
empirical formulae in common use in the literature.  This result
indicates that TF studies need reduced attenuation corrections by
typically 0.5 mag in B band at i=60$^\circ$, and 0.25 mag in I at the
same inclination. Inclination corrections based on more advanced RT
models of edge-on spiral galaxies, however, do fit some of our dustier
simulations very well and point to the need to include star-forming
regions as well as spirality in future efforts.

Our models also predict a significantly grayer attenuation curve than
found for both a simple foreground screen of dust and an attenuation curve derived from UV-bright star forming galaxies. We also
find that the attenuation curve becomes $\sim$40\% grayer with
inclination, something not considered in currently favored attenuation models. Both the large amount of edge-on variability of the
models and the gradual systematic increase in R$_{B,B-I}$ with
inclination are a strong caution for extrapolating empirically
determined attenuation curves to objects with high inclinations.

We have not been able to produce clumpy dust models with the realistic
high-latitude dust ``chimneys'' seen in the data, although it is
unlikely including them would significantly change our results due to
their low spatial frequency and small size. It is unclear whether the
lack of high-latitude dust in our models is due to a poor choice of
fitness function or the fractal geometry, although the fractal
algorithm is certainly not optimized to produce extended dust
``fingers''. We find that the shapelet coefficients of the data have
fewer extreme (large or small) coefficient values as well as a better
balance of power between odd and even coefficients than most of the
clumpy models; these observations will help us to improve our fitness
metric.

Finding a method for computing model fitnesses more sensitive to
small-scale structures would help improve the reproduction of dust
substructure regardless of the clumping formalism. Multi-stage fitting
processes may be useful to reduce the size of parameter space; global
parameters (like central brightness, scale-length, and spirality)
could be determined from smooth, lower resolution models covering the
whole galaxy then fed into a much higher resolution model where the
only free parameters relate to the dust density and clumping.

In producing quantitative RT modeling which includes spirality and
clumping we have made the first foray into a new generation of highly
detailed simulations that produce images capable of resembling real
galaxy morphologies at any inclination. Extending clumpy,
non-axisymmetric models based on our prototype to larger fields of
view and large wavelength baselines, we expect to be able to constrain
the 3D structure of spiral disks using physically realistic
simulations as well as more accurately predict thermal dust
re-emission in the mid-infrared. Additionally, the increase in resolution of far-IR images
provided by the {\it Herschel Space Observatory} will enable the next
generation of SED models to fit SEDs to multiple positions of individual
galaxies, where the signatures of non-axisymmetric structures will be
more difficult to average out. By creating a realistic form of spiral structure and dust clumping
we provide a blueprint for future work in this area. 

\acknowledgements{This research was supported by NSF AST-1009471. We
  thank Chris Howk and Bob Benjamin for useful discussions. We also thank an
  anonymous referee for many useful comments and suggestions which improved
  the quality of this work. We acknowledge use of the
  HyperLeda database (http://leda.univ-lyon1.fr). All of the data
  presented in this paper were obtained from the Multimission Archive
  at the Space Telescope Science Institute (MAST). STScI is operated
  by the Association of Universities for Research in Astronomy, Inc.,
  under NASA contract NAS5-26555. Support for MAST for non-HST data is
  provided by the NASA Office of Space Science via grant NNX09AF08G
  and by other grants and contracts.}


\begin{thebibliography}{}
\bibitem[Baes et al.(2003)]{Baes03} Baes, M., et al.\ 2003, \mnras, 343, 1081 
\bibitem[Baes et al.(2011)]{Baes11} Baes, M., Verstappen, J., 
De Looze, I., et al.\ 2011, \apjs, 196, 22 
\bibitem[Bahcall(1983)]{Bahcall83} Bahcall, J.~N.\ 1983, \apj, 267, 52 
\bibitem[Bershady(1995)]{Bershady95} Bershady, M.~A.\ 1995, \aj, 109, 87 
\bibitem[Bershady et al.(2010a)]{Bershady10a} Bershady, M.~A., 
Verheijen, M.~A.~W., Swaters, R.~A., Andersen, D.~R., Westfall, K.~B., 
\& Martinsson, T.\ 2010, \apj, 716, 198 
\bibitem[Bershady et al.(2010b)]{Bershady10b} Bershady, M.~A., 
Verheijen, M.~A.~W., Westfall, K.~B., Andersen, D.~R., Swaters, R.~A., 
\& Martinsson, T.\ 2010, \apj, 716, 234 
\bibitem[Bianchi(2008)]{Bianchi08} Bianchi, S.\ 2008, \aap, 490, 461 
\bibitem[Bianchi et al.(1996)]{Bianchi96} Bianchi, S., Ferrara, A., \&
  Giovanardi, C.\ 1996, \apj, 465, 127 
\bibitem[Bianchi \& Xilouris(2011)]{Bianchi11} Bianchi, S., \& Xilouris, E.~M.\ 2011, \aap, 531, L11 
\bibitem[Bosch(2010)]{Bosch10} Bosch, J.\ 2010, \aj, 140, 870 
\bibitem[Calzetti et al.(2000)]{Calzetti00} Calzetti, D., Armus, L., Bohlin, R.~C., Kinney, A.~L., Koornneef, J., \& Storchi-Bergmann, T.\ 2000, \apj, 533, 682 
\bibitem[Calzetti et al.(1994)]{Calzetti94} Calzetti, D., Kinney, 
A.~L., \& Storchi-Bergmann, T.\ 1994, \apj, 429, 582 
\bibitem[Cardelli et al.(1989)]{Cardelli89} Cardelli, J.~A., Clayton, G.~C., \& Mathis, J.~S.\ 1989, \apj, 345, 245 
\bibitem[Dalcanton et al.(2004)]{Dalcanton04} Dalcanton, J.~J., 
Yoachim, P., \& Bernstein, R.~A.\ 2004, \apj, 608, 189 
\bibitem[de Jong(1996a)]{deJong96} de Jong, R.~S.\ 1996, \aap, 313, 377 
\bibitem[de Jong(1996b)]{deJong96fits} de Jong, R.~S.\ 1996, Journal 
of Astronomical Data, 2, 1 
\bibitem[Domingue et al.(2000)]{Domingue00} Domingue, D.~L., Keel, 
W.~C., \& White, R.~E., III 2000, \apj, 545, 171 
\bibitem[Draine(2003)]{Draine03} Draine, B.~T.\ 2003, \apj, 598, 
1017 
\bibitem[Driver et al.(2008)]{Driver08} Driver, S.~P., Popescu, C.~C., Tuffs, R.~J., Graham, A.~W., Liske, J., \& Baldry, I.\ 2008, \apjl, 678, L101 
\bibitem[Elmegreen(1997)]{Elmegreen97} Elmegreen, B.~G.\ 1997, \apj, 477, 196 
\bibitem[Elmegreen \& Elmegreen(1982)]{Elmegreen82} Elmegreen, D.~M., \& Elmegreen, B.~G.\ 1982, \mnras, 201, 1021 
\bibitem[Elmegreen \& Falgarone(1996)]{Elmegreen96} Elmegreen, B.~G., \& Falgarone, E.\ 1996, \apj, 471, 816 
\bibitem[Freeman(1970)]{Freeman70} Freeman, K.~C.\ 1970, \apj, 
160, 811
\bibitem[Garcia-Burillo et al.(1992)]{GarciaBurillo92} Garcia-Burillo, S., Guelin, M., Cernicharo, J., \& Dahlem, M.\ 1992, \aap, 266, 21
\bibitem[Garcia-Burillo \& Guelin(1995)]{GarciaBurillo95} Garcia-Burillo, S., \& Guelin, M.\ 1995, \aap, 299, 657 
\bibitem[Gerard(1973)]{Gerard73} Gerard, E.\ 1973, \aap, 28, 95 
\bibitem[Giovanelli et al.(1994)]{Giovanelli94} Giovanelli, R., Haynes, M.~P., Salzer, J.~J., Wegner, G., da Costa, L.~N.,  \& Freudling, W.\ 1994, \aj, 107, 2036 
\bibitem[Gordon et al.(1997)]{Gordon97} Gordon, K.~D., Calzetti,  D., \& Witt, A.~N.\ 1997, \apj, 487, 625 
\bibitem[Holwerda et al.(2005)]{Holwerda05} Holwerda, B.~W., Gonzalez, R.~A.,
  Allen, R.~J., \& van der Kruit, P.~C.\ 2005, \aj, 129, 1396
\bibitem[Holwerda et al.(2009)]{Holwerda09} Holwerda, B.~W., Keel, 
W.~C., Williams, B., Dalcanton, J.~J., 
\& de Jong, R.~S.\ 2009, \aj, 137, 3000 
\bibitem[Howk \& Savage(1997)]{Howk97} Howk, J.~C., \& Savage, B.~D.\ 1997, \aj, 114, 2463 
\bibitem[Howk \& Savage(1999)]{Howk99} Howk, J.~C. \& Savage, B.~D.\ 1999, \aj, 117, 2077
\bibitem[Howley et al.(2008)]{Howley08} Howley, K.~M., Geha, M., Guhathakurta, P., Montgomery, R.~M., Laughlin, G., \& Johnston, K.~V.\ 2008, \apj, 683, 722 
\bibitem[Indebetouw et al.(2006)]{Indebetouw06} Indebetouw, R., Whitney, B.~A., Johnson, K.~E., \& Wood, K.\ 2006, \apj, 636, 362 
\bibitem[Joung \& Mac Low(2006)]{Joung06} Joung, M.~K.~R., \& Mac Low, M.-M.\ 2006, \apj, 653, 1266 
\bibitem[Kamphuis et al.(2007)]{Kamphuis07} Kamphuis, P., Holwerda, B.~W., Allen, R.~J., Peletier, R.~F., \& van der Kruit, P.~C.\ 2007, \aap, 471, L1 
\bibitem[Kelly \& McKay(2004)]{Kelly04} Kelly, B.~C., \& McKay, T.~A.\ 2004, \aj, 127, 625 
\bibitem[Kennicutt(1981)]{Kennicutt81} Kennicutt, R.~C., Jr.\ 1981, \aj, 86, 1847 
\bibitem[Keppel et al.(1991)]{Keppel91} Keppel, J.~W., Dettmar, R.-J., Gallagher, J.~S., III, \& Roberts, M.~S.\ 1991, \apj, 374, 507 
\bibitem[Kuchinski et al.(1998)]{Kuchinski98} Kuchinski, L.~E., Terndrup, D.~M., Gordon, K.~D., \& Witt, A.~N.\ 1998, \aj, 115, 1438 
\bibitem[Kuijken(2006)]{Kuijken06} Kuijken, K.\ 2006, \aap, 456, 827 
\bibitem[Kylafis \& Bahcall(1987)]{Kylafis87} Kylafis, N.~D., \& Bahcall,
  J.~N.\ 1987, \apj, 317, 637 
\bibitem[Lewis et al.(2009)]{Lewis09} Lewis, N.~K., Cook, 
T.~A., Wilton, K.~P., et al.\ 2009, \apj, 706, 306 
\bibitem[Mapelli et al.(2008)]{Mapelli08} Mapelli, M., Moore, B., \& Bland-Hawthorn, J.\ 2008, \mnras, 388, 697 
\bibitem[Massey et al.(2004)]{Massey04} Massey, R., Refregier, A., Conselice,
  C.~J., David, J., \& Bacon, J.\ 2004, \mnras, 348, 214 
\bibitem[Mathis et al.(1977)]{Mathis77} Mathis, J.~S., Rumpl, 
W., \& Nordsieck, K.~H.\ 1977, \apj, 217, 425 
\bibitem[Mathis et al.(2002)]{Mathis02} Mathis, J.~S., Whitney, B.~A., \& Wood, K.\ 2002, \apj, 574, 812 
\bibitem[Matthews \& Wood(2001)]{Matthews01} Matthews, L.~D., \& Wood, K.\ 2001, \apj, 548, 150 
\bibitem[Mihos et al.(1999)]{Mihos99} Mihos, J.~C., Spaans, M., 
\& McGaugh, S.~S.\ 1999, \apj, 515, 89 
\bibitem[Misiriotis et al.(2000)]{Misiriotis00} Misiriotis, A., Kylafis, N.~D., Papamastorakis, J., \& Xilouris, E.~M.\ 2000, \aap, 353, 117 
\bibitem[Misiriotis \& Bianchi(2002)]{Misiriotis02} Misiriotis, A., \& Bianchi, S.\ 2002, \aap, 384, 866 
\bibitem[Oosterloo et al.(2007)]{Oosterloo07} Oosterloo, T., 
Fraternali, F., \& Sancisi, R.\ 2007, \aj, 134, 1019 
\bibitem[\protect\citeauthoryear{Paturel et al.}{2003}]{Paturel03} Paturel G.,
  Petit C., Prugniel P., Theureau G., Rousseau J., Brouty M., Dubois P.,
  Cambr{\'e}sy L., 2003, A\&A, 412, 45 
\bibitem[Perrin et al.(1995)]{Perrin95} Perrin, J.-M., Darbon, S., \& Sivan, J.-P.\ 1995, \aap, 304, L21 
\bibitem[Pierini et al.(2002)]{Pierini02} Pierini, D., Majeed, 
A., Boroson, T.~A., \& Witt, A.~N.\ 2002, \apj, 569, 184 
\bibitem[Pierini et al.(2004)]{Pierini04} Pierini, D., Gordon, K.~D., Witt, A.~N., \& Madsen, G.~J.\ 2004, \apj, 617, 1022 
\bibitem[Popescu et al.(2000)]{Popescu00} Popescu, C.~C., Misiriotis, A., Kylafis, N.~D., Tuffs, R.~J., \& Fischera, J.\ 2000, \aap, 362, 138 
\bibitem[Popescu et al.(2011)]{Popescu11} Popescu, C.~C., Tuffs, R.~J., Dopita, M.~A., et al.\ 2011, \aap, 527, A109 
\bibitem[Rand et al.(1990)]{Rand90} Rand, R.~J., Kulkarni, 
S.~R., \& Hester, J.~J.\ 1990, \apjl, 352, L1 
\bibitem[Refregier(2003)]{Refregier03} Refregier, A.\ 2003, \mnras, 338, 35 
\bibitem[Sankrit \& Wood(2001)]{Sankrit01} Sankrit, R., \& Wood, K.\ 2001, \apj, 555, 532 
\bibitem[Schweizer(1976)]{Schweizer76} Schweizer, F.\ 1976, \apjs, 31, 313 
\bibitem[Sirianni et al.(2005)]{Sirianni05} Sirianni, M., et al.\ 2005, \pasp, 117, 1049 
\bibitem[Strong(1978)]{Strong78} Strong, A.~W.\ 1978, \aap, 66, 205 
\bibitem[Swaters et al.(1997)]{Swaters97} Swaters, R.~A., 
Sancisi, R., \& van der Hulst, J.~M.\ 1997, \apj, 491, 140 
\bibitem[Tuffs et al.(2004)]{Tuffs04} Tuffs, R.~J., Popescu, C.~C., V{\"o}lk, H.~J., Kylafis, N.~D., \& Dopita, M.~A.\ 2004, \aap, 419, 821 
\bibitem[Tully \& Fouque(1985)]{Tully85} Tully, R.~B., \& Fouque, P.\ 1985,
  \apjs, 58, 67 
\bibitem[Tully et al.(1998)]{Tully98} Tully, R.~B., Pierce, 
M.~J., Huang, J.-S., Saunders, W., Verheijen, M.~A.~W., 
\& Witchalls, P.~L.\ 1998, \aj, 115, 2264 
\bibitem[van der Kruit(1984)]{vanderKruit84} van der Kruit, P.~C.\ 1984, \aap, 140, 470 
\bibitem[Verheijen(2001)]{Verheijen01b} Verheijen, M.~A.~W.\ 2001, 
\apj, 563, 694 
\bibitem[Verheijen \& Sancisi(2001)]{Verheijen01a} Verheijen, M.~A.~W., \&
  Sancisi, R.\ 2001, \aap, 370, 765 
\bibitem[White(1979)]{White79} White, R.~L.\ 1979, \apj, 229, 954 
\bibitem[White et al.(2000)]{White00} White, R.~E., III, Keel, W.~C., \& Conselice, C.~J.\ 2000, \apj, 542, 761 
\bibitem[Witt(1977)]{Witt77} Witt, A.~N.\ 1977, \apjs, 35, 1 
\bibitem[Witt et al.(1992)]{Witt92} Witt, A.~N., Thronson, H.~A., Jr., \&
  Capuano, J.~M., Jr.\ 1992, \apj, 393, 611 
\bibitem[Witt \& Gordon(1996)]{Witt96} Witt, A.~N., \& Gordon, K.~D.\ 1996, \apj, 463, 681 
\bibitem[Witt \& Gordon(2000)]{Witt00} Witt, A.~N., \& Gordon, K.~D.\ 2000, \apj, 528, 799 
\bibitem[Wood et al.(1996)]{Wood96} Wood, K., Bjorkman, J.~E., 
Whitney, B.~A., \& Code, A.~D.\ 1996, \apj, 461, 828 
\bibitem[Wood \& Jones(1997)]{Wood97} Wood, K., \& Jones, T.~J.\ 1997, \aj, 114, 1405 
\bibitem[Wood \& Reynolds(1999)]{Wood99a} Wood, K., \& Reynolds, R.~J.\ 1999, \apj, 525, 799 
\bibitem[Wood et al.(1999)]{Wood99b} Wood, K., Crosas, M., 
\& Ghez, A.\ 1999, \apj, 516, 335 
\bibitem[Wood \& Loeb(2000)]{Wood00}  Wood, K., \& Loeb, A.\ 2000, \apj, 545, 86 
\bibitem[Wood et al.(2005)]{Wood05} Wood, K., Haffner, L.~M., 
Reynolds, R.~J., Mathis, J.~S., \& Madsen, G.\ 2005, \apj, 633, 295 
\bibitem[Wood et al.(2010)]{Wood10} Wood, K., Hill, A.~S., 
Joung, M.~R., Mac Low, M.-M., Benjamin, R.~A., Haffner, L.~M., Reynolds, 
R.~J., \& Madsen, G.~J.\ 2010, \apj, 721, 1397
\bibitem[Xilouris et al.(1998)]{Xilouris98} Xilouris, E.~M., Alton, P.~B., Davies, J.~I., Kylafis, N.~D., Papamastorakis, J., \& Trewhella, M.\ 1998, \aap, 331, 894 
\bibitem[Xilouris et al.(1999)]{Xilouris99} Xilouris, E.~M., Byun, Y.~I., Kylafis, N.~D., Paleologou, E.~V., \& Papamastorakis, J.\ 1999, \aap, 344, 868 
\bibitem[Yusef-Zadeh et al.(1984)]{Yusef-Zadeh84} Yusef-Zadeh, F., 
Morris, M., \& White, R.~L.\ 1984, \apj, 278, 186 
\end{thebibliography}
\end{document}